%% file: final-draft.tex
\documentclass[journal, 11pt,one column,twosides]{IEEEtran} 
\usepackage[utf8]{inputenc} 
\usepackage[T1]{fontenc}
\usepackage{url}
\usepackage{ifthen}
\usepackage{empheq} 
\usepackage{cite}
\usepackage{amsmath} 
\usepackage{graphicx}
\usepackage{caption}
\usepackage{subcaption}
\usepackage{color}
\usepackage{epsfig}

\usepackage{amssymb}
\usepackage{amsmath}
\usepackage{amsthm}
\usepackage{latexsym}
\usepackage{setspace}
\usepackage{bbm}
\usepackage{arydshln}
\usepackage{textcomp}
\usepackage{amsfonts}
\usepackage{blindtext}
\usepackage{enumerate}
\usepackage{upgreek}
\usepackage[bottom]{footmisc}
\usepackage{tikz}
\newcommand{\qeed}{\hfill $\blacksquare$}

\include{preamble}
\usepackage{secdot}
\allowdisplaybreaks
\sectiondot{subsection}
\sectiondot{subsubsection}
\usepackage{mleftright}

\usepackage{chngcntr}
\usepackage{fancyhdr}

\fancypagestyle{firstpage}
{
    \fancyhead[R]{To appear, {\it IEEE Transactions on Information Theory}, 2022}
}
\begin{document}

\newcommand\MC{{ \ - \!\!\circ\!\! - \ }}

\theoremstyle{theorem}
\newtheorem{theorem}{Theorem}
\newtheorem{corollary}[theorem]{Corollary}
\newtheorem{lemma}[theorem]{Lemma}
\newtheorem{proposition}[theorem]{Proposition}
\theoremstyle{definition}
\newtheorem{definition}{Definition}

\title{Distribution Privacy Under Function Recoverability} 

\author{Ajaykrishnan Nageswaran and Prakash Narayan$^\dag$ }
\maketitle
{\renewcommand{\thefootnote}{}
\footnotetext{
$^\dag$A. Nageswaran and P. Narayan are with the Department of
Electrical and Computer Engineering and the Institute for Systems
Research, University of Maryland, College Park, MD 20742, USA.
E-mail: \{ajayk, prakash\}@umd.edu. This work was supported by the U.S.
National Science Foundation under Grant CCF $1527354$.
}
}

\maketitle

\maketitle

\thispagestyle{firstpage}

\begin{abstract}
A user generates $n$ independent and identically distributed data
random variables with a probability mass function that must be guarded from a querier. The querier
must recover, with a prescribed accuracy, a given function of the data from each of $n$ independent and identically distributed query responses upon eliciting them from the user. The user chooses the data probability mass function
and devises the random query responses to maximize distribution privacy
as gauged by the (Kullback-Leibler) divergence between the former and the querier's best estimate of it based on the $n$ query responses. Considering an arbitrary function, a basic achievable lower
bound for distribution privacy is provided that does not depend on $n$ and corresponds to worst-case privacy. Worst-case privacy equals the logsum cardinalities of inverse atoms under the given function, with the number of summands decreasing as the querier recovers the function with improving accuracy. Next, upper (converse) and lower (achievability) bounds for distribution privacy, dependent on $n,$ are developed. The former improves upon
worst-case privacy and the latter does so under suitable assumptions; both converge to it as $n$ grows. The converse and achievability proofs identify explicit strategies for the user and the querier.
\end{abstract}

\section{Introduction}
\label{sec:intro}
A user generates data represented by independent and identically distributed (i.i.d.)  repetitions of a finite-valued random variable (rv) with an underlying
probability mass function (pmf) that the user selects
and seeks to keep private from a querier who wishes to compute a given function
of the data. For this purpose, the querier elicits user-provided i.i.d. query responses
that are suitably randomized versions of the data. The user devises
the query responses so as to allow the querier to recover the function value from
every query response with a prescribed accuracy, while maximizing privacy of the
data pmf. 

Specifically, the user chooses
a data pmf $P_X$ of the rv $X$ and produces $n\geq 1$ i.i.d.\footnotemark\footnotetext{For $n=1$, clearly ``i.i.d.'' is redundant.} query responses as the outputs of a stochastic matrix $W$, with inputs being $n$ i.i.d. repetitions of $X$, such that the querier can recover the function value from each query response with probability at least $\rho$, $0\leq\rho\leq 1$.
The querier picks an estimator $\widehat{P}_n$ for the pmf $P_X$ based on the $n$ query responses. Our notion of \textit{distribution $\rho$-privacy for $n$ query responses} entails the (Kullback-Leibler) divergence between $P_X$ and the querier's estimate of it being
maximized and minimized, respectively, with respect to $\left(W,P_X\right)$
and $\widehat{P}_n$. The order of optimization allows $\widehat{P}_n$
to depend on $W$, and $P_X$ on $\widehat{P}_n.$ This setting can be viewed also as that with $n$ queriers to each of whom the user provides a query response from which the function value can be recovered with probability not less than $\rho;$ the queriers then cooperate to estimate $P_X$ from their pooled $n$ i.i.d. query responses. 

A suggestive interpretation of distribution $\rho$-privacy entails Nature generating its secrets according to a pmf $P_X=P_X^*,$ say, that is hardest for a mortal querier to fathom under the function recoverability requirement above. On account of the continuity of $D\left(P_X\big|\big|\widehat{P}_n\right)$ with respect to $\left(P_X,\widehat{P}_n\right)$ when the support of $P_X$ is contained in that of $\widehat{P}_n$, a user -- constrained unlike Nature -- chooses a feasible $P_X$ that is proximate to $P_X^*$. Potential futuristic applications include: an AI-driven financial trader who reveals trading preferences through
daily actions (recoverability of function values) but seeks to guard
the workings of an underlying probabilistic algorithm (distribution privacy); and IoT sensors that must recover user commands for execution (recoverability) but without
the details of user habits being compromised (distribution privacy). An instance of the second category would occur when a user's predilections for Smart TV~\cite{Michele14} programs must not be compromised when program requests are made to a service provider. 

In the new problem formulation above, our main results are as follows. Considering an arbitrary function, we first provide a basic achievable lower
bound for distribution $\rho$-privacy that does not depend on $n$ and represents ``worst-case'' privacy. This worst-case privacy is characterized as a function of $\rho$ and equals the logsum cardinalities of inverse atoms under the given function, with the number of summands decreasing as $\rho$ increases from $0$ to $1.$ We introduce specialized strategies: ``sparse pmf'' and ``locally identical query response'' for the user and ``locally uniform estimator'' for the querier. Primitive forms of these strategies play a role in establishing the characterization of worst-case privacy. We then provide upper (converse) and lower (achievability) bounds for distribution $\rho$-privacy -- the former for every $n$ and the latter for all $n$ suitably large. These bounds are shown to be asymptotically tight, converging to worst-case privacy with increasing $n$ in the worthwhile regime $0.5<\rho\leq 1$. A key facilitating step in our converse proof is the recognition that with the querier's strategy restricted to a locally uniform estimator, the user can make do with a sparse pmf and locally identical query response without sacrificing distribution privacy. The roles of restriction and adequacy are reversed in the achievability proof. Significantly, the converse and achievability proofs spell out specifications of explicit user and querier actions. Preliminary versions of this work are in~\cite{Nages20} for binary-valued functions, and in~\cite{Nages21}.

An extensive body of prior work exists on distribution estimation in the context of data privacy 
(cf. e.g.,~\cite{Duncan86},~\cite{Fienberg98},\\\cite{Chan12},\cite{Hsu12},\cite{Bassily15} and references therein). Privacy constraints, when explicitly present, are dominantly in the sense of differential privacy (cf.~\cite{Dwork06, DworkSmith06}). In a series~\cite{Duchi16},~\cite{Kairouz16},~\cite{Ye17},~\cite{Pastore18}, samples of user data are generated according to a probability 
distribution from a given family of distributions. A randomized version of each of the samples is made available 
to a querier, with the randomization mechanism being differentially private of a given privacy level. The querier then forms an estimate of the user's distribution based on the differentially private query responses. Considering the minmax of
the expected $\ell_2$-distance between the user's distribution and the querier's estimate (maximum and minimum, respectively, over possible user distributions and querier estimators), its minimum is examined over all the differentially private randomization mechanisms of the 
given level. In Section~\ref{sec:disc}, we broach the idea of examining our present work in the context of this approach. In another line of work \cite{Cover72},~\cite{Krich98},~\cite{Braess02},~\cite{Braess04},~\cite{Paninski04}, \textit{sans privacy considerations} but relevant to ours, data samples are 
generated according to a distribution from a given set. The objective is for the user to select a distribution that resists estimation
by the best estimator under a divergence cost. Investigated accordingly are the
maximum and minimum, respectively, over user distributions and estimators of the expected divergence between the user distribution and the estimate. Also, see~\cite{Kamath15} for a similar minmax study under other loss measures. These approaches to distribution estimation, with or without privacy, do not require computation of a function of the underlying data.

Data privacy in various forms (rather than privacy of data distribution) is the subject of another vast body of work. Considerations include maximizing data utility under privacy constraints (for instance, differential privacy, privacy based on information measures, and data estimation error probability); examples can be found in~\cite{Hardt10},~\cite{Smith11},~\cite{Bassily15},~\cite{Asoodeh16},~\cite{Geng16},~\cite{Asoodeh18}. Likewise, data utility-privacy tradeoffs are analyzed also by maximizing privacy for a given level of utility~\cite{Rebollo10},~\cite{Calmon12},~\cite{Makhdoumi13},~\cite{Sankar13},~\cite{Huang17},~\cite{Liao18}. Our prior work~\cite{Nages19} is of the nature of the latter where, under an explicit constraint on function recoverability, data privacy is maximized. Specifically, for finite-valued data and query responses, upon limiting ourselves to privacy as a probability of error and recoverability as a (pointwise) conditional probability of error, we obtain utility-privacy tradeoffs for single and multiple query responses. Our present work is in this spirit: maximizing the privacy of data distribution under a function recoverability constraint.

Our model for distribution $\rho$-privacy is described in Section~\ref{sec:prelim} which then characterizes the resulting worst-case privacy, as demonstrated by an achievability proof. Section~\ref{sec:main_thm} defines specialized user and querier strategies, and states the converse and achievability theorems which are proved in Section~\ref{sec:proofs}. The concluding Section~\ref{sec:disc} provides a heuristic explanation of the characterization of distribution $\rho$-privacy, and cites unanswered questions including one that touches on local differential privacy.

\section{Preliminaries and Worst-Case Privacy} 
\label{sec:prelim}

A user generates data represented by i.i.d. rvs $X_1,\ldots,X_n$, $n\geq 1$,
with pmf $P_X$ and with $X_1$ taking values in a finite set $\cX$ of cardinality $|\cX|=r\geq 2$. Consider a given
mapping $f:\cX\rightarrow\cZ=\{0,1,\ldots,k-1\},$ $2\leq k\leq r$. Let $f^{-1}$ denote the corresponding preimage mapping with $f^{-1}(z)=\left\{x\in\cX:f(x)=z\right\}, \ z\in\cZ$. For realizations $X_1=x_1,\ldots,X_n=x_n$, a querier -- who does not know $x^n\triangleq \left(x_1,\ldots,x_n\right)$ or
$P_X$ -- wishes to compute $f(x_1),\ldots,f(x_n)$ from $\cZ$-valued rvs $Z_1,\ldots,Z_n$,
termed \textit{query responses} (QRs), that are provided by the user. Each QR $Z_t$,
$t=1,\ldots,n$, must satisfy the following recoverability condition.\footnotemark\footnotetext{As observed in~{\cite[p. 3473, towards the end of Section II]{Nages19}}, there is no loss of generality in~\eqref{eq:rho-recov},~\eqref{eq:rho-recovW} by considering the $\rho$-QR rvs $Z_1,\ldots,Z_n$ to be $\cZ$-valued. If $Z_t,\ t=1,\ldots,n$, had an alphabet larger than $\cZ$, the querier would {\it estimate} $f(X)$ based on $Z_t$. However, the user can emulate any such estimation strategy of the querier to produce another $\cZ$-valued $\rho$-QR.}

\begin{definition}
\label{def:rho-recov}
Given $0\leq\rho\leq 1,$ a QR $Z_t$ is 
\textit{$\rho$-recoverable} ($\rho$-QR) if 
\begin{equation}
\label{eq:rho-recov}
P\left(Z_t=f(x)\big|X_t=x\right)\geq 
\rho,\hspace{2mm}x\in\cX.
\end{equation} 
\noindent Condition~\eqref{eq:rho-recov} can be written equivalently 
in terms of a stochastic matrix $W:\cX\rightarrow\cZ$ with the 
requirement 
\begin{equation}
\label{eq:rho-recovW}
W\big(f\left(x\right)|x\big)\geq \rho,\hspace{2mm}x\in\cX
\end{equation}
\noindent and such a $W,$ too, will be termed a $\rho$-QR. Note that $\rho$-recoverability 
in~\eqref{eq:rho-recov},~\eqref{eq:rho-recovW} does not depend on $P_X$. 
\end{definition}

The $\rho$-QRs
$Z_1,\ldots,Z_n$ are assumed to satisfy
\begin{align}
P\left(Z^n=z^n|X^n=x^n\right)
&\triangleq P\left(Z_1=z_1,\ldots,Z_n=z_n|X_1=x_1,\ldots,X_n=x_n\right)\nonumber\\
&=\prod\limits_{t=1}^{n} P\left(Z_t=z_t|X_t=x_t\right)\nonumber\\
&=\prod\limits_{t=1}^{n} W\left(z_t|x_t\right),\ \ x^n\in\cX^n,z^n\in\cZ^n,\label{eq:Z_iid}
\end{align} 
\noindent whereupon since $X_1,\ldots,X_n$  are i.i.d., so too are $Z_1,\ldots,Z_n$, with pmf $\left(P_XW\right)(z)\triangleq\sum\limits_{x\in\cX}P_X(x)W(z|x),\ z\in\cZ$. The user chooses the pmf $P_X$ and the $\rho$-QRs $Z_1,\ldots,Z_n$ or equivalently $W$.
The querier observes $Z_1,\ldots,Z_n$
and seeks to estimate $P_X$ by means of a suitable estimator $\widehat{P}_n:\cZ^n\rightarrow \Delta_r$, where $\Delta_r$
is the $r$-dimensional simplex associated with $\cX$.

The measure of discrepancy between the pmf $P_X$ and the querier's estimate $\widehat{P}_n$ is
\begin{equation}
\label{eq:div}
\pi_n\left(\rho,W,P_X,\widehat{P}_n\right)\triangleq 
\mathbb{E}\left[D\left(P_X\big|\big|\widehat{P}_n\left(Z^n\right)\right)\right], \ \ \ 0\leq\rho\leq 1
\end{equation}
\noindent where $D(\cdot||\cdot)$ denotes (Kullback-Leibler) divergence\footnotemark\footnotetext{All logarithms and exponentiations are with respect to the base $2.$} and  expectation is with respect to the pmf
$P_XW$. The user and querier
devise $\left(W,P_X\right)$ and $\widehat{P}_n$, respectively, to maximize and minimize 
$\pi_n\left(\rho,W,P_X,\widehat{P}_n\right)$. Our notion of distribution privacy assumes
conservatively that the querier is cognizant of the user's choice of the randomized privacy 
mechanism $W$ which depends on $0\leq\rho\leq 1$; this dependence is not displayed explicitly in the right-side of~\eqref{eq:div} so as to help contain notational growth.

\begin{definition}
\label{def:dist_privacy}
For $0\leq\rho\leq 1$, \textit{distribution $\rho$-privacy} is
\begin{equation}
\label{eq:dist-rho-priv}
\pi_n(\rho)\triangleq \sup_{\substack{W:W\left( f 
\left(x\right) | x \right)\geq\rho\\x\in\cX}}\ \inf_{\widehat{P}_n:\cZ^n\rightarrow\Delta_r}
\ \sup_{P_X\in\Delta_r}
\pi_n\left(\rho,W,P_X,\widehat{P}_n\right), \ \ \ n\geq 1
\end{equation}
\noindent where $W$ is as in~\eqref{eq:rho-recovW} and $\pi_n\left(\rho,W,P_X,\widehat{P}_n\right)$ is given by~\eqref{eq:div}.
\end{definition}

\noindent \textit{Remarks}: 
\begin{enumerate}[(i)]
\item The order of maximizations and minimization in~\eqref{eq:dist-rho-priv}
accommodates the dependence of $\widehat{P}_n$ on $W$ (and $\rho$) in providing a conservative measure of
distribution privacy. On the other hand, privacy, if gauged by $\inf\limits_{\widehat{P}_n}\sup\limits_{W}\sup\limits_{P_X}$ in~\eqref{eq:dist-rho-priv},
would be larger, in general, but would not allow the querier to be aware of the privacy
mechanism $W$.
\item We note that $\pi_n(\rho)$ in~\eqref{eq:dist-rho-priv}, if defined instead in terms of 
$\sup\limits_{W}\sup\limits_{P_X}\inf\limits_{\widehat{P}_n}$, would equal zero unrealistically. Also, reversing the roles of $P_X$ and
$\widehat{P}_n\left(Z^n\right)$ in $D\left(\cdot||\cdot\right)$ in~\eqref{eq:div},~\eqref{eq:dist-rho-priv} leads
to an unrealistic $\pi_n(\rho)=\infty$.
\item Clearly, it suffices to restrict the querier's estimators $\widehat{P}_n$ in~\eqref{eq:dist-rho-priv} to those that satisfy $\widehat{P}_n\left(z^n\right)(x)>0, \ z^n\in\cZ^n, \ x\in\cX$. If the querier were to assign $\widehat{P}_n\left(z^n\right)(x)=0$ to any $x\in\cX,$ the user can choose $P_X(x)>0$ for that $x$ (since by~\eqref{eq:dist-rho-priv}, $P_X$ can depend on $\widehat{P}_n$), thereby rendering $\pi_n(\rho)=\infty$.
\end{enumerate}

A justification is in order of our model above and choice of divergence as the measure of distribution privacy in~\eqref{eq:div}. First, from a purely heuristic standpoint, for a fixed $\rho$, any meaningful privacy measure should display the qualitative feature that the associated distribution privacy is nondecreasing with decreasing ``atomicity'' of a given mapping $f:\cX\rightarrow\cZ.$ In other words, the fewer and larger the atoms induced in $\cX$ by $f^{-1},$ the better is the ability of the user to conceal a pmf $P_X$ from the querier. As will be seen below, the concept of distribution $\rho$-privacy defined in terms of divergence in~\eqref{eq:div} brings out this behaviour in precise terms and quantifies its dependence on $\rho$ and $n$. In fact, our main results in Theorems~\ref{thm:nonasymp_bnd},~\ref{thm:main-conv} and~\ref{thm:main-achiev} below depend on $f:\cX\rightarrow\cZ$ only through the sizes $\left|f^{-1}(j)\right|, \ j=0,1,\ldots,k-1.$ Thus, our divergence formulation is divulgent and also eminently tractable. While other measures of discrepancy between distributions could have been used, any reasonable choice ought to yield answers that do not veer significantly from our divergence-based results that bear out  heuristics. We emphasize that our model has features that have been biased deliberately against the user so as to make for a conservative (i.e., diminished) extent of privacy. The recoverability requirement in~\eqref{eq:rho-recov},~\eqref{eq:rho-recovW} is imposed stringently for every $t=1,\ldots,n$, rather than for only over a block of length $n$ $\rho$-QRs; the latter, in the limit $n\rightarrow\infty$, would ask only for {\it asymptotic} recoverability. Next, as assumed in~\eqref{eq:Z_iid}, the $\rho$-QR $W$ is fixed for $t=1,\ldots,n$, whereby $Z_1,\ldots,Z_n$ are rendered i.i.d. Moreover, as mentioned before Definition 2, the querier is allowed knowledge of the $\rho$-QR $W$. If the user were permitted  time-varying $\rho$-QRs
$W_t, \ t=1,\ldots,n$, it remains open whether a suitably modified definition of distribution $\rho$-privacy could lead to privacy enhancement.\\ 

Two elementary attributes of $\pi_n(\rho)$ are contained in
\vspace{0.1cm}
\begin{proposition}
\label{prop:1}
For $0\leq\rho\leq 1$, $\pi_n(\rho)$ is nonincreasing in $n\geq 1$. Furthermore,
\begin{equation}
\label{eq:pin_ub}
\pi_n(\rho)\leq \log r, \ \ \ \ n\geq 1.
\end{equation}
\end{proposition}
\vspace{0.1in}
\noindent \textbf{Proof}: To show that $\pi_{n+1}(\rho)\leq \pi_n(\rho), \ n\geq 1,$ observe by~\eqref{eq:div},~\eqref{eq:dist-rho-priv} that in $\pi_{n+1}(\rho)$, for every fixed $W$,
\begin{equation}
    \label{eq:n+1-n}
    \inf_{\widehat{P}_{n+1}:\cZ^{n+1}\rightarrow\Delta_r} \ \sup_{P_X} \ 
    \mathbb{E}\left[D\left(P_X\big|\big|\widehat{P}_{n+1}\left(Z^{n+1}\right)\right)\right]\leq
    \inf_{\widehat{P}_n:\cZ^{n+1}\rightarrow\Delta_r} \ \sup_{P_X} \ 
    \mathbb{E}\left[D\left(P_X\big|\big|\widehat{P}_n\left(Z^{n+1}\right)\right)\right]
\end{equation}
\noindent where, with an abuse of notation, a restricted estimator $\widehat{P_{n}}:\cZ^{n+1}\rightarrow\Delta_r$ yields the same estimate for all $z^{n+1}\in\cZ^{n+1}$ with common $z^n$ (thereby ignoring $z_{n+1}$). Then, noting that the expectation in the right-side of~\eqref{eq:n+1-n} with respect to $z^{n+1}\in\cZ^{n+1}$ is effectively over $z^n\in\cZ^n$, we get from~\eqref{eq:n+1-n} that 
\begin{equation}
    \label{eq:n-n}
    \inf_{\widehat{P}_{n+1}} \ \sup_{P_X} \ 
    \mathbb{E}\left[D\left(P_X\big|\big|\widehat{P}_{n+1}\left(Z^{n+1}\right)\right)\right]\leq
    \inf_{\widehat{P_{n}}} \ \sup_{P_X} \ 
    \mathbb{E}\left[D\left(P_X\big|\big|\widehat{P}_{n}\left(Z^{n}\right)\right)\right].
\end{equation}
Taking $\sup\limits_W$ on both sides of~\eqref{eq:n-n} yields $\pi_{n+1}(\rho)\leq \pi_n(\rho)$.

Turning to~\eqref{eq:pin_ub}, upon choosing $\widehat{P}_n\left(z^n\right)(x)=1/r,  \ z^n\in\cZ^n, \  x\in\cX,$ we get from~\eqref{eq:div},~\eqref{eq:dist-rho-priv} that
\begin{align*}
\pi_n(\rho)&\leq\sup_W \ \sup_{P_X} \ \sum_{z^n\in\cZ^n}\left(P_XW\right)^n\left(z^n\right)
\left(\log r - H\left(P_X\right)\right)\\
&=\log r - \inf_{P_X} \ H\left(P_X\right)=\log r.
\end{align*}\qeed\\ \\

Given $z^n\in\cZ^n$, let $Q^{(n)}=Q^{(n)}\left(z^n\right)$ be its $n$-type, i.e., the empirical pmf on $\cZ$ associated with $z^n$ (cf. e.g.,~\cite{Csi06}). For a given $n$-type $Q^{(n)}$ on $\cZ$, let $\mathcal{T}_{Q^{(n)}}$ be the set of all sequences
in $\cZ^n$ of type $Q^{(n)}$. Let $\mathcal{Q}^{(n)}$ be the set of all $n$-types on $\cZ$. Denote $\left(P_XW\right)^n\left(\mathcal{T}_{Q^{(n)}}\right)\triangleq \sum\limits_{z^n\in\mathcal{T}_{Q^{(n)}}}\left(P_XW\right)^n(z^n)$. As shown next, it is adequate to consider querier estimators $\widehat{P}_n:\mathcal{Q}^{(n)}\rightarrow\Delta_r$ that are based on the type $Q^{(n)}$ of $z^n$ in $\cZ^n$, with said type serving, in effect, as a sufficient statistic. Then, a convenient 
representation for $\pi_n(\rho)$ in~\eqref{eq:dist-rho-priv} is provided by

\begin{lemma}
\label{lem:estim_type}
For $0\leq\rho\leq 1$, 
\begin{equation}
\label{eq:lemm-type-equiv}
\pi_n(\rho)=\sup_{W}\ \inf_{\widehat{P}_n}\ 
\sup_{P_X} \sum\limits_{Q^{(n)}\in\mathcal{Q}^{(n)}} 
\left(P_XW\right)^n\left(\mathcal{T}_{Q^{(n)}}\right) D\left(P_X\big|\big|\widehat{P}_n\left(Q^{(n)}\right)\right)
\end{equation}
\noindent with $\widehat{P}_n\left(Q^{(n)}\right)$ representing identical estimates in $\Delta_r$ for all $z^n\in\mathcal{T}_{Q^{(n)}}$.
\end{lemma}

\noindent\textbf{Proof:} Observe that for fixed $W,\widehat{P}_n,P_X$, 
\begin{equation}
\label{eq:rho-priv-type}
\pi_n\left(\rho,W,P_X,\widehat{P}_n\right)=
\sum\limits_{Q^{(n)}\in\mathcal{Q}^{(n)}} 
\sum\limits_{z^n\in\mathcal{T}_{Q^{(n)}}} 
\left(P_XW\right)^n\left(z^n\right) D\left(P_X\big|\big|\widehat{P}_n\left(z^n\right)\right).
\end{equation}
\noindent For a fixed $Q^{(n)}$, since $(P_XW)^n(z^n)$ is the same for all $z^n\in\mathcal{T}_{Q^{(n)}}$, 
if $\widehat{P}_n(z^n)$ were to vary across $z^n\in\mathcal{T}_{Q^{(n)}}$, the querier can pick
that $\tilde{z^n}$, say, in $\mathcal{T}_{Q^{(n)}}$ for which $D\left(P_X\big|\big|\widehat{P}_n\left(\tilde{z^n}\right)\right)$ is smallest over $\mathcal{T}_{Q^{(n)}}$ and use
$\widehat{P}_n\left(\tilde{z^n}\right)$ as the estimate of $P_X$ for all
$z^n\in\mathcal{T}_{Q^{(n)}}$, denoting it by $\widehat{P}_n\left(Q^{(n)}\right)$; this will only serve
to decrease the right-side of~\eqref{eq:rho-priv-type}, bearing in mind the $\inf$ with 
respect to $\widehat{P}_n$ in the left-side of~\eqref{eq:dist-rho-priv}. Then the right-side of~\eqref{eq:rho-priv-type}
becomes
\begin{equation}
\label{eq:bin-priv-type}
\sum\limits_{Q^{(n)}\in\mathcal{Q}^{(n)}} 
\left(P_XW\right)^n\left(\mathcal{T}_{Q^{(n)}}\right) D\left(P_X\big|\big|\widehat{P}_n\left(Q^{(n)}\right)\right)
\end{equation}
\noindent leading to~\eqref{eq:lemm-type-equiv}. \qeed\\ \\

We close this section with an achievability result that affords a basic lower bound for $\pi_n(\rho)$ as a function of $\rho$, and also a characterization of $\pi_n(\rho)$ for low values of $\rho$; none of these bounds depends on $n$. This lower bound will be lent additional significance in Section~\ref{sec:main_thm} by the converse and achievability results of Theorems~\ref{thm:main-conv} and~\ref{thm:main-achiev}, respectively. Also, the choice of a ``sparse'' user pmf $P_X$ and a ``locally uniform'' pmf as the querier's estimate in the proof of the following result will motivate the concepts of a ``$k$-sparse pmf'' in Definition 5 and ``locally uniform estimator'' in Definition 3 below. 

For $P_X$ in $\Delta_r$, denote the derived pmf $P^l_X\left(f^{-1}\right)$ on $\cZ$, $l=1,\ldots,k$, by
\begin{equation}
\label{eq:px_f}
P_X^l\left(f^{-1}\right)\triangleq\left(\underbrace{\frac{P_X\left(\bigcup\limits_{l'=0}^{l-1}f^{-1}(l')\right)}{l},\ldots,\frac{P_X\left(\bigcup\limits_{l'=0}^{l-1}f^{-1}(l')\right)}{l}}_{l\text{ repetitions}}, P_X\left(f^{-1}(l)\right),
\ldots,P_X\left(f^{-1}(k-1)\right)\right).
\end{equation}
\noindent In particular, for $l=1$ we denote 
\begin{equation} 
\label{eq:PXf1PXf}
P_X^1\left(f^{-1}\right)\triangleq P_X\left(f^{-1}\right)=\left(P_X\left(f^{-1}(0)\right),P_X\left(f^{-1}(1)\right),\ldots,P_X\left(f^{-1}(k-1)\right)\right).\end{equation} 
\noindent Also, for $l=1,\ldots,k$, let $\cZ_l$ denote a generic $l$-sized subset of $\cZ$, with $\cZ_k=\cZ$. 

We define 
\begin{equation}
    \label{eq:priv_guaren}
    \omega(\rho)\triangleq
\begin{cases}
     \log \left|\bigcup\limits_{j\in\cZ_k}f^{-1}(j)\right| =\log \left|f^{-1}(\cZ)\right| =\log |\cX|= \log r, &0\leq\rho\leq \frac{1}{k}\\
    \max\limits_{\cZ_l\subset \cZ} \ \log \left|\bigcup\limits_{j\in\cZ_l}f^{-1}(j)\right| = \max\limits_{\cZ_l\subset \cZ} \ \log \sum\limits_{j\in\cZ_l} \left|f^{-1}(j)\right|, &\frac{1}{l+1}<\rho\leq \frac{1}{l}, \ 1\leq l\leq k-1,
\end{cases}
\end{equation}
\noindent which, under the assumption
\begin{equation}
    \label{eq:priv_nonasymp_lbb}
|f^{-1}(0)|\geq |f^{-1}(1)|\geq\ldots\geq |f^{-1}(k-1)|
\end{equation}
\noindent yields, for $1/k<\rho\leq 1,$ the simplification 
\begin{equation}
\label{eq:omega-nondec}
\max\limits_{\cZ_l\subset \cZ} \ \log \sum\limits_{j\in\cZ_l} \left|f^{-1}(j)\right|=  \log \sum\limits_{j=0}^{l-1} \left|f^{-1}(j)\right|, \ \ \ 1\leq l\leq k-1.
\end{equation}
\noindent It is verified readily from~\eqref{eq:priv_guaren} that $\omega(\rho)$ is nonincreasing in $0\leq\rho\leq 1$.
\vspace{.6cm}

We show next that $\omega(\rho)$ bears the significance of ``worst-case'' distribution $\rho$-privacy.
\begin{theorem}
\label{thm:nonasymp_bnd}
For each $n\geq 1$, $\pi_n(\rho)$ is nonincreasing in $0\leq\rho\leq 1$, and
\begin{equation}
\label{eq:priv_nonasymp_lb}
\pi_n(\rho) \geq \omega(\rho), \ \ \ \ 0\leq\rho\leq 1
\end{equation}
\noindent and 
\begin{equation}
\label{eq:priv_nonasymp_ub}
\pi_n(\rho)=\omega(\rho)=\log r,\ \ \ \ 0\leq\rho\leq \frac{1}{k}.
\end{equation}
\end{theorem}
\vspace{0.15in}
\noindent\textit{Remark}: By Theorem~\ref{thm:nonasymp_bnd}, for the ``single-shot'' case $n=1,$ $\pi_1(\rho)=\log r, \ 0\leq\rho\leq 1/k,$ and $\pi_1(\rho)\geq\omega(\rho), \ 1/k<\rho\leq 1.$ However, a full characterization of $\pi_1(\rho)$ for $1/k<\rho\leq 1$ remains open.
\vspace{0.1in}\\
\noindent\textbf{Proof:} For each $n\geq 1$, it is obvious by~\eqref{eq:dist-rho-priv} that $\pi_n(\rho)$ is nonincreasing
in $0\leq\rho\leq 1$.

Turning to~\eqref{eq:priv_nonasymp_lb}, we shall show that
\begin{equation}
\label{eq:priv_nonasymp_lba}
\pi_n(\rho)\geq \max\limits_{\cZ_l\subseteq \cZ} \ \log\left|\bigcup_{j\in\cZ_l}f^{-1}(j)\right|, \ \ 0\leq\rho\leq \frac{1}{l}, \ \ 1\leq l\leq k
\end{equation}
\noindent from which~\eqref{eq:priv_nonasymp_lb} is deduced readily.

Assume~\eqref{eq:priv_nonasymp_lbb} without loss of essential generality. Fix $l\in\{1,\ldots,k\}.$ For $0\leq\rho\leq 1/l$, the user selects\\ $W_l:\cX\rightarrow\cZ$ as
\begin{equation}
\label{eq:Wl}
W_l(j|x)=\begin{cases}
    \frac{1}{l}, \ \ &x\in\bigcup\limits_{l'=0}^{l-1}f^{-1}(l'), \  j=0,1,\ldots,l-1\\
    1, \ \ &x\in\bigcup\limits_{l'=l}^{k-1}f^{-1}(l'), \   j=f(x).
\end{cases}
\end{equation}
\noindent Clearly, $P_XW_l=P^l_X\left(f^{-1}\right)$ in $\Delta_k$ (see~\eqref{eq:px_f}). 
Expressing $\sup\limits_{P_X\in\Delta_r}$ in~\eqref{eq:dist-rho-priv} as
\[
\sup\limits_{\left(\alpha,\alpha_l,\ldots,\alpha_{k-1}\right)\in\Delta_{k-l+1}}\         \sup\limits_{P_X\in\Delta_r:P_X^l\left(f^{-1}\right)=\left(
\underbrace{\tfrac{\alpha}{l},\ldots,\tfrac{\alpha}{l}}_{l\text{ repetitions}}
,\alpha_l,\ldots,\alpha_{k-1}\right)}
\]
\noindent we obtain from~\eqref{eq:dist-rho-priv}, noting that the expectation in~\eqref{eq:div} is with respect to 
\begin{equation}
\label{eq:PXWl-f-1}
P_XW_l=P_X^l\left(f^{-1}\right)=\underbar{$\alpha$}^l\triangleq\left(
\underbrace{\tfrac{\alpha}{l},\ldots,\tfrac{\alpha}{l}}_{l\text{ repetitions}}
,\alpha_l,\ldots,\alpha_{k-1}\right)\end{equation}
\noindent that
\begin{align}
\pi_n(\rho)&\geq \inf_{\widehat{P}_n}\ 
\sup\limits_{\left(\alpha,\alpha_l,\ldots,\alpha_{k-1}\right)\in\Delta_{k-l+1}}\         \sup\limits_{P_X:P_X^l\left(f^{-1}\right)=\underbar{$\alpha$}^l} \ 
\mathbb{E}_{\underbar{$\alpha$}^l}\left[D\left(P_X\big|\big|\widehat{P}_n\left(Z^n\right)\right)\right]\nonumber\\
&\geq   
\sup\limits_{\left(\alpha,\alpha_l,\ldots,\alpha_{k-1}\right)\in\Delta_{k-l+1}}\ \inf_{\widehat{P}_n} \          \sup\limits_{P_X:P_X^l\left(f^{-1}\right)=\underbar{$\alpha$}^l} \ 
\mathbb{E}_{\underbar{$\alpha$}^l}\left[D\left(P_X\big|\big|\widehat{P}_n\left(Z^n\right)\right)\right]\nonumber\\
&\geq   
\sup\limits_{\left(\alpha,\alpha_l,\ldots,\alpha_{k-1}\right)\in\Delta_{k-l+1}}\ \inf_{R\in\Delta_r} \          \sup\limits_{P_X:P_X^l\left(f^{-1}\right)=\underbar{$\alpha$}^l}  \ 
D\left(P_X\big|\big|R\right).\label{eq:lemm-final-eqn}
\end{align}     
\noindent 
Now, observe in~\eqref{eq:lemm-final-eqn} that for a fixed $\left(\alpha,\alpha_l,\ldots,\alpha_{k-1}\right)\in\Delta_{k-l+1}$, a ``sparse'' pmf $P_X\in\Delta_r$ of limited support size $k-l+1$ with probabilities $\alpha,\alpha_l,\ldots,\alpha_{k-1}$, respectively, on (single support) symbols in each of $\bigcup\limits_{l'=0}^{l-1} f^{-1}(l'),f^{-1}(l),$\\$\ldots,f^{-1}(k-1)$ satisfies the constraint $P_X^l\left(f^{-1}\right)=\underbar{$\alpha$}^l=\left(\alpha/l,\ldots,\alpha/l,\alpha_l,\ldots,\alpha_{k-1}\right).$
Furthermore, such a $P_X$ with these support symbols being the lowest $R$-probability symbols in $\bigcup\limits_{l'=0}^{l-1}f^{-1}(l'),f^{-1}(l),\ldots,f^{-1}(k-1)$, respectively, will serve to maximize $D\left(P_X||R\right)$. 
Accordingly, the pmf $R\in\Delta_r$ that maximizes said lowest probabilities (without knowledge of $P_X$)
and thereby minimizes $D\left(P_X||R\right)$, is a ``locally uniform'' pmf, viz. 
\[R(x)=\begin{cases}
\frac{\beta}{\left|\bigcup\limits_{l'=0}^{l-1}f^{-1}(l^{'})\right|}, \ \ &x\in \bigcup\limits_{l'=0}^{l-1}f^{-1}(l')\\
\frac{\beta_j}{\left|f^{-1}(j)\right|}, \ \ &x\in f^{-1}(j), \ j=l,\ldots,k-1
\end{cases}
\]
\noindent for some $\left(\beta,\beta_l,\ldots,\beta_{k-1}\right)\in\Delta_{k-l+1}$. Then in~\eqref{eq:lemm-final-eqn}, for a fixed $\left(\alpha,\alpha_l,\ldots,\alpha_{k-1}\right)\in\Delta_{k-l+1}$,
\begin{align}
&\inf_{R\in\Delta_r}\ \sup_{P_X:P_X^{l}\left(f^{-1}\right)
=\underbar{$\alpha$}^l}D\left(P_X\big|\big|R\right)\nonumber\\&=\inf_{\left(\beta,\beta_l,\ldots,\beta_{k-1}\right)\in\Delta_{k-l+1}} \ 
\alpha\log\frac{\alpha}{\frac{\beta}{\left|\bigcup\limits_{l'=0}^{l-1}f^{-1}(l')\right|}}+\sum_{j=l}^{k-1}\alpha_j\log\frac{\alpha_j}{\frac{\beta^j}{\left|f^{-1}(j)\right|}}\nonumber\\
&=\inf_{\left(\beta,\beta_l,\ldots,\beta_{k-1}\right)\in\Delta_{k-l+1}} \alpha\log \left|\bigcup\limits_{l'=0}^{l-1}f^{-1}(l')\right|+ \sum_{j=l}^{k-1}
\alpha_j\log \left|f^{-1}(j)\right|\nonumber\\ 
&\hspace{3.5cm}+ D\left(\left(\alpha,\alpha_l,\ldots,\alpha_{k-1}\right)||\left(\beta,\beta_l,\ldots,\beta_{k-1}\right)\right)\nonumber\\
&=\alpha\log \left|\bigcup\limits_{l'=0}^{l-1}f^{-1}(l')\right|+ \sum_{j=l}^{k-1}
\alpha_j\log \left|f^{-1}(j)\right| \label{eq:lemm-final-eq1}
\end{align}
\noindent with the minimum attained by $\left(\beta,\beta_l,\ldots,\beta_{k-1}\right)=\left(\alpha,\alpha_l,\ldots,\alpha_{k-1}\right)$. Combining~\eqref{eq:lemm-final-eqn} and~\eqref{eq:lemm-final-eq1},
\begin{align}
\pi_n(\rho)&\geq\sup_{\left(\alpha,\alpha_l,\ldots,\alpha_{k-1}\right)\in\Delta_{k-l+1}} \  \alpha\log \left|\bigcup\limits_{l'=0}^{l-1}f^{-1}(l')\right|+ \sum_{j=l}^{k-1}
\alpha_j\log \left|f^{-1}(j)\right|\nonumber\\
&=\log \left|\bigcup\limits_{l'=0}^{l-1}f^{-1}(l')\right|\label{eq:thm1-fin}
\end{align}
\noindent with the maximum attained by 
\begin{equation}
\label{eq:alpha-values}
\alpha=1, \alpha_l=\cdots=\alpha_{k-1}=0
\end{equation} 
\noindent upon recalling~\eqref{eq:priv_nonasymp_lbb}. This establishes~\eqref{eq:priv_nonasymp_lba} with the obvious replacement by the right-side therein of the right-side of~\eqref{eq:thm1-fin}.

Turning to~\eqref{eq:priv_nonasymp_ub},
observe that with $l=k$ and $0\leq\rho\leq 1/k$, we get from~\eqref{eq:thm1-fin} that $\pi_n(\rho)\geq \log r$. With Proposition~\ref{prop:1},~\eqref{eq:priv_nonasymp_ub} follows.
\qeed
\vspace{0.2cm}

The results of Theorem~\ref{thm:nonasymp_bnd} are interpreted as follows. Theorem~\ref{thm:nonasymp_bnd} guarantees a worst-case (i.e., $n$-free) level of distribution $\rho$-privacy $\omega(\rho),$ $0\leq\rho\leq 1.$ Specifically, for $1\leq l\leq k-1$ determined by $\rho$ according to~\eqref{eq:priv_guaren}, $\pi_n(\rho)$ is at least $\omega(\rho)=\max\limits_{\cZ_l\subset \cZ} \ \log \sum\limits_{j\in\cZ_l} \left|f^{-1}(j)\right|,$ i.e., the logsum of the sizes of the $l$ largest atoms in the $f^{-1}$-partition of $\cX.$ This guaranteed $\rho$-privacy is achieved by the user's choice of $P_X$ as a point mass on any symbol in the union of these $l$ atoms (cf. passage following~\eqref{eq:lemm-final-eqn}, and~\eqref{eq:alpha-values}) and $\rho$-QR $W_l$ as in~\eqref{eq:Wl}. Then, the resulting pmf $P_XW_l$ on $\cZ$ has (restricted) support on the $f$-images in $\cZ$ of these $l$ atoms in $\cX$, and is uniform among them~\eqref{eq:PXWl-f-1}. This helps explain the form of $\omega(\rho).$ We note at this point that Theorem~\ref{thm:main-achiev} below will describe an achievability scheme whose privacy, for large and suitable but finite $n$, can exceed $\omega(\rho)$ while tending to it as $n\rightarrow\infty.$ Moreover, Theorem~\ref{thm:nonasymp_bnd}, together with Theorem~\ref{thm:main-conv}, will establish that $\lim\limits_n \ \pi_n(\rho)=\omega(\rho)$ for $0.5<\rho\leq 1.$

\section{Converse and Achievability Theorems}
\label{sec:main_thm}

Our main Theorems~\ref{thm:main-conv} and~\ref{thm:main-achiev} constitute, respectively, converse and achievability results for distribution $\rho$-privacy, and yield $n$-dependent upper and lower bounds for $\pi_n(\rho).$ Instrumental to their proofs are user and querier strategies that employ special constructs. Among the querier's estimators $\widehat{P}_n:
\cQ^{(n)}\rightarrow\Delta_r$, $n\geq 1$, pertinent to our converse and achievability
proofs for $\pi_n(\rho)$, respectively, will be classes of ``locally uniform estimators'' and
``smooth estimators.'' Furthermore, from the user's standpoint, ``$k$-sparse pmfs $P_{X}^{sp}$''
and ``locally identical $\rho$-QRs $W^{lo}$'' are material in the converse proof.
\vspace{0.075in}
\begin{definition} Let
$\beta^{(n)}=\left\{\beta^{(n)}\left(Q^{(n)}\right)\right\}_{Q^{(n)}\in\cQ^{(n)}}$, $n\geq 1$, be a set of pmfs in $\Delta_k$
indexed by $Q^{(n)} \in \cQ^{(n)}$ with each member pmf being of form $\beta^{(n)}\left(Q^{(n)}\right)=\left(\beta^{(n)}_0\left(Q^{(n)}\right),\beta^{(n)}_1\left(Q^{(n)}\right),\ldots,\beta^{(n)}_{k-1}\left(Q^{(n)}\right)\right)$.
A \textit{locally uniform estimator} $\widehat{P}_n^{\beta^{(n)}}:\cQ^{(n)}\rightarrow
\Delta_r$ is defined for each $Q^{(n)}\in \cQ^{(n)}$ by
\[\widehat{P}_n^{\beta^{(n)}}\left(Q^{(n)}\right)(x)=
\frac{\beta_j^{(n)}\left(Q^{(n)}\right)}{\left|f^{-1}(j)\right|}, \ \ x\in f^{-1}(j), \ j\in \cZ.
\]
\end{definition}
\vspace{0.1cm}
\begin{definition}
Consider any partition of $\cZ$ into $k'\leq k$ atoms and, with an abuse of notation, label the atoms by $\cZ'=\cZ'(k')=\{0,1,\ldots,k'-1\}.$ Let $Q'^{(n)}$ be an $n$-type on $\cZ',$ $\mathcal{T}_{Q'^{(n)}}$ the set of all sequences in $\cZ'^n$ of type $Q'^{(n)},$ and $\mathcal{Q}'^{(n)}$ the set of all $n$-types on $\cZ'.$ A \textit{smooth estimator} $\widehat{P}_n:\cQ'^{(n)}\rightarrow\Delta_r$, $n\geq 1$, is such that for some $\gamma_n>0,$ $ \hat{\gamma}_n>0$ and $c_n>0, \ n\geq 1,$ with 
\begin{equation}
\label{eq:smooth_estim_lims}
\lim\limits_n \gamma_n= 0, \ \ \  \lim\limits_n \hat{\gamma}_n=0, \ \ \   \lim\limits_n c_n=0 \ \ \  \text{and} \ \ \ \frac{\hat{\gamma}_n}{c_n}=o\left(\frac{1}{n}\right) \ \ \text{implying} \ \  \lim\limits_n \frac{\hat{\gamma}_n}{c_n}=0,
\end{equation}
\noindent it holds for $Q'^{(n)}\neq Q''^{(n)}$ on $\cZ'$ with var$\left(Q'^{(n)},Q''^{(n)}\right)\leq \gamma_n$ that
var$\left(\widehat{P}_n\left(Q'^{(n)}\right),\widehat{P}_n\left(Q''^{(n)}\right)\right) \leq \hat{\gamma}_n$, where var$(\cdot,\cdot)$ denotes variational
distance (in $\Delta_{k'}$ or $\Delta_r$); and for each $Q'^{(n)}\in \mathcal{Q}'^{(n)}$,
$\widehat{P}_n\left(Q'^{(n)}\right)(x)\geq c_n, \ x\in\cX.$ Denote the class of all such estimators by $\mathcal{S}_n=\mathcal{S}_n(k'), \ n\geq 1$.
\end{definition}
\vspace{0.07in}
\textit{Remark}: A smooth estimator has the feature that QRs with neighbouring types lead to proximate
pmf estimates by the querier. Its second feature of full support $\cX$ is motivated by Remark (iii) following Definition 2.

\begin{definition}
A \textit{$k$-sparse pmf} $P_{X}^{sp}$ on $\cX$ is defined by
\[
P_{X}^{sp}(x)=\alpha_j,\ \ \text{for some $x\in f^{-1}(j)$}, \ j\in \cZ,
\]
\noindent and for some $k$-pmf $\left(\alpha_0,\alpha_1,\ldots,\alpha_{k-1}\right)$ in $\Delta_k.$
\end{definition}
\vspace{0.1cm}
\begin{definition} (i) A \textit{locally identical $\rho$-QR} $W^{lo}:\cX\rightarrow\cZ$ has the form: for each $j\in\cZ$, $W^{lo}(\cdot|x)$ is identical for $x\in f^{-1}(j)$, i.e., $W^{lo}$ has identical rows for all $x$ in $f^{-1}(j)$, $j\in\cZ$. Associated with each such $W^{lo}$ is a stochastic matrix $V=V\left(W^{lo}\right):\cZ\rightarrow\cZ$ given by
\[
V\left(W^{lo}\right)\left(j|j'\right)=W^{lo}(j|x) \ \text{for every $x\in f^{-1}\left(j'\right)$}, \ j,j'\in\cZ;
\]
\noindent in particular $V\left(W^{lo}\right)(j|j)\geq\rho, \ j\in\cZ.$ \\ 
\noindent(ii) Let $\mathcal{V}(\rho)$ be the set of all stochastic matrices $V:\cZ\rightarrow\cZ$ with $V(j|j)\geq \rho, \ j\in\cZ$. For each $V\in\mathcal{V}(\rho)$, set
\begin{equation}
\label{eq:deltak-V}
\Delta_k(V)\triangleq \left\{\underbar{$\alpha$}\in\Delta_k:\underbar{$\alpha$}=\underbar{$\beta$} V \ \text{for some $\underbar{$\beta$}\in\Delta_k$}\right\}.
\end{equation}
\end{definition}
\vspace{0.1in}
\noindent\textit{Remarks}:  (i) For $W^{lo}:\cX\rightarrow\cZ$ and $V:\cZ\rightarrow\cZ$ as above and for any $P_X$ in $\Delta_r,$ it follows that $P_XW^{lo}=P_X\left(f^{-1}\right)V,$ both in $\Delta_k.$\\
(ii) For $V\in\mathcal{V}(\rho), \ 0.5<\rho\leq 1$, $V$ is diagonally-dominated so that $V^{-1}$ exists~{\cite[Theorem $3.3.9$]{Rao00}}.\vspace{0.1in}

The following (information geometric) notion will be pertinent for our converse Theorem~\ref{thm:main-conv}. For $Q^{(n)}$ in $\cQ^{(n)},$ and $V\in\cV(\rho), \ 0.5<\rho\leq 1,$ let   
\begin{equation}
    \label{eq:Reverse_Iproj}
    \widetilde{Q}\left(Q^{(n)}\right)\triangleq \arg \min\limits_{Q\in\Delta_k(V)} \ D\left(Q^{(n)}||Q\right)
\end{equation}
\noindent be the reverse I-projection of $Q^{(n)}$ on $\Delta_k(V);$ and for all $\rho>0,$ the minimum exists by~{\cite[Theorem $3.4$]{Csi04}} since $\Delta_k(V)$ is a closed convex set in $\mathbb{R}^{k}$ and contains at least one pmf with support equal to $\cZ$ as $V(j|j)\geq\rho, \ j\in\cZ.$ Noting by~\eqref{eq:deltak-V} that $\widetilde{Q}\left(Q^{(n)}\right)V^{-1}$ lies in $\Delta_k,$ let
$\kappa_n\left(Q^{(n)}\right)$ be its positivized\footnotemark\footnotetext{The positivization serves to avoid zeros in $\kappa_n\left(Q^{(n)}\right).$} version in $\Delta_k$ defined as
\begin{equation}
    \label{eq:kappa}
    \kappa_n\left(Q^{(n)}\right)(j)=\frac{n\left(\widetilde{Q}\left(Q^{(n)}\right)V^{-1}\right)(j)+1}{n+k}, \ \ j\in\cZ.
\end{equation}
\noindent Observe that $\kappa_n\left(Q^{(n)}\right)$ is in $\Delta_k$ has full support $\cZ$. 

If the user chooses a locally identical $\rho$-QR $W:\cX\rightarrow\cZ$ with $V=V(W):\cZ\rightarrow\cZ$ (see Definition 6 (i)), then $Z_1,\ldots,Z_n$ are i.i.d. with (common) pmf $P_XW=P_X(f^{-1})V$ (see \eqref{eq:PXf1PXf}) and $P_X(f^{-1})V$ belongs to $\Delta_k(V)$ (see Definition 6 (ii)). The querier, with full knowledge of $V,$ and having observed a sequence $z^n$ in $\cZ^n$ of type $Q^{(n)},$ forms a maximum likelihood estimate of the query response pmf $P_X(f^{-1})V$ as $\widetilde{Q}\left(Q^{(n)}\right)$~\eqref{eq:Reverse_Iproj}. Therefore, if the querier is restricted to using a locally uniform estimator (see Definition 3), a natural choice for $\beta^{(n)}$ as an attendant proxy for $P_X(f^{-1})$ is $\kappa_n$ given by~\eqref{eq:kappa} and this plays a role in our converse result below.

We now state Theorems~\ref{thm:main-conv} and~\ref{thm:main-achiev}. Hereafter, we make Assumption~\eqref{eq:priv_nonasymp_lbb} without loss of essential generality; this assumption is made only for the sake of notational convenience. In particular, the upper bound in Theorem~\ref{thm:main-conv} for $\pi_n(\rho),$ $0.5<\rho\leq 1,$ tends to the lower bound in Theorem~\ref{thm:nonasymp_bnd} as $n\rightarrow\infty$. Theorem~\ref{thm:main-achiev} gives a lower bound for $\pi_n(\rho), \ 1/k<\rho\leq 1,$ that approaches, as $n\rightarrow\infty,$ worst-case privacy in Theorem~\ref{thm:nonasymp_bnd}. {\it A notable characteristic of Theorems~\ref{thm:nonasymp_bnd},~\ref{thm:main-conv} and~\ref{thm:main-achiev} is that for all $0\leq\rho\leq 1,$ the asymptotically optimal limits in $n$ of $\pi_n(\rho)$ are in terms of the logsum cardinalities of inverse atoms (images) under $f,$ with the number of summands decreasing as $\rho$ increases.} 

In the range $0.5<\rho\leq 1,$ a converse (upper) bound for $\pi_n(\rho), \ n\geq 1,$ and thereby for $\lim\limits_n \ \pi_n(\rho),$ is given by

\begin{theorem}
\label{thm:main-conv}
For $0.5<\rho\leq 1$ and every $n\geq 1,$ 
\begin{gather}
\pi_n(\rho)\leq \omega(\rho)+\Gamma_n(\rho)=\  \max\limits_{j\in\cZ} \ \log \ \left|f^{-1}(j)\right|+\Gamma_n(\rho), \ \ \ \ \text{where}\label{eq:omega-gamma}\\
  \Gamma_n(\rho)\triangleq
    \sup_{V\in\mathcal{V}(\rho)} \ 
    \sup_{\underbar{$\alpha$}\in\Delta_k(V)} \ \sum\limits_{Q^{(n)}\in\mathcal{Q}^{(n)} }\ \underbar{$\alpha$}^n \left(\mathcal{T}_{Q^{(n)}}\right) \ D\left(\underbar{$\alpha$} V^{-1}||\kappa_n\left(Q^{(n)}\right)\right). \label{eq:main-res}
\end{gather}
 \noindent Furthermore, 
 \begin{equation}
 \label{eq:main-thm-lim}
 \lim\limits_n \ \Gamma_n(\rho)=0 \ \ \text{and} \ \ \lim\limits_n \pi_n(\rho)= \max\limits_{j\in\cZ} \ \log \ \left|f^{-1}(j)\right|.
 \end{equation}
 \end{theorem}
 \vspace{0.2cm}
\noindent \textit{Remarks}: (i) In~\eqref{eq:omega-gamma}, note that $\omega(\rho)=\max\limits_{j\in\cZ} \ \log\big|f^{-1}(j)\big|$ by~\eqref{eq:priv_guaren}; and additionally under~\eqref{eq:priv_nonasymp_lbb}, $\omega(\rho)= \log\big|f^{-1}(0)\big|$.\\
(ii) In general, a closed-form expression is not available for the reverse I-projection $\widetilde{Q}\left(Q^{(n)}\right)$ in~\eqref{eq:Reverse_Iproj};
an iterative method for computing it is described in~{\cite[Example $5.1$]{Csi04}}. Hence, $\kappa_n\left(Q^{(n)}\right)$ in~\eqref{eq:kappa} and $\Gamma_n(\rho)$ in~\eqref{eq:main-res} lack explicit expressions.
 
The following achievability result is for $1/k<\rho\leq 1;$ for $0 \leq \rho\leq 1/k,$ Theorem~\ref{thm:nonasymp_bnd} above already characterizes $\pi_n(\rho).$  
\begin{theorem}
\label{thm:main-achiev}
\noindent Let $1/k<\rho\leq 1.$ For appropriate locally identical $\rho$-QRs $W^{lo}=W^{lo}(\rho)$, it holds that for the (restricted) class of smooth estimators for the querier
\begin{multline}
 \inf_{\widehat{P}_n\in\mathcal{S}_n} \ \sup_{P_X} \ \pi_n\left(
\rho,W^{lo},P_X,\widehat{P}_n\right)\\
 \geq \left(\omega(\rho)+\Lambda_n(\rho)\right)\lambda_n(\rho)= \left(\log\sum_{j=0}^{l-1}\big|f^{-1}(j)\big|+\Lambda_n(\rho)\right)\lambda_n(\rho), \ \ \ \ \frac{1}{l+1}<\rho\leq \frac{1}{l}, \ \ 1\leq l\leq k-1,\label{eq:achiev_thm_main}
\end{multline}
\noindent for all $n$ large enough, where for $l=l(\rho)\leq{\left\lfloor\frac{k}{2}\right\rfloor},$ 
\begin{equation}
\label{eq:Lambda-def1}
\Lambda_n(\rho)\triangleq
\begin{cases}\log\left(1+\frac{\sum\limits_{j=l}^{\left\lfloor\frac{k}{l}\right\rfloor l-1}\big |f^{-1}(j)\big|}{e\sum\limits_{j=0}^{l-1}\big |f^{-1}(j)\big|}
\frac{\min\left\{\frac{\left\lceil nl\left(\frac{1-l\rho}{\left\lfloor\frac{k}{l}\right\rfloor l-l}\right)\right\rceil}{n},\ l\rho\right\}-l\left(\frac{1-l\rho}{\left\lfloor\frac{k}{l}\right\rfloor l-l}\right)}{l\rho-l\left(\frac{1-l\rho}{\left\lfloor\frac{k}{l}\right\rfloor l-l}\right)}
\right)-
    \frac{\hat{\gamma}_n}{c_n}, \ \ &{\sum\limits_{j=l}^{\left\lfloor\frac{k}{l}\right\rfloor  l-1}\big |f^{-1}(j)\big|}\leq{\sum\limits_{j=0}^{l-1}\big |f^{-1}(j)\big|}\\
    \log\left(\frac{\sum\limits_{j=l}^{\left\lfloor\frac{k}{l}\right\rfloor l-1}\big |f^{-1}(j)\big|}{\sum\limits_{j=0}^{l-1}\big |f^{-1}(j)\big|}\right)\frac{\min\left\{\frac{\left\lceil nl\left(\frac{1-l\rho}{\left\lfloor\frac{k}{l}\right\rfloor l-l}\right)\right\rceil}{n},\ l\rho\right\}-l\left(\frac{1-l\rho}{l\left\lfloor\frac{k}{l}\right\rfloor-l}\right)}{l\rho-l\left(\frac{1-l\rho}{\left\lfloor\frac{k}{l}\right\rfloor l-l}\right)}-
    \frac{\hat{\gamma}_n}{c_n}, \ \ &{\sum\limits_{j=l}^{\left\lfloor\frac{k}{l}\right\rfloor l-1}\big |f^{-1}(j)\big|}>{\sum\limits_{j=0}^{l-1}\big |f^{-1}(j)\big|}
    \end{cases} \end{equation}
\noindent and for $l=l(\rho)>{\left\lfloor\frac{k}{2}\right\rfloor},$ 
    \begin{equation}
    \label{eq:Lambda-def2}
    \Lambda_n(\rho)\triangleq
\log\left(1+\frac{\big |f^{-1}(l)\big|}{e\sum\limits_{j=0}^{l-1}\big |f^{-1}(j)\big|}
\frac{\frac{\left\lceil n(1-l\rho)\right\rceil}{n}-(1-l\rho)}{l\rho}
\right)-
    \frac{\hat{\gamma}_n}{c_n}    
    \end{equation}
    \noindent and for all $l=l(\rho),$ 
\begin{equation}
    \label{eq:lambda-def}
\lambda_n(\rho)\triangleq 1-  \frac{3e^{\left(\frac{4\left(\left\lfloor\frac{k}{l}\right\rfloor+k-\left\lfloor\frac{k}{l}\right\rfloor l -1\right)\sqrt{\zeta}}{5\sqrt{e}}\right)}}{n^{\zeta}} \ \  \ \text{with $\zeta>1$}.
\end{equation}
\noindent Furthermore
\begin{equation}
\label{eq:achiev_thm_limres}
\lim\limits_n \ \Lambda_n(\rho)=0, \ \ \ \lim\limits_n \ \lambda_n(\rho)=1, \ \ \ \ \frac{1}{k}<\rho\leq 1
\end{equation}
\noindent and
\begin{equation}
    \label{eq:pi_n_lblimit}
\lim_n  \  \inf_{\widehat{P}_n\in\mathcal{S}_n} \ \sup_{P_X} \ \pi_n\left(
\rho,W^{lo},P_X,\widehat{P}_n\right)\geq\omega(\rho)= 
\log\sum_{j=0}^{l-1}\big|f^{-1}(j)\big|, \ \ \ \  \frac{1}{l+1}<\rho\leq \frac{1}{l}, \ \ 1\leq l \leq k-1.
\end{equation}
\end{theorem}
\vspace{0.2in}
\noindent \textit{Remarks}: (i) The proof of Theorem~\ref{thm:main-achiev} will show achievability with the user's choice of $P_X$ taking the form of appropriate sparse pmfs. Note that the left-side of~\eqref{eq:achiev_thm_main} serves as a lower bound for $\pi_n(\rho)$ for the class of smooth querier's estimators.\\
(ii) In~\eqref{eq:achiev_thm_main} and~\eqref{eq:pi_n_lblimit}, $\omega(\rho)=\log\sum\limits_{j=0}^{l-1}\big|f^{-1}(j)\big|$ by~\eqref{eq:priv_guaren} and~\eqref{eq:priv_nonasymp_lbb}.\\
(iii) Our result in Theorem~\ref{thm:main-achiev} must be qualified. In~\eqref{eq:achiev_thm_main}, the $\log$ terms in~\eqref{eq:Lambda-def1},~\eqref{eq:Lambda-def2} can equal $0$ for some values of $n$. For a larger set of $n$s, $\Lambda_n(\rho)=O\left(\log\left(1+\frac{1}{n}\right)\right)=O\left(\frac{1}{n}\right)$. Also, $1-\lambda_n(\rho)=O\left(\frac{1}{n^{\zeta}}\right), \ \zeta>1$. Consequently, the right-side of~\eqref{eq:achiev_thm_main} strictly exceeds $\omega(\rho)$ for all large and  suitable (but for those from the mentioned set) $n$.
\vspace{0.2in}

The proofs of Theorems~\ref{thm:main-conv} and~\ref{thm:main-achiev} are provided in Sections~\ref{subsec:ub} and~\ref{subsec:lb}, respectively.\\ 

We close this section by interpreting the results of Theorems~\ref{thm:nonasymp_bnd},~\ref{thm:main-conv} and~\ref{thm:main-achiev} when particularized to $f:\cX\rightarrow\cZ=\cX$ being an invertible mapping. Then a $\rho$-QR $W:\cX\rightarrow\cX$ is an $r\times r$-stochastic matrix with diagonal elements $\geq\rho$. Loosely speaking,\\
(a) for $0\leq\rho\leq 0.5$, it is clear  that no accurate estimation of $P_X$ from $Z_1,\ldots,Z_n$ -- in the sense of the right-side of~\eqref{eq:dist-rho-priv} tending to $0$ as $n\rightarrow\infty$ -- is possible by the querier;\\
(b) on the other hand, for $0.5<\rho\leq 1$, strongly consistent estimation of $P_X$ by the querier is possible.\\In this context, by Theorem~\ref{thm:nonasymp_bnd}, for all $n\geq 1$,
\begin{equation}
\label{eq:omega_invertible}
    \pi_n(\rho)\begin{cases}
    =\omega(\rho)=\log r, \ \ &0\leq\rho\leq \frac{1}{r}\\
    \geq \omega(\rho)=\log l, \ \ &\frac{1}{l+1}<\rho\leq\frac{1}{l}, \ \ 1\leq l \leq r-1
    \end{cases}
\end{equation}
\noindent which, since $\omega(\rho)>0$ for $0\leq\rho\leq 0.5$ by~\eqref{eq:omega_invertible}, reinforces (a) above. Next, $\omega(\rho)=0$ for $0.5<\rho\leq 1$ by~\eqref{eq:omega_invertible}, and Theorem~\ref{thm:main-conv} gives that for every $n\geq 1$,
\[
\pi_n(\rho)\leq\Gamma_n(\rho)
\]
\noindent by~\eqref{eq:omega-gamma},~\eqref{eq:main-res}, where the inner and outer suprema in~\eqref{eq:main-res} are over all $\underbar{$\alpha$}$ in the row space of $W$ and all $\rho$-QRs $W:\cX\rightarrow\cX$, respectively. Also, $\lim\limits_n \ \pi_n(\rho)=0$ by~\eqref{eq:main-thm-lim}, in keeping with (b) above. However, by Theorem~\ref{thm:main-achiev}, for large and suitable but finite $n$, a positive distribution $\rho$-privacy of at least $\Lambda_n(\rho)\lambda_n(\rho)>0$ can be achieved, in effect owing to the querier being unable to estimate $P_X$ accurately from $Z_1,\ldots,Z_n$. Here, $\Lambda_n(\rho)$ and $\lambda_n(\rho)$ are specialized from~\eqref{eq:Lambda-def1},~\eqref{eq:Lambda-def2} and~\eqref{eq:lambda-def}, respectively, with $k=r$ and $\left|f^{-1}(j)\right|=1, \ j=0,\ldots,r-1$.

\section{Proofs of Theorems~\ref{thm:main-conv} and ~\ref{thm:main-achiev}}
\label{sec:proofs}

\subsection{Technical Lemmas}
\vspace{0.1cm}
The following technical Lemmas~\ref{lem:local_unif} and~\ref{lem:div_var-dis} are pertinent to Theorems~\ref{thm:main-conv} and~\ref{thm:main-achiev}, respectively. Their proofs are relegated to Appendix~\ref{app:tech_lemmas}.
\begin{lemma}
\label{lem:local_unif}
Consider a $k$-partition $\mathcal{A}=\left(A_0,A_1,\ldots,A_{k-1}\right)$ of $\cX$ with $A_j\neq \emptyset, \ j\in\cZ.$ Let $P$ be a pmf on $\cX$ and $P(\mathcal{A})=\left(P(A_0),P(A_1),\ldots,P(A_{k-1})\right)$ the corresponding pmf in $\Delta_k.$ Fix $\underbar{$\beta$} = \left(\beta_0,\beta_1,\ldots,\beta_{k-1}\right)\in\Delta_k$ and let $Q$ be a pmf on $\cX$ given by
\[
Q(x)=
\frac{\beta_j}{|A_j|}, \ x\in A_j, \ j\in\cZ.
\]
\noindent Then
\[
    D(P||Q)\leq D\left(P(\mathcal{A})\big|\big|\underbar{$\beta$}\right)
    +\sum_{j\in\cZ} \ P(A_j)\log |A_j|
\]
\noindent with equality iff $P$ is a $k$-point mass with
\begin{equation}\label{eq:k-point}
    P(x_j')= 
P(A_j) \ \ \ \text{for some $x_j'\in A_j, \ j\in\cZ.$}
\end{equation}
\end{lemma}
\vspace{0.1cm}
\begin{lemma}
\label{lem:div_var-dis}
Consider pmfs $P,Q$ and $Q_o$ on $\cX$ such that $support \ (P) 
\subseteq support \ (Q) \subseteq support \ (Q_o)$. Then
\[
D\left(P||Q\right)\geq D\left(P||Q_o\right)-\frac{\text{\normalfont{var}}\left(Q,Q_o\right)}{Q_o^{min}}
\]
\noindent where $Q_o^{min}$ is the smallest nonzero value of $Q_o$.
\end{lemma}

\subsection{Proof of Theorem~\ref{thm:main-conv}}
\label{subsec:ub}
\vspace{0.1in}
Since the querier's estimator $\widehat{P}_n:\cZ^n\rightarrow\Delta_r$ of $P_X$ is based on the $\cZ$-valued observations $Z_1,\ldots,Z_n$, a reasonable procedure entails the estimation of $P_X$ in two steps, without sacrificing the essence of the infimum in~\eqref{eq:dist-rho-priv}. In a first step, $\widehat{P}_n$ estimates $P_X\left(f^{-1}\right)$ from $Z_1,\ldots,Z_n$. Next, $\widehat{P}_n$ estimates $P_X$ by uniformizing  $P_X\left(f^{-1}\right)$ over symbols in each inverse atom under $f$; any nonuniform assignment of $P_X\left(f^{-1}\right)$ would be undesirable as it would enable the user to put the entire $P_X$-probability on the lowest $\widehat{P}_n$- probability symbol in an inverse atom. This suggests the essential optimality in~\eqref{eq:dist-rho-priv} of locally uniform estimators.

Proceeding with this reasoning, a crucial facilitating step is to show that when the querier uses a locally uniform estimator, the user's actions 
can be limited to $k$-sparse pmfs and locally identical $\rho$-QRs {\it without loss of distribution privacy.}
\begin{lemma}
\label{lem:WV-reduction}
Fix $0\leq\rho\leq 1$. For every $n\geq 1$ and $\beta^{(n)}=\left\{\beta^{(n)}\left(Q^{(n)}\right)\right\}_{Q^{(n)}\in\cQ^{(n)}}$,
\begin{equation}
\label{eq:WV-reduc}
\sup_W \ 
\inf_{\widehat{P}_n^{\beta^{(n)}}}\ \sup_{P_X} \ \pi_n\left(\rho,W,P_X,\widehat{P}_n^{\beta^{(n)}}\right)
=\sup_{W^{lo}} \ 
\inf_{\widehat{P}_n^{\beta^{(n)}}}\ \sup_{P_X^{sp}} \ \pi_n\left(\rho,W^{lo},P_X^{sp},\widehat{P}_n^{\beta^{(n)}}\right).
\end{equation}
\end{lemma}
\noindent\textbf{Proof}: Since the suprema in the right-side of~\eqref{eq:WV-reduc} are over restricted
sets, it suffices to show that~\eqref{eq:WV-reduc} holds with ``$\leq.$'' Specifically, we show that
for every $P_X$ and $W$ there exist $P_X^{sp}$ and $W^{lo}$ such that
\begin{equation}
    \label{eq:WV-reduc-pf}
    \pi_n\left(\rho,W,P_X,\widehat{P}_n^{\beta^{(n)}}\right)\\
\leq \pi_n\left(\rho,W^{lo},P_X^{sp},\widehat{P}_n^{\beta^{(n)}}\right).
\end{equation}
\noindent To this end, let 
\begin{equation}
\label{eq:PX_b}
P_{X}^{sp}(x)=P_X\left(f^{-1}\left(j\right)\right),\ \ \text{for some $x\in f^{-1}\left(j\right)$}, \ j\in \cZ \  \left(\text{see Definition $5$}\right),
\end{equation}
\noindent and let $W^{lo}:\cX\rightarrow\cZ$ be specified as follows:\\
- for $j$ with $P_X\left(f^{-1}(j)\right)>0$: for each $x\in f^{-1}(j)$
\begin{equation}
\label{eq:WW}
    W^{lo}(j'|x) =\sum_{x'\in f^{-1}(j)}\frac{P_X(x')}{P_X\left(f^{-1}(j)\right)}W(j'|x'), \ \ \ j'\in\cZ
\end{equation}
- for $j$ with $P_X\left(f^{-1}(j)\right)=0$: for each $x\in f^{-1}(j)$
\begin{equation}  
W^{lo}(j'|x) =
\begin{cases} 
\rho, \ \ &j'=j\\
\frac{1-\rho}{k-1}, \ \ &j'\neq j \in\cZ.
\end{cases}\label{eq:WWb}
\end{equation}
\noindent From~\eqref{eq:PX_b},~\eqref{eq:WW} and~\eqref{eq:WWb}, it is clear that for each $j\in\cZ$,
\begin{align}
    \left(P_XW\right)(j)
    &=\sum_{x\in\cX} \ P_X(x) W(j|x)\nonumber\\
& =   \sum\limits_{j'\in \cZ: P_X\left(f^{-1}(j')\right)>0} \ 
    \sum_{x\in f^{-1}(j')} \ P_X(x)W(j|x)\nonumber\\
    &=\sum\limits_{j'\in \cZ: P_X\left(f^{-1}(j')\right)>0} \  P_X\left(f^{-1}(j')\right) \ 
    \sum_{x\in f^{-1}(j')} \ \frac{P_X(x)}{P_X\left(f^{-1}(j')\right)}W(j|x)\nonumber\\
   &=\sum\limits_{j'\in \cZ: P_X\left(f^{-1}(j')\right)>0} P_X^{sp}(x') \ W^{lo}(j|x'), \ \ \ \ \text{for some $x'\in f^{-1}(j')$}\nonumber\\
 &=   \left(P_X^{sp}W^{lo}\right)(j)\label{eq:WWb-reduc}
\end{align}
where the fourth equality above uses~\eqref{eq:PX_b} and~\eqref{eq:WW}. Then, using~\eqref{eq:bin-priv-type} and~\eqref{eq:PX_b} -~\eqref{eq:WWb-reduc},
\begin{equation}
    \label{eq:lemubb}
    \pi_n\left(\rho,W,P_X,\widehat{P}_n^{\beta^{(n)}}\right)
    =\sum\limits_{Q^{(n)}\in\mathcal{Q}^{(n)}} 
\left(P_X^{sp}W^{lo}\right)^n\left(\mathcal{T}_{Q^{(n)}}\right) D\left(P_X\big|\big|\widehat{P}_n^{\beta^{(n)}}\left(Q^{(n)}\right)
\right).
\end{equation}
Moreover, by Lemma~\ref{lem:local_unif},
\begin{align}
    D\left(P_X\big|\big|\widehat{P}_n^{\beta^{(n)}}\left(Q^{(n)}\right)\right)
    &\leq
    D\left(P_X\left(f^{-1}\right)\big|\big|\beta^{(n)}\left(Q^{(n)}\right)\right)+\sum_{j\in\cZ} \ P_X\left(f^{-1}(j)\right)\log |f^{-1}(j)|\nonumber\\
    &=D\left(P_X^{sp}\big|\big|\widehat{P}_n^{\beta^{(n)}}\left(Q^{(n)}\right)\right)\label{eq:lemlbb}
\end{align}
\noindent by~\eqref{eq:PX_b} and Definition 3. By~\eqref{eq:lemubb} and~\eqref{eq:lemlbb}, and recalling
\eqref{eq:bin-priv-type}
\begin{align*}
\pi_n\left(\rho,W,P_X,\widehat{P}_n^{\beta^{(n)}}\right)
  &  \leq\sum\limits_{Q^{(n)}\in\mathcal{Q}^{(n)}} 
\left(P_X^{sp}W^{lo}\right)^n\left(\mathcal{T}_{Q^{(n)}}\right)D\left(P_X^{sp}\big|\big|\widehat{P}_n^{\beta^{(n)}}\left(Q^{(n)}\right)\right)\\
  &=\pi_n\left(\rho,W^{lo},P_X^{sp},\widehat{P}_n^{\beta^{(n)}}\right),
\end{align*}
\noindent which is~\eqref{eq:WV-reduc-pf}.
\qeed
\vspace{0.15in}

Turning to Theorem~\ref{thm:main-conv}, note by~\eqref{eq:dist-rho-priv} that
upon restricting the querier's choice to locally uniform estimators and using Lemma~\ref{lem:WV-reduction}
\begin{equation}
    \label{eq:lemm-ub}
    \pi_n(\rho)\leq \sup_{W^{lo}} \ \inf_{\widehat{P}_n^{\beta^{(n)}}} \ \sup_{P_X^{sp}} \ 
\pi_n\left(\rho,W^{lo},P_X^{sp},\widehat{P}_n^{\beta^{(n)}}\right).
\end{equation}
\noindent For fixed $W^{lo}$ and $P^{sp}_X$, let
\begin{equation}
    \label{eq:alpha_PW}
    \underbar{$\alpha$}=P_X^{sp}W^{lo}=P_X^{sp}\left(f^{-1}\right)V\left(W^{lo}\right)
\end{equation}
\noindent where $V\left(W^{lo}\right):\cZ\rightarrow\cZ$ is as in Definition 6 (i). Since $\rho>0.5,$
$V\left(W^{lo}\right)^{-1}$ exists (see Remark (ii) following Definition 6).
Then, upon fixing $\widehat{P}_n^{\beta^{(n)}}$, too, using~\eqref{eq:bin-priv-type} and~\eqref{eq:alpha_PW} we get
\begin{align}
\pi_n\left(\rho,W^{lo},P^{sp}_X,\widehat{P}_n^{\beta^{(n)}}\right)
&=\sum\limits_{Q^{(n)}\in\mathcal{Q}^{(n)}} 
\underbar{$\alpha$}^n\left(\mathcal{T}_{Q^{(n)}}\right)D\left(P_X^{sp}\big|\big|\widehat{P}_n^{\beta^{(n)}}\left(Q^{(n)}\right)\right)\nonumber\\
&=\sum\limits_{Q^{(n)}\in\mathcal{Q}^{(n)}} 
\underbar{$\alpha$}^n\left(\mathcal{T}_{Q^{(n)}}\right)D\left(P_X^{sp}\left(f^{-1}\right)\big|\big|\beta^{(n)}\left(Q^{(n)}\right)\right)\nonumber\\&+\sum_{j\in\cZ}P_X^{sp}\left(f^{-1}(j)\right)\log |f^{-1}(j)|\label{eq:lemm_use1}
\end{align}
\noindent by equality in Lemma~\ref{lem:local_unif}. Next, note from~\eqref{eq:alpha_PW} that
\begin{equation}
    P_X^{sp}\left(f^{-1}\right)=\underbar{$\alpha$} \ \left( V\left(W^{lo}\right)\right)^{-1}.\label{eq:lemm_use2}
\end{equation}
\noindent Then, from~\eqref{eq:lemm_use1} and~\eqref{eq:lemm_use2},
\begin{align}
\pi_n\left(\rho,W^{lo},P^{sp}_X,\widehat{P}_n^{\beta^{(n)}}\right)&\leq
\max_{j\in\cZ} \ \log |f^{-1}(j)|\nonumber\\&+\sum\limits_{Q^{(n)}\in\mathcal{Q}^{(n)}}
\underbar{$\alpha$}^n\left(\mathcal{T}_{Q^{(n)}}\right)D\left(\underbar{$\alpha$} \ \left( V\left(W^{lo}\right)\right)^{-1}\Big|\Big|\beta^{(n)}\left(Q^{(n)}\right)\right). \label{eq:lemm_use3}
\end{align}
\noindent
Hence, in~\eqref{eq:lemm-ub} upon using~\eqref{eq:alpha_PW} -~\eqref{eq:lemm_use3},
\begin{align*}
    \pi_n(\rho)&\leq \max_{j\in\cZ} \ \log |f^{-1}(j)|\\
&  + \sup_{W^{lo}} \ \inf_{\beta^{(n)}\left(Q^{(n)}\right)} \ \sup_{\underbar{$\alpha$}\in\Delta_k\left(V\left(W^{lo}\right)\right)}\sum\limits_{Q^{(n)}\in\mathcal{Q}^{(n)}} 
\underbar{$\alpha$}^n\left(\mathcal{T}_{Q^{(n)}}\right)D\left(\underbar{$\alpha$} \ \left( V\left(W^{lo}\right)\right)^{-1}\Big|\Big|\beta^{(n)}\left(Q^{(n)}\right)\right)\\
&=\max_{j\in\cZ} \ \log |f^{-1}(j)|\\&+  \sup_{\substack{V:\cZ\rightarrow\cZ\\V\left(j'|j'\right)\geq\rho, \ j'\in\cZ}}   \ \inf_{\beta^{(n)}\left(Q^{(n)}\right)} \ \sup_{\underbar{$\alpha$}\in\Delta_k(V)}\sum\limits_{Q^{(n)}\in\mathcal{Q}^{(n)}} 
\underbar{$\alpha$}^n\left(\mathcal{T}_{Q^{(n)}}\right)D\left(\underbar{$\alpha$} \  V^{-1}\big|\big|\beta^{(n)}\left(Q^{(n)}\right)\right).\end{align*}
\noindent Upon choosing $\beta^{(n)}\left(Q^{(n)}\right)=\kappa_n\left(Q^{(n)}\right)$ (see~\eqref{eq:kappa}) and by Definition 6 (ii), which defines $\cV(\rho),$ we get
\[
\pi_n(\rho)\leq\max_{j\in\cZ} \ \log |f^{-1}(j)|+  \sup_{V\in\cV(\rho)}   \  \sup_{\underbar{$\alpha$}\in\Delta_k(V)}\sum\limits_{Q^{(n)}\in\mathcal{Q}^{(n)}} 
\underbar{$\alpha$}^n\left(\mathcal{T}_{Q^{(n)}}\right)D\left(\underbar{$\alpha$} \  V^{-1}\big|\big|\kappa_n\left(Q^{(n)}\right)\right)\]
\noindent which is~\eqref{eq:main-res}.

Next, to show~\eqref{eq:main-thm-lim}, observe that
\[
\Gamma_n(\rho)=
    \sup_{V\in\mathcal{V}(\rho)} \ 
    \sup_{\alpha\in\Delta_k(V)} \ \mathbb{E}_{\underbar{$\alpha$}}\  \left[ D\left(\underbar{$\alpha$} V^{-1}||\kappa_n\left(T_n\right)\right)\right]
\]
\noindent where the $\cQ^{(n)}$-valued rv $T_n$ has underlying pmf \underbar{$\alpha$}$\in\Delta_k(V).$ Continuing
\begin{align}
\Gamma_n(\rho)&=\sup_{V\in\mathcal{V}(\rho)} \ 
    \sup_{\underline{\alpha}\in\Delta_k(V)} \ \mathbb{E}_{\underbar{$\alpha$}}\  \left[\sum_{j\in\cZ}\left(\underbar{$\alpha$}V^{-1}\right)(j)\log\frac{\left(\underbar{$\alpha$}V^{-1}\right)(j)}{\kappa_n(T_n)(j)}\right]\nonumber\\
    &=\sup_{V\in\mathcal{V}(\rho)} \ 
    \sup_{\underline{\alpha}\in\Delta_k(V)} \ \sum_{j\in\cZ} \ \mathbb{E}_{\underbar{$\alpha$}}\  \left[\left(\underbar{$\alpha$}V^{-1}\right)(j)\log\frac{\left(\underbar{$\alpha$}V^{-1}\right)(j)}{\kappa_n(T_n)(j)}\right]\nonumber\\
    &\leq\sum_{j\in\cZ} \ \sup_{V\in\mathcal{V}(\rho)} \ 
    \sup_{\underline{\alpha}\in\Delta_k(V)} \ \mathbb{E}_{\underbar{$\alpha$}}\  \left[\left(\underbar{$\alpha$}V^{-1}\right)(j)\log\frac{\left(\underbar{$\alpha$}V^{-1}\right)(j)}{\kappa_n(T_n)(j)}\right].
    \label{eq:gamma_split}
\end{align}
\noindent Denoting the rvs in $[\cdots]$ in~\eqref{eq:gamma_split} above by
\[
    \Phi_n^j(\underbar{$\alpha$},V) = \Phi_n^j\left(\kappa_n(T_n)(j),\underbar{$\alpha$},V\right) \triangleq \ \left(\underbar{$\alpha$}V^{-1}\right)(j)\log\frac{\left(\underbar{$\alpha$}V^{-1}\right)(j)}{\kappa_n(T_n)(j)}, \ \ \ \ j\in\cZ,
\]
\noindent we show in Appendix~\ref{app:gamma_pf} that
\begin{equation}
    \label{eq:conv_final}
    \lim_n \  \sup_{V\in\cV(\rho)} \ \sup_{\underbar{$\alpha$}\in\Delta_k(V)} \ \mathbb{E}\left[\big|\Phi_n^j(\underbar{$\alpha$},V)\big|\right]=0, \ \ \ j\in\cZ.
\end{equation}
\noindent Then, by~\eqref{eq:gamma_split} and~\eqref{eq:conv_final}, the first assertion in~\eqref{eq:main-thm-lim} obtains. The second assertion in~\eqref{eq:main-thm-lim} follows from~\eqref{eq:gamma_split},~\eqref{eq:conv_final} and~\eqref{eq:priv_guaren},~\eqref{eq:priv_nonasymp_lb} with $l=1$.\qeed
\subsection{Proof of Theorem~\ref{thm:main-achiev}}
\label{subsec:lb}
\vspace{0.1cm}
\begin{figure}[h]
    \centering
    \includegraphics[scale=.75]{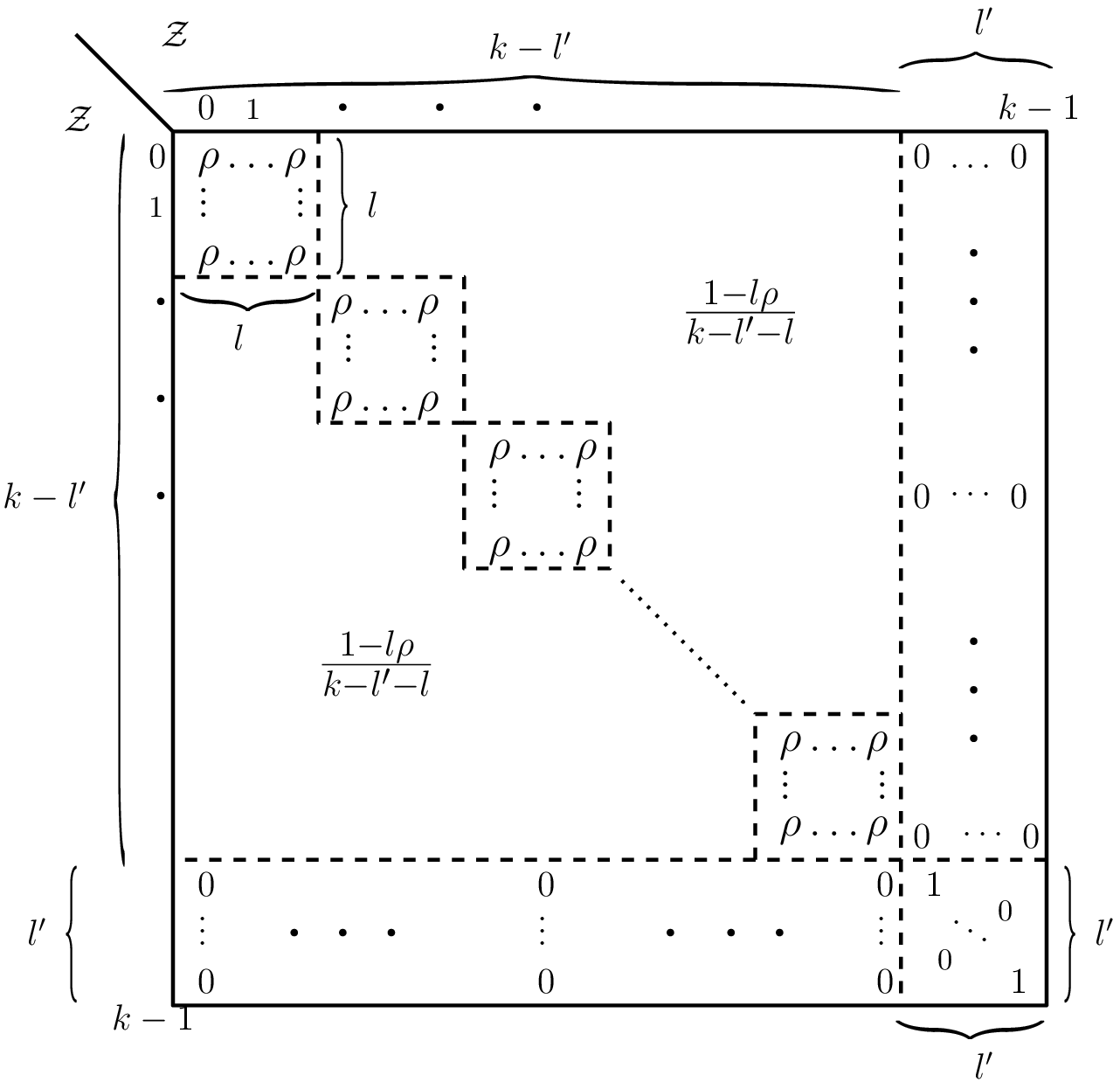}
    \caption{$V_1:\cZ\rightarrow\cZ$}
    \label{fig:V}
\end{figure}
Fix $1/k<\rho\leq 1.$ As in the statement of the theorem, $l=l(\rho)$ is determined by
\begin{equation}
    \label{eq:rho_bd}
    \frac{1}{l+1}<\rho\leq\frac{1}{l}, \ \ \ 1\leq l \leq k-1.
\end{equation}
\noindent We consider separately the cases $l\leq\left\lfloor\frac{k}{2}\right\rfloor$ and $l>\left\lfloor\frac{k}{2}\right\rfloor.$ 

The proof proceeds in the following four steps for each of the cases $l\leq\left\lfloor\frac{k}{2}\right\rfloor$ and\footnotemark\footnotetext{For $k=2,$ only the case $l\leq\left\lfloor\frac{k}{2}\right\rfloor$ occurs.} $l>\left\lfloor\frac{k}{2}\right\rfloor:$ 
\begin{enumerate}[{\it 1.}]
    \item description of chosen locally identical $\rho$-QRs $V_1:\cZ\rightarrow\cZ$ and $V_2:\cZ\rightarrow\cZ$ for the two cases,\\ respectively (see Definition 6 (i));
    \item reduction in the choice of querier estimators induced by $V_1$ and $V_2;$
    \item selection of a set of sparse pmfs with suitable range cardinality;
    \item establishment of the sufficiency of locally uniform querier estimators and identification of a specific \\such estimator.
\end{enumerate}
\noindent These steps are described next with some of the details provided in Appendix~\ref{app:achiev_pf}.\\

\noindent {\bf Case} $l\leq\left\lfloor\frac{k}{2}\right\rfloor:$\\ \\
{\it Step 1}: The user selects $P_X$ (to be specified later) and a locally-identical $\rho$-QR $W_1^{lo}=W_1^{lo}(\rho):\cX\rightarrow\cZ$ described next in terms of an associated stochastic matrix $V_1=V_1\left(W_1^{lo}\right):\cZ\rightarrow\cZ$. It is assumed that the rows of $W_1^{lo}$ are arranged in order, respectively, according to $f^{-1}(0),f^{-1}(1),\ldots,f^{-1}(k-1);$ this entails no loss of generality. Set
\begin{equation}
    \label{eq:s_srho}
l'=l'(\rho)\triangleq k-\left\lfloor\frac{k}{l}\right\rfloor l.
\end{equation}
\noindent Clearly $0\leq l'<l.$ Then, as illustrated in Fig.~\ref{fig:V}, $V_1:\cZ\rightarrow\cZ$ is chosen as follows:
\begin{itemize}
    \item the top-left $(k-l')\times(k-l')$-subblock of $V_1$ consists of $\left\lfloor\frac{k}{l}\right\rfloor=\frac{k-l'}{l}$ diagonal blocks of $l\times l$-matrices with all entries equal to $\rho,$ and with the remaining entries being $\frac{1-l\rho}{k-l'-l};$
    \item the bottom-right $l'\times l'$-subblock is an identity matrix;
    \item the bottom-left $l'\times(k-l')$-subblock and the top-right $(k-l')\times l'$-subblock consist of zeros.
\end{itemize}
\noindent In the specification of $V_1$ above, note that $k-l'-l\geq 1$ since 
\begin{equation}
\label{eq:k/l-lb}
\left\lfloor\frac{k}{l}\right\rfloor \geq \left\lfloor\frac{k}{\left\lfloor\frac{k}{2}\right\rfloor}\right\rfloor\geq \left\lfloor\frac{k}{\frac{k}{2}}\right\rfloor=2
\ \ \ \text{so that} \ \ \
k-l'-l=\left\lfloor\frac{k}{l}\right\rfloor l-l\geq 2l-l=l\geq 1.
\end{equation}
\noindent The rationale for our specific choice of $V_1:\cZ\rightarrow\cZ$ is guided by two features. First, it is advantageous for the user if $V_1$ has as few distinct rows as possible. Second, each diagonal element must be at least $\rho,$ by the $\rho$-recoverability constraint. Thus, the chosen $V_1:\cZ\rightarrow\cZ$ has $\left\lfloor\frac{k}{l}\right\rfloor$ blocks of $l$ rows that are identical (and distinct among such blocks), with $\left\lfloor\frac{k}{l}\right\rfloor$ $l\times l$ $\rho$-blocks along the diagonal, except for boundary fillers.\\ \\
{\it Step 2}: With $V_1:\cZ\rightarrow\cZ$ as above and for any $P_X$ in $\Delta_r,$ $P_X(f^{-1})V_1\in\Delta_k$ has identical entries in each of $\left\lfloor\frac{k}{l}\right\rfloor=\frac{k-l'}{l}$ blocks (each with $l$ entries); and possibly $l'$ distinct entries $P_X\left(f^{-1}(k-l')\right),\ldots,P_X\left(f^{-1}(k-1)\right).$ Accordingly, consider a reduced set resulting from $\cZ,$ namely
\begin{equation}
\label{eq:red_set}
\cZ'\triangleq\left\{0,1,\ldots,\frac{k-l'}{l}-1,\frac{k-l'}{l},\ldots\frac{k-l'}{l}+l'-1\right\}
\end{equation}
\noindent obtained by merging those symbols in $\cZ$ that lie within each of the mentioned blocks, and thereby of diminished cardinality
\begin{equation}
    \label{eq:card_k}
    k'=k'(\rho)\triangleq\frac{k-l'}{l}+l' = \left\lfloor\frac{k}{l}\right\rfloor+l'\leq k
\end{equation}
\begin{figure}[h]
    \centering
    \includegraphics[scale=.45]{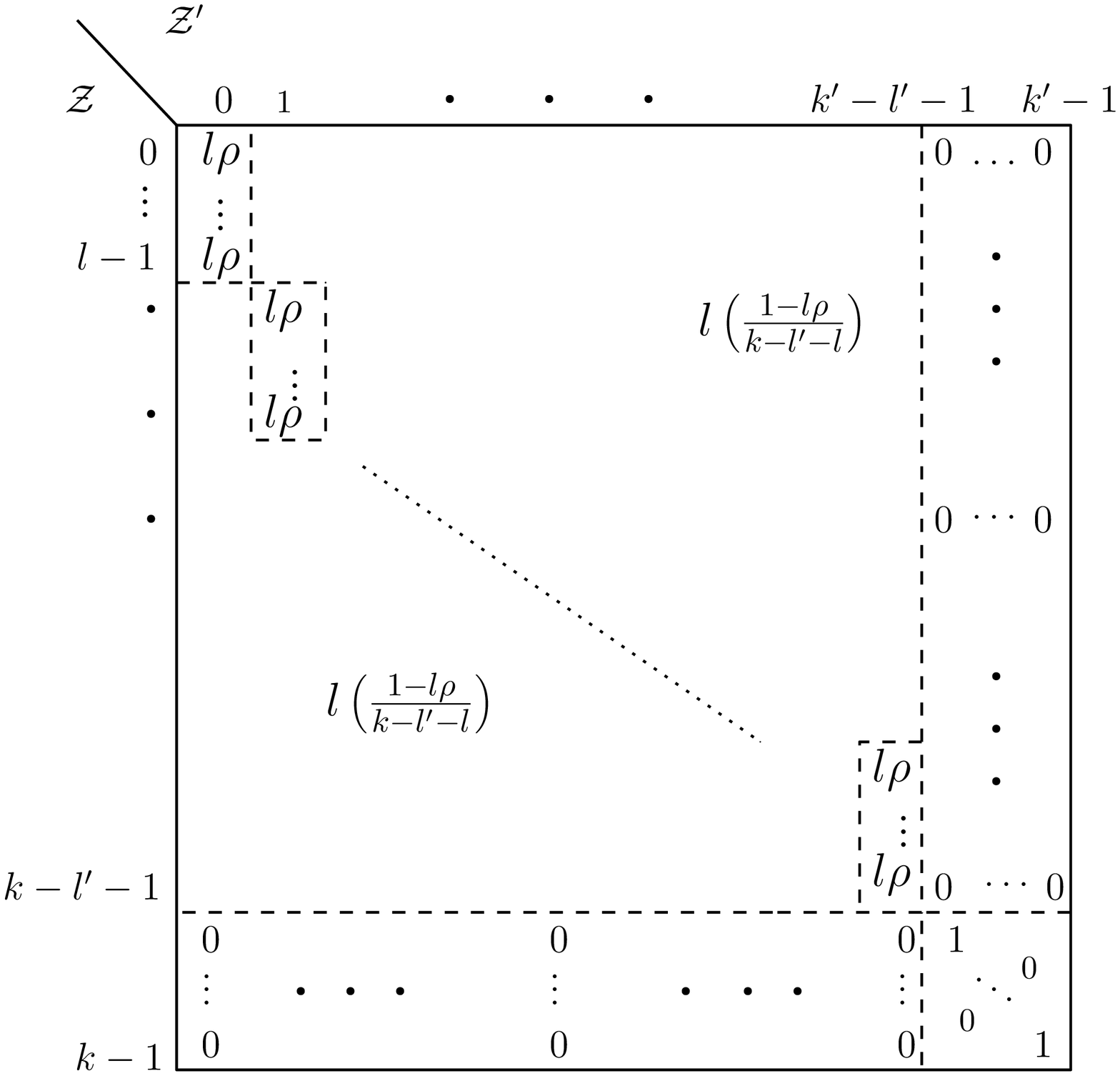}
    \caption{$V'_1:\cZ\rightarrow\cZ'$}
    \label{fig:V'}
\end{figure}
\noindent on which $P_X\left(f^{-1}\right)V_1$ can have possibly different probability values. The resulting merged probabilities on $\cZ'$ are obtained as $P_X\left(f^{-1}\right)V_1',$ where $V_1'=V_1'(V_1):\cZ\rightarrow\cZ'$ is obtained by merging blocks of $l$ columns of $V_1$ (and their elements). Then, $V_1':\cZ\rightarrow\cZ'$ is as illustrated in Fig.~\ref{fig:V'}, and is described as follows.\\ \\
For $0\leq j\leq k-1$ and $0\leq j'\leq k'-1,$
\begin{equation}
\label{eq:Vtilde_rho}
V'_1(j'|j)=\begin{cases}
    l\rho, \ \ \ &j'=0,1,\ldots,k'-l'-1, \ \ \ j=j'\cdot l,\ldots,(j'+1)l-1\\
    l\left(\frac{1-l\rho}{k-l'-l}\right),  \ \ \ &j'=0,1,\ldots,k'-l'-1, \ \ \ j\neq j'\cdot l,\ldots,(j'+1)l-1, \ \ j\leq k-l'-1\\ 
    1,  \ \ \ &(j'=k'-l',j=k-l'),\ldots,(j'=k'-1,j=k-1)\\
    0, \ \ \ &\text{otherwise}.
    \end{cases}
\end{equation}
\noindent For the user's choice of $V_1=V_1\left(W_1^{lo}\right):\cZ\rightarrow\cZ$ as above, its effect on the querier's estimation of $P_X$ is governed by $V'_1=V'_1(V_1):\cZ\rightarrow\cZ'$ in~\eqref{eq:Vtilde_rho}. Let $\cQ'^{(n)}$ denote the set of all types on $\cZ'^n.$ Then precisely, referring to~\eqref{eq:bin-priv-type}, we claim that
\begin{multline}
\label{eq:priv_red_smooth}
    \inf_{\widehat{P}_n:\cQ^{(n)}\rightarrow\Delta_r} \ \sup_{P_X} \ \sum\limits_{Q^{(n)}\in\cQ^{(n)}} \ \left(P_X\left(f^{-1}\right)V_1\right)^n  \left(\cT_{Q^{(n)}}\right)  D\left(P_X\big|\big| \widehat{P}_n\left(Q^{(n)}\right)\right)\\=
        \inf_{\widehat{P}_n:\cQ'^{(n)}\rightarrow\Delta_r} \ \sup_{P_X} \ \sum\limits_{Q'^{(n)}\in\cQ'^{(n)}} \ \left(P_X\left(f^{-1}\right)V_1'\right)^n  \left(\cT_{Q'^{(n)}}\right)  D\left(P_X\big|\big| \widehat{P}_n\left(Q'^{(n)}\right)\right)
\end{multline}
\noindent with an obvious abuse of notation of $\widehat{P}_n$ in the right-side. To this end, observe that all $z^n\in\cT_{Q^{(n)}}$ for some (fixed) $Q^{(n)}\in\cQ^{(n)}$ result in (possibly different) $z'^n\in{\cZ'}^n$ but of a common type $Q'^{(n)}\in\cQ'^{(n)},$ by the merge described in the passage preceding~\eqref{eq:red_set}; furthermore, different $Q^{(n)}\in\cQ^{(n)}$ can map into the same $Q'^{(n)}\in\cQ'^{(n)}.$ Using this observation and mimicking the proof of Lemma~\ref{lem:estim_type}, the claim in~\eqref{eq:priv_red_smooth} follows.

Hereafter in this proof, we restrict attention in~\eqref{eq:priv_red_smooth} to smooth estimators $\widehat{P}_n:\cQ'^{(n)}\rightarrow\Delta_r.$ Then, recalling~\eqref{eq:bin-priv-type}, we get from~\eqref{eq:priv_red_smooth} that
\begin{align}
   & \inf_{\widehat{P}_n\in\mathcal{S}_n(k')} \ \sup_{P_X} \ \pi_n\left(
\rho,W_1^{lo},P_X,\widehat{P}_n\right)\nonumber\\&= \inf_{\widehat{P}_n\in\mathcal{S}_n(k')} \ \sup_{P_X} \ \sum_{Q'^{(n)}\in\cQ'^{(n)}} \ \left(P_X(f^{-1})V_1'\right)^n \left(\mathcal{T}_{Q'^{(n)}}\right) \ D\left(P_X\big|\big|\widehat{P}_n\left(Q'^{(n)}\right)\right)\nonumber\\
    &=\inf_{\widehat{P}_n\in\mathcal{S}_n(k')} \ \sup_{\underline{\alpha}\in\Delta_{k'}} \  \sup_{P_X:P_X(f^{-1})V_1'=\underline{\alpha}} \ \sum_{Q'^{(n)}\in\cQ'^{(n)}} \ \underline{\alpha}^n \left(\mathcal{T}_{Q'^{(n)}}\right) \ D\left(P_X\big|\big|\widehat{P}_n\left(Q'^{(n)}\right)\right)\label{eq:alph_achiec_pf}.
\end{align}
\noindent The sum in the right-side of~\eqref{eq:alph_achiec_pf} is bounded below further as follows: for each $\underline{\alpha}\in\Delta_{k'},$ we restrict attention to those types $Q'^{(n)}\in\cQ'^{(n)}$ that are close to types $Q'^{(n)}(\underline{\alpha})$ which approximate $\underline{\alpha}.$ Then,
\begin{multline}
\label{eq:alph_achiec_pf1}
    \sum_{Q'^{(n)}\in\cQ'^{(n)}} \ \underline{\alpha}^n \left(\mathcal{T}_{Q'^{(n)}}\right) \ D\left(P_X\big|\big|\widehat{P}_n\left(Q'^{(n)}\right)\right)\\
    \geq     \sum_{Q'^{(n)}\in\cQ'^{(n)}: \ \text{var}\left(Q'^{(n)},Q'^{(n)}(\underline{\alpha})\right)\leq\gamma_n} \ \underline{\alpha}^n \left(\mathcal{T}_{Q'^{(n)}}\right) \ D\left(P_X\big|\big|\widehat{P}_n\left(Q'^{(n)}\right)\right)
\end{multline}
\noindent where
\begin{equation}
\label{eq:qna}
Q'^{(n)}(\underline{\alpha})(j')=\frac{\left\lceil n\underline{\alpha}(j')\right\rceil}{n}, \ \ j'=1,\ldots,k'-1
\ \ \ \ \text{and} \ \ \ \  
Q'^{(n)}(\underline{\alpha})(0)=1-\sum\limits_{j'\neq 0}\frac{\left\lceil n\underline{\alpha}(j')\right\rceil}{n}
\end{equation}
\noindent and $\gamma_n$ will be specified below. For $Q'^{(n)}(\underline{\alpha})$ in~\eqref{eq:qna} to be a pmf in $\Delta_{k'},$ it suffices for $\underline{\alpha}\in\Delta_{k'}$ and $n$ to satisfy
\[
1-\sum\limits_{j'\neq 0} \ \frac{\left\lceil n\underline{\alpha}(j')\right\rceil}{n}\geq 0
\]
\noindent  which, in turn, is implied if
\begin{equation}
\label{eq:Q'n-pmf}
 1-\sum\limits_{j'\neq 0}\frac{ n\underline{\alpha}(j')+1}{n}=\underline{\alpha}(0)-\frac{k'-1}{n}\geq 0.
\end{equation}
\noindent In Appendix~\ref{app:achiev_pf}, we shall show that $\underline{\alpha}\in\Delta_{k'}$ can be restricted further and $n$ chosen large enough with
 \begin{equation}
    \label{eq:N0}
    n\geq N_0(k')=2k',
\end{equation}
\noindent so that~\eqref{eq:Q'n-pmf} holds (without any dependence of $N_0$ on $\underline{\alpha}$).

Since $\widehat{P}_n$ is a smooth estimator in $\mathcal{S}_n(k'),$ var$\left(Q'^{(n)},Q'^{(n)}(\underline{\alpha})\right)\leq\gamma_n$ implies\\ var$\left(\widehat{P}_n\left(Q'^{(n)}\right),\widehat{P}_n\left(Q'^{(n)}(\underline{\alpha})\right)\right)\leq\hat{\gamma}_n$ and $\widehat{P}_n\left(Q'^{(n)}(\underline{\alpha})\right)(x)\geq c_n>0, \ x\in\cX$ (see Definition 4).\\ Then in the right-side
of~\eqref{eq:alph_achiec_pf1}, by Lemma~\ref{lem:div_var-dis}
\begin{align}
    D\left(P_X\big|\big|\widehat{P}_n\left(Q'^{(n)}\right)\right)&\geq 
D\left(P_X\big|\big|\widehat{P}_n\left(Q'^{(n)}(\underline{\alpha})\right)\right)-\frac{\text{var}\left(\widehat{P}_n\left(Q'^{(n)}\right),\widehat{P}_n\left(Q'^{(n)}(\underline{\alpha})\right)\right)}{\min\limits_{x\in\cX} \ \widehat{P}_n\left(Q'^{(n)}(\underline{\alpha})\right) (x)}\nonumber\\
&\geq D\left(P_X\big|\big|\widehat{P}_n\left(Q'^{(n)}(\underline{\alpha})\right)\right)-\frac{\hat{\gamma}_n}{c_n}.\label{eq:alph_achiec_pf2}
\end{align}
\noindent Hence in~\eqref{eq:alph_achiec_pf1}, using~\eqref{eq:alph_achiec_pf2},
\begin{multline}
    \label{eq:alph_achiec_pf3}
    \sum_{Q'^{(n)}\in\cQ'^{(n)}} \ \underline{\alpha}^n \left(\mathcal{T}_{Q'^{(n)}}\right) \ D\left(P_X\big|\big|\widehat{P}_n\left(Q'^{(n)}\right)\right)\\\geq \left[
    D\left(P_X\big|\big|\widehat{P}_n\left(Q'^{(n)}(\underline{\alpha})\right)\right)-
    \frac{\hat{\gamma}_n}{c_n}
    \right] \ \sum_{Q'^{(n)}\in\cQ'^{(n)}: \ \text{var}\left(Q'^{(n)},Q'^{(n)}(\underline{\alpha})\right)\leq\gamma_n} \ \underline{\alpha}^n \left(\mathcal{T}_{Q'^{(n)}}\right).
\end{multline}
\noindent Next, in the right-side of~\eqref{eq:alph_achiec_pf3}, with $T_n'$ denoting a $\cQ'^{(n)}$-valued rv with underlying pmf $\underline{\alpha}\in\Delta_{k'},$ we get
\begin{align}
    \sum_{Q'^{(n)}\in\cQ'^{(n)}: \ \text{var}\left(Q'^{(n)},Q'^{(n)}(\underline{\alpha})\right)\leq\gamma_n} \ \underline{\alpha}^n \left(\mathcal{T}_{Q'^{(n)}}\right)&=P\left(\text{var}\left(T_n',Q'^{(n)}(\underline{\alpha})\right)\leq\gamma_n\right)\label{eq:hoeffd_lb1}\\
    &=1-P\left(\text{var}\left(T_n',Q'^{(n)}(\underline{\alpha})\right)>\gamma_n\right)\nonumber\\
    &\geq 1-P\left(\text{var}\left(T_n',\underline{\alpha}\right)+\text{var}\left(\underline{\alpha},Q'^{(n)}(\underline{\alpha})\right)\geq\gamma_n\right)\label{eq:triangular1}\\
    &\geq 1-P\left(\text{var}\left(T_n',\underline{\alpha}\right)\geq\gamma_n-\frac{2(k'-1)}{n}\right)\label{eq:triangular2}
\end{align}
\noindent where~\eqref{eq:triangular1} is by the triangle inequality for var$(\cdot,\cdot),$ and~\eqref{eq:triangular2} holds since $\text{var}\left(\underline{\alpha},Q'^{(n)}(\underline{\alpha})\right)\leq \frac{2(k'-1)}{n}$ by~\eqref{eq:qna}. Denoting 
\begin{equation}
    \label{eq:eps-gamma}
    \epsilon_n=\gamma_n-\frac{2(k'-1)}{n},
\end{equation}
\noindent we obtain by~{\cite[Lemma $3$]{Devroye83}} that 
\begin{equation}
    \label{eq:hoeff_dev}
 P\left(\text{var}\left(T_n',\underline{\alpha}\right)\geq\epsilon_n\right)\leq 3e^{\left(-n\frac{\epsilon_n^2}{25}\right)}
\end{equation}
\noindent for all $n$ such that
\begin{equation}
\label{eq:eps_condn}
\epsilon_n>0 \ \ \ \ 
\text{and}\ \ \ \
n\epsilon_n^2\geq 20k'.
\end{equation}
\noindent Now, pick 
\begin{equation}
    \label{eq:gamma_lognn}
    \gamma_n=5\sqrt{
    \frac{\zeta\ln 2\log n}{n}}, \ \ \ n\geq 1\text{ and }\zeta>1
\end{equation}
\noindent so that $\lim\limits_n \ \gamma_n=0.$ Then~\eqref{eq:eps_condn} holds for all $n\geq N_1(k')$ determined by
\[
5\sqrt{n\zeta\ln 2 \log n}>2(k'-1),
\]
\noindent and for all $n\geq N_2(k')$ determined by
\[
25\zeta\ln 2\log n + \frac{4(k'-1)^2}{n}-20(k'-1)\sqrt{\frac{\zeta\ln 2\log n}{n}}\geq 20k'.
\]
\noindent Then for $n\geq \max\{N_1(k'),N_2(k')\}$,~\eqref{eq:hoeff_dev} holds, and thereby by~\eqref{eq:eps-gamma} and~\eqref{eq:gamma_lognn},
\begin{equation}
P\left(\text{var}\left(T_n',\underline{\alpha}\right)\geq\gamma_n-\frac{2(k'-1)}{n}\right)\leq \frac{3e^{\left(\frac{-4(k'-1)^2}{25n}\right)}e^{\left(\frac{4(k'-1)}{5}\sqrt{\frac{\zeta\ln 2\log n}{n}}\right)}}{n^{\zeta}} \leq\frac{3e^{\left(\frac{4(k'-1)\sqrt{\zeta}}{5\sqrt{e}}\right)}}{n^{\zeta}} 
\end{equation}
\noindent so that in~\eqref{eq:hoeffd_lb1}
\begin{equation}
\label{eq:hoeff_final}
    P\left(\text{var}\left(T_n',Q'^{(n)}(\underline{\alpha})\right)\leq\gamma_n\right) \geq 1-  \frac{3e^{\left(\frac{4(k'-1)\sqrt{\zeta}}{5\sqrt{e}}\right)}}{n^{\zeta}},
\end{equation}
\noindent where the right-side above is nonnegative for $n\geq N_3(k').$ Upon gathering~\eqref{eq:alph_achiec_pf},~\eqref{eq:alph_achiec_pf1},~\eqref{eq:alph_achiec_pf3},~\eqref{eq:hoeffd_lb1} and~\eqref{eq:hoeff_final}, we get that for all $n\geq \max\{N_0(k'),N_1(k'),N_2(k'),N_3(k')\},$
\begin{multline}
\label{eq:pi_n_hoeffding}
    \inf_{\widehat{P}_n\in\mathcal{S}_n(k')} \ \sup_{P_X} \ \pi_n\left(
\rho,W_1^{lo},P_X,\widehat{P}_n\right)\\ \geq \inf_{\widehat{P}_n\in\mathcal{S}_n(k')} \ \sup_{\underline{\alpha}\in\Delta_{k'}} \  \sup_{P_X:P_X(f^{-1})V_1'=\underline{\alpha}} \ \left[ D\left(P_X\big|\big|\widehat{P}_n\left(Q'^{(n)}(\underline{\alpha})\right)\right)-
    \frac{\hat{\gamma}_n}{c_n}\right] \left(1-  \frac{3e^{\left(\frac{4(k'-1)\sqrt{\zeta}}{5\sqrt{e}}\right)}}{n^{\zeta}}\right). 
\end{multline}\\

{\it Step 3}: It remains to reduce the right-side of~\eqref{eq:pi_n_hoeffding} to~\eqref{eq:achiev_thm_main},~\eqref{eq:Lambda-def1},~\eqref{eq:lambda-def}. The main steps are outlined below and the details are given in Appendix~\ref{app:achiev_pf}. First, for $\underline{\alpha}\in\Delta_{k'},$ a straightforward manipulation using~\eqref{eq:Vtilde_rho} shows that $P_X(f^{-1})V_1'=\underline{\alpha}$ can be written as
\begin{gather}
\hspace{-13.8cm}l\rho P_X\left(\bigcup\limits_{j=j'l}^{j'(l+1)-1}f^{-1}(j)\right)\nonumber\\+
    \left(1-P_X\left(\bigcup\limits_{j=j'l}^{j'(l+1)-1}f^{-1}(j)\right)-
    P_X\left(\bigcup\limits_{j=k'-l'}^{k'-1}f^{-1}(j)\right)\right)l\left(\frac{1-l\rho}{k-l'-l}\right)=\underline{\alpha}(j'), \ \ \ j'=0,1,\ldots,k'-l'-1,\label{eq:PXV'_condn1}\\
    \text{and }\hspace{.5cm} P_X\left(f^{-1}(j')\right)=\underline{\alpha}(j'), \ \ \ j'=k'-l',\ldots,k'-1.\label{eq:PXV'_condn}
\end{gather}
\noindent Combining~\eqref{eq:PXV'_condn1} and~\eqref{eq:PXV'_condn}, we get
\begin{equation}
\label{eq:PXV'_condn2}
     P_X\left(\bigcup\limits_{j=j'l}^{j'(l+1)-1}f^{-1}(j)\right)=\frac{\underline{\alpha}(j')-l\left(\frac{1-l\rho}{k-l'-l}\right)\left(1-\sum\limits_{j=k'-l'}^{k'-1}\underline{\alpha}(j)\right)}{l\rho-l\left(\frac{1-l\rho}{k-l'-l}\right)}, \ \ \  j'=0,1,\ldots,k'-l'-1.
\end{equation}
\noindent Then in~\eqref{eq:pi_n_hoeffding}, fixing $\widehat{P}_n\in\mathcal{S}_n(k')$ and $\underline{\alpha}\in\Delta_{k'},$
\begin{equation}
\label{eq:PXV_equiv}
     \sup_{P_X:P_X(f^{-1})V_1'=\underline{\alpha}} \  D\left(P_X\big|\big|\widehat{P}_n\left(Q'^{(n)}(\underline{\alpha})\right)\right)=\sup_{P_X:P_X\sim~ \eqref{eq:PXV'_condn},~\eqref{eq:PXV'_condn2}} \  D\left(P_X\big|\big|\widehat{P}_n\left(Q'^{(n)}(\underline{\alpha})\right)\right)
\end{equation}
\noindent where $P_X\sim \eqref{eq:PXV'_condn},\eqref{eq:PXV'_condn2}$ connotes $P_X$ consistent with~\eqref{eq:PXV'_condn} and~\eqref{eq:PXV'_condn2}. Next, consider a derived mapping $f'^{-1}:\cZ'\rightarrow\cX$ defined in terms of $f^{-1}:\cZ\rightarrow\cX$ as follows:
\begin{equation}
    \label{eq:f'_map}
    f'^{-1}(j')=\begin{cases}
 \bigcup\limits_{j=j'l}^{j'(l+1)-1}f^{-1}(j), \ \ \ &j'=0,1,\ldots,k'-l'-1\\
 f^{-1}(k-k'+j'), \ \ \ &j'=k'-l',\ldots,k'-1
    \end{cases}
\end{equation}
\noindent and define a $k'$-sparse pmf $P_X^{sp}$ on $\cX$ as in Definition 5 with $k', \ j'\in\cZ'$ and $f'^{-1}$ in lieu of $k, \ j\in\cZ$ and $f^{-1}$ therein. The right-side of~\eqref{eq:PXV_equiv} is bounded below further by a restriction to $k'$-sparse pmfs $P_X^{sp}$ on $\cX$ whose support symbols are the lowest $\widehat{P}_n\left(Q'^{(n)}(\underline{\alpha})\right)$-probability symbols within $f'^{-1}(j'), \ j'=0,1,\ldots,k'-1.$ Additionally, pick $\underline{\alpha}\in\Delta_{k'}$ in~\eqref{eq:pi_n_hoeffding} with $\underline{\alpha}(j')=0, \ j'=k'-l',\ldots,k'-1.$ Then a straightforward substitution in~\eqref{eq:pi_n_hoeffding} using~\eqref{eq:PXV_equiv} and~\eqref{eq:f'_map} yields that
\begin{multline}
\label{eq:div-log}
\inf_{\widehat{P}_n\in\mathcal{S}_n(k')} \ \sup_{\underline{\alpha}\in\Delta_{k'}} \  \sup_{P_X:P_X\sim\eqref{eq:PXV'_condn}, \ \eqref{eq:PXV'_condn2}} \  D\left(P_X\big|\big|\widehat{P}_n\left(Q'^{(n)}(\underline{\alpha})\right)\right)\\    
\geq\inf_{\widehat{P}_n\in\mathcal{S}_n(k')}  \sup_{\substack{\underline{\alpha}\in\Delta_{k'}: \\ 
\underline{\alpha}(j')=0, \ j'=k'-l',\ldots,k'-1}} \\ \sum_{j'=0}^{k'-l'-1} \ \frac{\underline{\alpha}(j')-l\left(\frac{1-l\rho}{k-l'-l}\right)}{l\rho-l\left(\frac{1-l\rho}{k-l'-l}\right)}\log \left(\frac{\underline{\alpha}(j')-l\left(\frac{1-l\rho}{k-l'-l}\right)}{l\rho-l\left(\frac{1-l\rho}{k-l'-l}\right)}\frac{1}{\min\limits_{x\in f'^{-1}(j')}\ \widehat{P}_n\left(Q'^{(n)}(\underline{\alpha})\right)(x)} \right),
\end{multline}
\noindent where the coefficient of each $\log$ term is in $[0,1].$ \\ \\{\it Step 4}: In~\eqref{eq:div-log}, observe that for every $\widehat{P}_n:\cQ'^{(n)}\rightarrow\cX,$ there exists a locally uniform estimator (see Definition 3, with $k'$ replacing $k$ and $f'^{-1}$ replacing $f^{-1}$), depending on
$\widehat{P}_n,$ and specified by
\[
\beta^{(n)}=\left\{\beta^{(n)}\left(Q'^{(n)}\right)\right\}_{Q'^{(n)}\in\cQ'^{(n)}}, \ \ \ \ n\geq 1,
\]
\noindent with $\beta^{(n)}\left(Q'^{(n)}\right)=\left(\beta_0^{(n)}\left(Q'^{(n)}\right),\beta_1^{(n)}\left(Q'^{(n)}\right),\ldots,\beta_{k'-1}^{(n)}\left(Q'^{(n)}\right)\right)\in\Delta_{k'}$ and
\begin{align*}
    \widehat{P}_n^{\beta^{(n)}}\left(Q'^{(n)}\right)(x)&=\frac{\beta_{j'}^{(n)}\left(Q'^{(n)}\right)}{\big|f'^{-1}(j')\big|}, \ \ \ \ x\in f'^{-1}(j'), \ \ j'\in\cZ'\nonumber\\
    &=\frac{\widehat{P}_n\left(Q'^{(n)}\right)\left(f'^{-1}(j')\right)}{\big|f'^{-1}(j')\big|}, \ \ \ \ x\in f'^{-1}(j'), \ \ j'\in\cZ'\label{eq:unif_estim}
\end{align*}
\noindent and with the obvious property that
\[
\frac{1}{\min\limits_{x\in f'^{-1}(j')}\ \widehat{P}_n\left(Q'^{(n)}(\underline{\alpha})\right)(x)}\geq 
\frac{\big|f'^{-1}(j')\big|}{\widehat{P}_n\left(Q'^{(n)}(\underline{\alpha})\right)\left(f'^{-1}(j')\right)}
, \ \ \ \ j'\in\cZ'.\]
\noindent Hence, $\inf\limits_{\widehat{P}_n}$ in the right-side of~\eqref{eq:div-log} can be restricted to $\inf\limits_{\widehat{P}_n^{\beta^{(n)}}},$ and becomes
\begin{multline}
\label{eq:unif_form}
\inf_{\widehat{P}_n^{\beta^{(n)}}\in\mathcal{S}_n(k')} \ \sup_{\substack{\underline{\alpha}\in\Delta_{k'}:\\ 
\underline{\alpha}(j')=0, \ j'=k'-l',\ldots,k'-1}} \\ \sum_{j'=0}^{k'-l'-1} \ \frac{\underline{\alpha}(j')-l\left(\frac{1-l\rho}{k-l'-l}\right)}{l\rho-l\left(\frac{1-l\rho}{k-l'-l}\right)}\log \left(\frac{\underline{\alpha}(j')-l\left(\frac{1-l\rho}{k-l'-l}\right)}{l\rho-l\left(\frac{1-l\rho}{k-l'-l}\right)}\frac{\big|f'^{-1}(j')\big|}{\widehat{P}_n^{\beta^{(n)}}\left(Q'^{(n)}(\underline{\alpha})\right)\left(f'^{-1}(j')\right)} \right).
\end{multline}
\noindent Finally, a further lower bound for~\eqref{eq:unif_form} in Appendix~\ref{app:achiev_pf}, taken together with~\eqref{eq:pi_n_hoeffding}, yields
\begin{align}
\label{eq:1}
\pi_n(\rho)
& \geq \left(\log\big|f'^{-1}(0)\big|+\Lambda_n(\rho)\right)\lambda_n(\rho)
\end{align}
\noindent where
\begin{equation}
\label{eq:2}
\Lambda_n(\rho)\triangleq
\begin{cases}\log\left(1+\frac{\sum\limits_{j'=1}^{k'-l'-1}\big |f'^{-1}(j')\big|}{e\big |f'^{-1}(0)\big|}
\frac{\min\left\{\frac{\left\lceil nl\left(\frac{1-l\rho}{k-l'-l}\right)\right\rceil}{n}, \ l\rho\right\}-l\left(\frac{1-l\rho}{k-l'-l}\right)}{l\rho-l\left(\frac{1-l\rho}{k-l'-l}\right)}\right)
-
    \frac{\hat{\gamma}_n}{c_n}, \ \ &{\sum\limits_{j'=1}^{k'-l'-1}\big |f'^{-1}(j')\big|}\leq{\big |f'^{-1}(0)\big|}\\
    \log\left(\frac{\sum\limits_{j'=1}^{k'-l'-1}\big |f'^{-1}(j')\big|}{\big |f'^{-1}(0)\big|}\right)\frac{\min\left\{\frac{\left\lceil nl\left(\frac{1-l\rho}{k-l'-l}\right)\right\rceil}{n},\ l\rho\right\}-l\left(\frac{1-l\rho}{k-l'-l}\right)}{l\rho-l\left(\frac{1-l\rho}{k-l'-l}\right)}-
    \frac{\hat{\gamma}_n}{c_n}, \ \ &{\sum\limits_{j'=1}^{k'-l'-1}\big |f'^{-1}(j')\big|}>{\big |f'^{-1}(0)\big|}
    \end{cases}\end{equation}
    \noindent and 
    \begin{equation}
    \label{eq:3}
        \lambda_n(\rho)\triangleq 1-  \frac{3e^{\left(\frac{4(k'-1)\sqrt{\zeta}}{5\sqrt{e}}\right)}}{n^{\zeta}}.
    \end{equation}
\noindent From~\eqref{eq:f'_map}, the passage following it, and~\eqref{eq:div-log}, we have that the user-selected $P_X$ is a $k'$-sparse pmf with associated mapping $f'$ described by~\eqref{eq:f'_map}. Additional details of the specific $k'$-sparse pmf chosen by the user are provided in Appendix~\ref{app:achiev_pf}.

Thus, for the case $l\leq \left\lfloor\frac{k}{2}\right\rfloor,$~\eqref{eq:achiev_thm_main} with~\eqref{eq:Lambda-def1},~\eqref{eq:lambda-def} follow from~\eqref{eq:1},~\eqref{eq:2},~\eqref{eq:3} upon recalling~\eqref{eq:s_srho},~\eqref{eq:card_k} and~\eqref{eq:f'_map}.
\begin{figure}
    \centering
    \includegraphics[scale=0.55]{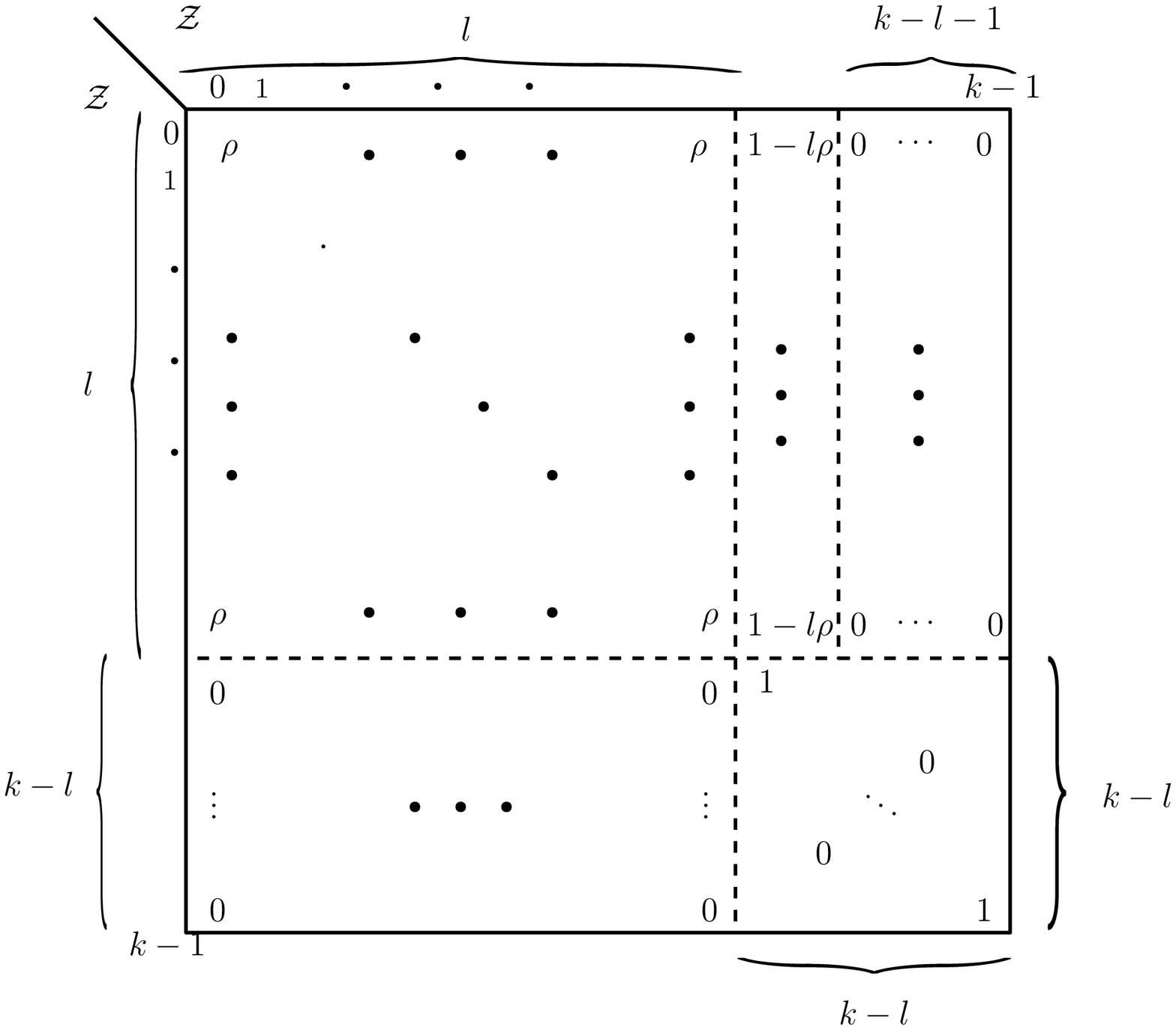}
    \caption{$V_2:\cZ\rightarrow\cZ$}
    \label{fig:V_2}
\end{figure}
\vspace{0.15cm}\\

\noindent{\bf Case} $l>\left\lfloor\frac{k}{2}\right\rfloor:$\\ \\
{\it Step 1}: The user selects a locally-identical $\rho$-QR $W_2^{lo}=W_2^{lo}(\rho):\cX\rightarrow\cZ$ described in terms of an associated stochastic matrix $V_2=V_2\left(W_2^{lo}\right):\cZ\rightarrow\cZ$ and under the assumption that the rows of $W_2^{lo}$ are arranged in order, as in the previous case, according to $f^{-1}(0),f^{-1}(1),\ldots,f^{-1}(k-1).$ The user-selected $P_X$ will be specified later. Let $l'=l'(\rho), \  \cZ'$ and $k'=k'(\rho)$ be as in~\eqref{eq:s_srho},~\eqref{eq:red_set} and~\eqref{eq:card_k}, respectively. Note that
\[
\frac{k}{l}\leq\frac{k}{\left\lfloor\frac{k}{2}\right\rfloor+1}\leq \frac{k}{\frac{k+1}{2}}=\frac{2}{1+\frac{1}{k}}< 2 \ \ \ \text{and} \ \ \  \frac{k}{l}\geq \frac{k}{k-1}\geq 1 \ \ \  \text{implies} \ \ \  \left\lfloor\frac{k}{l}\right\rfloor=1.
\] \noindent Hence, $l'=k-l\geq 1$ and $k'=1+k-l.$ As illustrated in Fig.~\ref{fig:V_2}, $V_2:\cZ\rightarrow\cZ$ is chosen as follows:
\begin{itemize}
    \item all the entries of the top-left $l\times l$-subblock are $\rho;$ 
    \item the bottom-right $(k-l)\times (k-l)$-subblock is an identity matrix;
    \item the bottom-left $(k-l)\times l$-subblock and the top-right $l\times (k-l-1)$-subblock consist of zeros;
   \item the remaining entries are $1-l\rho.$
\end{itemize}
\noindent The rationale for this structure of $V_2:\cZ\rightarrow\cZ$ is similar to that for the case $l\leq\left\lfloor\frac{k}{2}\right\rfloor$ (see passage following~\eqref{eq:k/l-lb}), noting that $\left\lfloor\frac{k}{l}\right\rfloor=1$ gives a single $\rho$-block.\\ \\
\noindent {\it Step 2}: Consider the stochastic matrix $V'_2=V'_2(V_2):\cZ\rightarrow\cZ',$ obtained by merging the first $l$ columns of $V_2$, and described next.\\ \\
For $0\leq j\leq k-1$ and $0\leq j'\leq k'-1,$
\begin{equation}
\label{eq:Vtilde2_rho}
V_2'(j'|j)=\begin{cases}
    l\rho, \ \ \ &j'=0, \ \ \ j=0,1,\ldots,l-1\\
    1-l\rho,  \ \ \ &j'=1, \ \ \ j=0,1,\ldots,l-1\\
    1,  \ \ \ &(j'=1,j=l),\ldots,(j'=k'-1,j=k-1)\\
    0, \ \ \ &\text{otherwise}.
    \end{cases}
\end{equation}
\noindent Using identical arguments as in the case $l\leq \left\lfloor\frac{k}{2}\right\rfloor$ with $V_1$ and $V_1'$ replaced by $V_2$ and $V_2',$ respectively, the claim~\eqref{eq:priv_red_smooth} holds and we get
\begin{align*}
        &\inf_{\widehat{P}_n\in\mathcal{S}_n(k')} \ \sup_{P_X} \ \pi_n\left(
\rho,W_2^{lo},P_X,\widehat{P}_n\right)\\&= \inf_{\widehat{P}_n\in\mathcal{S}_n(k')} \ \sup_{P_X} \ \sum_{Q'^{(n)}\in\cQ'^{(n)}} \ \left(P_X(f^{-1})V'_2\right)^n \left(\mathcal{T}_{Q'^{(n)}}\right) \ D\left(P_X\big|\big|\widehat{P}_n\left(Q'^{(n)}\right)\right)\\
    &=\inf_{\widehat{P}_n\in\mathcal{S}_n(k')} \ \sup_{\underline{\alpha}\in\Delta_{k'}} \  \sup_{P_X:P_X(f^{-1})V'_2=\underline{\alpha}} \ \sum_{Q'^{(n)}\in\cQ'^{(n)}} \ \underline{\alpha}^n \left(\mathcal{T}_{Q'^{(n)}}\right) \ D\left(P_X\big|\big|\widehat{P}_n\left(Q'^{(n)}\right)\right).
\end{align*}
\noindent Following the same steps from~\eqref{eq:alph_achiec_pf} -~\eqref{eq:pi_n_hoeffding}, we get that for all $n\geq\max\{N_0(k'),N_1(k'),N_2(k'),N_3(k')\}$
\begin{multline}
\label{eq:pi_n_hoeffding1}
    \inf_{\widehat{P}_n\in\mathcal{S}_n(k')} \ \sup_{P_X} \ \pi_n\left(
\rho,W_2^{lo},P_X,\widehat{P}_n\right)\\ \geq \inf_{\widehat{P}_n\in\mathcal{S}_n(k')} \ \sup_{\underline{\alpha}\in\Delta_{k'}} \  \sup_{P_X:P_X(f^{-1})V_2'=\underline{\alpha}} \ \left[ D\left(P_X\big|\big|\widehat{P}_n\left(Q'^{(n)}(\underline{\alpha})\right)\right)-
    \frac{\hat{\gamma}_n}{c_n}\right] \left(1-  \frac{3e^{\left(\frac{4(k'-1)\sqrt{\zeta}}{5\sqrt{e}}\right)}}{n^{\zeta}}\right),
\end{multline}
\noindent where $N_0(k'),N_1(k'),N_2(k'),N_3(k')$ are described in~\eqref{eq:N0},~\eqref{eq:eps_condn} -~\eqref{eq:pi_n_hoeffding}.\\ \\
{\it Step 3}: In the right-side of~\eqref{eq:pi_n_hoeffding1},
\begin{multline}
\label{eq:ee}
    \sup_{\underline{\alpha}\in\Delta_{k'}} \  \sup_{P_X:P_X(f^{-1})V_2'=\underline{\alpha}} \  D\left(P_X\big|\big|\widehat{P}_n\left(Q'^{(n)}(\underline{\alpha})\right)\right)\\ \geq 
        \sup_{\substack{\underline{\alpha}\in\Delta_{k'}:\\
        \underline{\alpha}(j')=0, \ j'=2,\ldots,k'-1}} \  \sup_{P_X:P_X(f^{-1})V_2'=\underline{\alpha}} \  D\left(P_X\big|\big|\widehat{P}_n\left(Q'^{(n)}(\underline{\alpha})\right)\right).
\end{multline}
\noindent If $\underline{\alpha}\in\Delta_{k'}$ is such that\footnotemark  $\underline{\alpha}(j')=0, \ j'=2,\ldots,k'-1,$ then\footnotetext{For $l=k-1,$ i.e., $l'=1, \ k'=2,$ there are no constraints  on $\underline{\alpha}.$} $P_X(f^{-1})V_2'=\underline{\alpha}$, using~\eqref{eq:Vtilde2_rho}, gives
\begin{align}
    P_X\left(f'^{-1}(0)\right)&=\frac{\underline{\alpha}(0)}{l\rho}\label{eq:e1}\\
    P_X\left(f'^{-1}(1)\right)&=\frac{\underline{\alpha}(1)-(1-l\rho)}{l\rho}\label{eq:e2}\\
    P_X\left(f'^{-1}(j')\right)&=0, \ \ \ \  j'=2,\ldots,k'-1,\label{eq:e3}
\end{align}
\noindent where $f'^{-1}:\cZ'\rightarrow\cX$ is defined in~\eqref{eq:f'_map}. Then in the right-side of~\eqref{eq:ee},
\begin{multline}
\label{eq:ri}
\sup_{\substack{\underline{\alpha}\in\Delta_{k'}:\\
        \underline{\alpha}(j')=0, \ j'=2,\ldots,k'-1}} \  \sup_{P_X:P_X(f^{-1})V_2'=\underline{\alpha}} \  D\left(P_X\big|\big|\widehat{P}_n\left(Q'^{(n)}(\underline{\alpha})\right)\right)\\
        =\sup_{\substack{\underline{\alpha}\in\Delta_{k'}:\\
        \underline{\alpha}(j')=0, \ j'=2,\ldots,k'-1}} \  \sup_{P_X:P_X\sim~\eqref{eq:e1},~\eqref{eq:e2},~\eqref{eq:e3}} \  D\left(P_X\big|\big|\widehat{P}_n\left(Q'^{(n)}(\underline{\alpha})\right)\right).
\end{multline}
\noindent The right-side of~\eqref{eq:ri} is bounded below further by a restriction to $k'$-sparse pmfs $P_X^{sp}$ on $\cX$ whose support symbols are the lowest $\widehat{P}_n\left(Q'^{(n)}(\underline{\alpha})\right)$-probability symbols within $f'^{-1}(j'), \ j'=0,1,\ldots,k'-1,$ and we get
\begin{multline}
\label{eq:qq}
     \sup_{\substack{\underline{\alpha}\in\Delta_{k'}:\\
        \underline{\alpha}(j')=0, \ j'=2,\ldots,k'-1}} \  \sup_{P_X:P_X\sim~\eqref{eq:e1},~\eqref{eq:e2},~\eqref{eq:e3}} \  D\left(P_X\big|\big|\widehat{P}_n\left(Q'^{(n)}(\underline{\alpha})\right)\right)\\\geq 
             \sup_{\substack{\underline{\alpha}\in\Delta_{k'}:\\
        \underline{\alpha}(j')=0, \ j'=2,\ldots,k'-1}} \ \frac{\underline{\alpha}(0)}{l\rho}\log \left(\frac{\underline{\alpha}(0)}{l\rho}\frac{1}{\min\limits_{x\in f'^{-1}(0)}\ \widehat{P}_n\left(Q'^{(n)}(\underline{\alpha})\right)(x)} \right)
 \\
+ \frac{\underline{\alpha}(1)-(1-l\rho)}{l\rho}\log \left(\frac{\underline{\alpha}(1)-(1-l\rho)}{l\rho}\frac{1}{\min\limits_{x\in f'^{-1}(1)}\ \widehat{P}_n\left(Q'^{(n)}(\underline{\alpha})\right)(x)} \right).
 \end{multline}
\noindent Using~\eqref{eq:qq} in~\eqref{eq:pi_n_hoeffding1}, we obtain
\begin{multline*}
\inf_{\widehat{P}_n\in\mathcal{S}_n(k')} \ \sup_{\underline{\alpha}\in\Delta_{k'}} \  \sup_{P_X:P_X(f^{-1})V_2'=\underline{\alpha}} \  D\left(P_X\big|\big|\widehat{P}_n\left(Q'^{(n)}(\underline{\alpha})\right)\right) \geq \\
\inf_{\widehat{P}_n\in\mathcal{S}_n(k')} \
             \sup_{\substack{\underline{\alpha}\in\Delta_{k'}:\\
        \underline{\alpha}(j')=0, \ j'=2,\ldots,k'-1}}  \frac{\underline{\alpha}(0)}{l\rho}\ \log \left(\frac{\underline{\alpha}(0)}{l\rho}\frac{1}{\min\limits_{x\in f'^{-1}(0)}\ \widehat{P}_n\left(Q'^{(n)}(\underline{\alpha})\right)(x)} \right)\\
+ \ \frac{\underline{\alpha}(1)-(1-l\rho)}{l\rho}\log \left(\frac{\underline{\alpha}(1)-(1-l\rho)}{l\rho}\frac{1}{\min\limits_{x\in f'^{-1}(1)}\ \widehat{P}_n\left(Q'^{(n)}(\underline{\alpha})\right)(x)} \right).
\end{multline*}\\ \\
\noindent {\it Step 4}: Using the same reasoning as in the case $l\leq\left\lfloor\frac{k}{2}\right\rfloor,$ we can restrict $\inf\limits_{\widehat{P}_n\in\mathcal{S}_n(k')}$ above to locally uniform estimators $\inf\limits_{\widehat{P}_n^{\beta^{(n)}}\in\mathcal{S}_n(k')}$ (see Definition 3) so that
\begin{multline}
\label{eq:V2lb}
\inf_{\widehat{P}_n^{\beta^{(n)}}\in\mathcal{S}_n(k')} \              \sup_{\substack{\underline{\alpha}\in\Delta_{k'}:\\
        \underline{\alpha}(j')=0, \ j'=2,\ldots,k'-1}} \ \frac{\underline{\alpha}(0)}{l\rho}\log \left(\frac{\underline{\alpha}(0)}{l\rho}\frac{\big|f'^{-1}(0)\big|}{ \widehat{P}_n^{\beta^{(n)}}\left(Q'^{(n)}(\underline{\alpha})\right)\left(f'^{-1}(0)\right)} \right)
 \\
+ \frac{\underline{\alpha}(1)-(1-l\rho)}{l\rho}\log \left(\frac{\underline{\alpha}(1)-(1-l\rho)}{l\rho}\frac{\big|f'^{-1}(1)\big|}{ \widehat{P}_n^{\beta^{(n)}}\left(Q'^{(n)}(\underline{\alpha})\right)\left(f'^{-1}(1)\right)} \right).
\end{multline}
\noindent A further lower bound for~\eqref{eq:V2lb} in Appendix~\ref{app:achiev_pf}, taken together with~\eqref{eq:pi_n_hoeffding1}, yields~\eqref{eq:1} where
\begin{equation}
\label{eq:22}
    \Lambda_n(\rho)\triangleq
\log\left(1+\frac{\big |f'^{-1}(1)\big|}{e\big |f'^{-1}(0)\big|}
\frac{\frac{\left\lceil n(1-l\rho)\right\rceil}{n}-(1-l\rho)}{l\rho}
\right)-
    \frac{\hat{\gamma}_n}{c_n}    
    \end{equation}
    \noindent and 
\begin{equation}
\label{eq:33}
\lambda_n(\rho)\triangleq 1-  \frac{3e^{\left(\frac{4\left(k' -1\right)\sqrt{\zeta}}{5\sqrt{e}}\right)}}{n^{\zeta}}.
\end{equation}
\noindent From~\eqref{eq:V2lb}, we have that the user-selected $P_X$ is a $k'$-sparse pmf with associated mapping $f'$ described by~\eqref{eq:f'_map}. Additional details of the specific $k'$-sparse pmf chosen by the user are provided in Appendix~\ref{app:achiev_pf}.

Hence, when $l> \left\lfloor\frac{k}{2}\right\rfloor,$ we get~\eqref{eq:achiev_thm_main} along with~\eqref{eq:Lambda-def2},~\eqref{eq:lambda-def}  from~\eqref{eq:1},~\eqref{eq:22},~\eqref{eq:33} upon recalling~\eqref{eq:s_srho},~\eqref{eq:card_k} and~\eqref{eq:f'_map}.
\vspace{0.15cm}

Finally, in both cases $l\leq\left\lfloor\frac{k}{2}\right\rfloor$ and $l>\left\lfloor\frac{k}{2}\right\rfloor,$~\eqref{eq:achiev_thm_limres} follows from~\eqref{eq:Lambda-def1} -~\eqref{eq:lambda-def} and~\eqref{eq:smooth_estim_lims}; in particular, in the former case, observe in~\eqref{eq:Lambda-def1} that
\begin{align*}
    l\left(\frac{1-l\rho}{\left\lfloor\frac{k}{l}\right\rfloor l-l}\right)&=l\left(\frac{1-l\rho}{k-l'-l}\right), \ \ \ \text{by~\eqref{eq:s_srho}}\\
    &\leq l\rho, \ \ \ \text{by~\eqref{eq:lrho-inequality} (in Appendix~\ref{app:achiev_pf}).}
\end{align*}
\noindent Also,~\eqref{eq:pi_n_lblimit} is immediate from~\eqref{eq:achiev_thm_main} and~\eqref{eq:achiev_thm_limres}.
\qeed

\section{Discussion}
\label{sec:disc}

A minimum level of distribution $\rho$-privacy equal to worst-case privacy $\omega(\rho), \ 0\leq\rho\leq 1,$ is guaranteed by Theorem~\ref{thm:nonasymp_bnd} since $\pi_n(\rho)\geq\omega(\rho)$ for every $n\geq 1,$ with $\omega(\rho)$ being achievable. Under Assumption~\eqref{eq:priv_nonasymp_lbb}, $\omega(\rho)=\log\sum\limits_{j=0}^{l-1}\big|f^{-1}(j)\big|,$ where $l$ is determined by $\rho$ (see~\eqref{eq:priv_guaren}). For low recoverability, i.e., $\rho\leq 1/k,$ the user can pick a $\rho$-QR $W:\cX\rightarrow\cZ$ with all entries $=1/k,$ which renders $P_XW$ to be the uniform pmf on $\cZ$ for any $P_X.$ Clearly, the querier's best estimate for $P_X$ is the uniform pmf on $\cX,$ with a resulting distribution $\rho$-privacy of $\log r,$ which also equals $\omega(\rho)$ for $\rho\leq 1/k.$ On the other hand, for high recoverability, i.e., $\rho>0.5,$ for the user's choice of any pmf $P_X$ and $\rho$-QR $W,$ standard estimation methods show that the querier can estimate exactly $P_X\left(f^{-1}\right)$ in $\Delta_k$ (see~\eqref{eq:PXf1PXf}) from $Z^n$ as $n\rightarrow\infty.$ Then, for $\rho>0.5,$ distribution $\rho$-privacy informally equals $\inf\limits_{g:\Delta_k\rightarrow\Delta_r} \ \sup\limits_{P_X\in\Delta_r} \ D\left(P_X\big|\big|g\left(P_X\left(f^{-1}\right)\right)\right),$ where 
$g$ is an ``estimator'' of $P_X$ on the basis of $P_X\left(f^{-1}\right).$ It is shown in Appendix~\ref{app:asymp-rho0.5} that
\begin{equation}
    \label{eq:asymp_rho0.5}
    \inf_{g:\Delta_k\rightarrow\Delta_r} \ \sup_{P_X\in\Delta_r} \ D\left(P_X\big|\big|g\left(P_X\left(f^{-1}\right)\right)\right)=\max_{j\in\cZ} \ \log\big|f^{-1}(j)\big|
\end{equation}
\noindent thereby explaining the value of $\omega(\rho)$ for $\rho>0.5$ in~\eqref{eq:priv_guaren} with $l=1.$ Intermediate increasing values of $\rho$ in $(1/k,0.5]$ give $\omega(\rho)=\log\sum\limits_{j=0}^{l-1}\big|f^{-1}(j)\big| $ with $l$ decreasing from $k-1$ to $2$ in~\eqref{eq:priv_guaren},~\eqref{eq:omega-nondec}.
 
Theorem~\ref{thm:main-achiev} shows how achievable distribution $\rho$-privacy can be improved beyond worst-case privacy $\omega(\rho)$ for large and suitable but finite $n$. The underlying heuristic that governs appropriate user and querier strategies is as follows. Under the function $\rho$-recoverability constraint, the user picks a sparse pmf for the data and a locally identical $\rho$-QR that serve to smear the resulting pmf on $\cZ$ to be nearly uniform, at least over a subset corresponding to the images of the largest atoms in $\cX$ induced by $f^{-1}.$ The querier thereby is able to recover the function value with probability at least $\rho,$ but the attendant estimate of the data is forced to be nearly uniform over such atoms. The resulting gain in achievable distribution $\rho$-privacy over $\omega(\rho)$ is specified by Theorem~\ref{thm:main-achiev}.

The preceding observations show why distribution $\rho$-privacy defined in terms of divergence in~\eqref{eq:div} leads to useful insights in Theorems~\ref{thm:nonasymp_bnd} and~\ref{thm:main-achiev}, complemented by the converse Theorem~\ref{thm:main-conv} that is valid in the practically interesting regime $\rho>0.5.$ 

However, our analysis techniques suffer from shortcomings, too, suggesting room for improvement. Specifically, our approach fails to deliver a converse when $1/k<\rho\leq 0.5$ (owing to a potential noninvertibility of $V^{-1}$ for $\rho\leq 0.5;$ see Remark (ii) after Definition 6). On a related note, Theorems~\ref{thm:nonasymp_bnd} and~\ref{thm:main-conv} imply that 
\[
\lim_n \ \pi_n(\rho)=\omega(\rho), \ \ 0\leq \rho\leq 1/k \ \ \text{and} \ \ 0.5<\rho\leq 1.
\]
\noindent It remains unknown if the limit above holds also for $1/k<\rho\leq 0.5$. Next, in the proof of Theorem~\ref{thm:main-achiev}, our approach resorts to a smooth estimator for the querier (see Definition 4, especially its properties in~\eqref{eq:smooth_estim_lims}) in order to evade a divergence-borne fly in the ointment in~\eqref{eq:alph_achiec_pf}-\eqref{eq:alph_achiec_pf2}. However, this raises the question of whether the converse Theorem~\ref{thm:main-conv}, if restricted to such smooth querier's estimators, could lead to an increased upper bound than in~\eqref{eq:omega-gamma},~\eqref{eq:main-res}. This, too, remains unanswered. Furthermore, the additional restriction placed on large $n$ in Theorem~\ref{thm:main-achiev}, as mentioned in Remark (iii) after the statement of the theorem, is a weakness. We conjecture that the $\rho$-QRs $W_1^{lo}:\cX\rightarrow\cZ$ and $W_2^{lo}:\cX\rightarrow\cZ$ in the proof of Theorem~\ref{thm:main-achiev} in Section~\ref{subsec:lb} are adequate even without this restriction.

We close with a suggested framework for examining our work in the context of differential privacy and affiliated approaches to data distribution estimation~\cite{Duchi16},~\cite{Kairouz16},~\cite{Ye17},~\cite{Pastore18}. Consider a database that hosts multiple users' data that, in our setting, constitutes a data vector. Differential privacy stipulates that altering a data
vector slightly leads only to a near-imperceptible change in
the corresponding probability distribution of the output of the privacy mechanism, i.e., query responses that are randomized functions of data vectors. An altered formulation would seek to adhere to the function recoverability requirement and additionally guarantee differential privacy for the data vector, while simultaneously maximizing distribution privacy. The user selects the data pmf and the randomization mechanism, the latter enabling function recoverability while simultaneously being locally differentially private ($\epsilon$-LDP) ~\cite{Duchi16},~\cite{Kairouz16},~\cite{Ye17},~\cite{Pastore18}. Specifically, $\epsilon$-LDP, $\epsilon>0$, requires a QR $W:\cX\rightarrow\cZ$ to satisfy 
\begin{equation}
\label{eq:LDP}
\frac{W(j|x')}{W(j|x)}\leq e^{\epsilon}, \ \ \ \ x,x'\in\cX, \ \ j\in\cZ.
\end{equation}
\noindent For each $j\in\cZ$, $\rho$-recoverability in~\eqref{eq:rho-recovW} and~\eqref{eq:LDP} with $x'\in f^{-1}(j)$ imply that
\[
W(j|x)\geq \begin{cases}
\rho, \ \ &x\in f^{-1}(j)\\
\frac{\rho}{e^{\epsilon}}, \ \ &x\notin f^{-1}(j).
\end{cases}
\]
\noindent Furthermore, we must have
\[
1=\sum_{j\in\cZ} \ W(j|x)\geq \rho+(k-1)\frac{\rho}{e^{\epsilon}}, \ \ \ x\in\cX
\]
\noindent whence
\begin{equation}
    \label{eq:rho-eps}
    \rho\leq\frac{1}{1+(k-1)e^{-\epsilon}}<1.
\end{equation}
\noindent Thus, an $\epsilon$-LDP QR $W:\cX\rightarrow\cZ$ will be a $\rho$-QR for $\rho$ as in~\eqref{eq:rho-eps}, implying that $f$ cannot be $\rho$-recoverable for $\rho$ violating~\eqref{eq:rho-eps}. Also, it is significant that the $\epsilon$-LDP requirement~\eqref{eq:LDP} would preclude the $\rho$-QRs used in the proofs of Theorems~\ref{thm:nonasymp_bnd} and~\ref{thm:main-achiev} (with zeros in the associated stochastic matrices). In effect, a characterization of distribution privacy in~\eqref{eq:dist-rho-priv} under the additional requirement that the $\rho$-QR be locally differentially private of a given level, is open.



\appendices
\section{Proofs of Lemmas~\ref{lem:local_unif} and~\ref{lem:div_var-dis}}
\label{app:tech_lemmas}
\noindent {\bf Proof of Lemma~\ref{lem:local_unif}}:

Assume first that
\[0<\beta_j< 1 \ \ \text{and} \ \ 0<P(A_j)<1, \ \ j\in\cZ.\]
\noindent Then
\begin{align*}
    D(P||Q)&=\sum\limits_{j\in\cZ} \ \sum\limits_{x\in A_j} \ P(x)\log\frac{P(x)}{\frac{\beta_j}{|A_j|}}\\
   &=\sum\limits_{j\in\cZ} \ \sum\limits_{x\in A_j} \ P(x)\log\frac{P(x)}{\beta_j}+ \sum\limits_{j\in\cZ}P(A_j) \log |A_j|\\
   &=\sum\limits_{j\in\cZ} \ \sum\limits_{x\in A_j} \ P(x)\log\frac{P(x)}{\beta_j}\frac{P(A_j)}{P(A_j)}+ \sum\limits_{j\in\cZ}P(A_j) \log |A_j|\\
   &=\sum\limits_{j\in\cZ} \ \sum\limits_{x\in A_j} \ P(x)\log\frac{P(x)}{P(A_j)}+
   \sum\limits_{j\in\cZ} \ P(A_j)\log\frac{P(A_j)}{\beta_j}+\sum\limits_{j\in\cZ}P(A_j) \log |A_j|\\
 &= -\left[H(P)-H\left(P(\mathcal{A})\right)\right]+  D\left(P(\mathcal{A})\big|\big|\underbar{$\beta$}\right)
    +\sum_{j\in\cZ} \ P(A_j)\log |A_j|\\
    &\leq D\left(P(\mathcal{A})\big|\big|\underbar{$\beta$}\right)
    +\sum_{j\in\cZ} \ P(A_j)\log |A_j|.
\end{align*}
\noindent Equality holds above iff $H\left(P(\mathcal{A})\right)= H(P)$ which is tantamount to~\eqref{eq:k-point}.

Next, if $P(A_j)=0$ for some $j\in\cZ$, the claim holds upon replacing $\sum\limits_{j\in\cZ}$ throughout by
$\sum\limits_{j\in\cZ: P(A_j)\neq 0}$. Finally, if $\beta_j=0$ for some $j\in\cZ$, the only nontrivial case (when the right-side does not become $\infty$) is when $\beta_j=0$ implies $P(A_j)=0$; then the claim holds with the same replacement as above. \qeed

\vspace{0.4cm}
\noindent {\bf Proof of Lemma~\ref{lem:div_var-dis}}:

We have
\begin{align*}
    D\left(P||Q\right)-D\left(P||Q_o\right)&=\sum\limits_{x\in support \ \left(Q_o\right)} \ P(x) \ \log\frac{Q(x)}{Q_o(x)}\\
    &\leq\sum\limits_{x\in support \ \left(Q_o\right)} \ P(x) \ \left(\frac{Q(x)}{Q_o(x)}-1\right)\\
    &=\sum\limits_{x\in support \ \left(Q_o\right)} \ P(x) \ \left(\frac{Q(x)-Q_o(x)}{Q_o(x)}\right).
\end{align*}
Hence,
\begin{align*}
\left|    D\left(P||Q\right)-D\left(P||Q_o\right)\right|&\leq\sum\limits_{x\in support \ \left(Q_o\right)} \ \frac{P(x)}{Q_o(x)} \ \left|Q(x)-Q_o(x)\right|\\
    &\leq \ \frac{1}{Q_o^{min}}\ \text{\normalfont{var}}\left(Q,Q_o\right)
\end{align*}
so that
\[
-\frac{\text{\normalfont{var}}\left(Q,Q_o\right)}{Q_o^{min}}\leq
D\left(P||Q\right)-D\left(P||Q_o\right)\leq
\frac{\text{\normalfont{var}}\left(Q,Q_o\right)}{Q_o^{min}}
\]
\noindent whence
\[
D\left(P||Q\right)\geq D\left(P||Q_o\right)-\frac{\text{\normalfont{var}}\left(Q,Q_o\right)}{Q_o^{min}}.
\]
\qeed

\section{Proof of~\eqref{eq:conv_final}}
\label{app:gamma_pf}

To establish~\eqref{eq:conv_final}, we write for any $\epsilon>0$,
\begin{align}
    &\sup_{V\in\cV(\rho)} \ \sup_{\underline{\alpha}\in\Delta_k(V)} \ \mathbb{E}\left[\left|\Phi_n^j\left(\underline{\alpha},V\right)\right|\right]\nonumber\\&\leq \sup_{V\in\cV(\rho)} \ \sup_{\underline{\alpha}\in\Delta_k(V)} \ \mathbb{E}\left[\left|\Phi_n^j\left(\underline{\alpha},V\right)\right|\mathbbm{1}\left(\left|\Phi_n^j\left(\underline{\alpha},V\right)\right|>\epsilon\right)\right]+ \sup_{V\in\cV(\rho)} \ \sup_{\underline{\alpha}\in\Delta_k(V)} \ \mathbb{E}\left[\left|\Phi_n^j\left(\underline{\alpha},V\right)\right|\mathbbm{1}\left(\left|\Phi_n^j\left(\underline{\alpha},V\right)\right|\leq\epsilon\right)\right]\nonumber\\
    &\leq \sup_{V\in\cV(\rho)} \ \sup_{\underline{\alpha}\in\Delta_k(V)} \ \mathbb{E}\left[\left|\Phi_n^j\left(\underline{\alpha},V\right)\right|\mathbbm{1}\left(\left|\Phi_n^j\left(\underline{\alpha},V\right)\right|>\epsilon\right)\right]+ \epsilon\nonumber\\
        &\leq \sup_{V\in\cV(\rho)} \ \sup_{\underline{\alpha}\in\Delta_k(V)} \ \log(n+k) P\left(\left|\Phi_n^j\left(\underline{\alpha},V\right)\right|>\epsilon\right)+ \epsilon\label{eq:conv_proof_eps}
\end{align}
\noindent since 
\begin{align*}
\left|\Phi_n^j\left(\underline{\alpha},V\right)\right|&\leq \left|\left(\underbar{$\alpha$}V^{-1}\right)(j)\log\frac{\left(\underbar{$\alpha$}V^{-1}\right)(j)}{\kappa_n(T_n)(j)}\right|\\
&\leq 1. \log\frac{1}{1/(n+k)}=\log (n+k), \ \ \text{by~\eqref{eq:kappa}.}
\end{align*}
\noindent Since $\epsilon>0$ is arbitrary,~\eqref{eq:conv_final} will follow from~\eqref{eq:conv_proof_eps} if for any $\epsilon>0,$
\begin{equation*}
    \lim_n \  \log(n+k)  \sup_{V\in\cV(\rho)} \ \sup_{\underline{\alpha}\in\Delta_k(V)} \  P\left(\left|\Phi_n^j\left(\underline{\alpha},V\right)\right|>\epsilon\right) = 0.
\end{equation*}
\noindent For $V\in\cV(\rho)$, let $\nu_j$ denote the $j^{th}$ column of $V^{-1}, \ j\in\cZ.$ Noting that $\left(\underbar{$\alpha$}V^{-1}\right)(j)=\underbar{$\alpha$}\nu_j,$ we write
\begin{align}
&\log(n+k)  \sup_{V\in\cV(\rho)} \ \sup_{\underline{\alpha}\in\Delta_k(V)} \     P\left(\big|\Phi_n^j\left(\underbar{$\alpha$},V\right)\big|>\epsilon\right)\nonumber\\&=\log(n+k) \sup_{V\in\cV(\rho)} \ \sup_{\underline{\alpha}\in\Delta_k(V)} \ P\left(\left|\underbar{$\alpha$}\nu_j\log\frac{\underbar{$\alpha$}\nu_j}{\kappa_n(T_n)(j)}\right|>\epsilon\right)\nonumber\\
    &\leq \log(n+k) \sup_{V\in\cV(\rho)} \ \sup_{\underline{\alpha}\in\Delta_k(V)} \ P\left(\underbar{$\alpha$}\nu_j\log\frac{\underbar{$\alpha$}\nu_j}{\kappa_n(T_n)(j)}>\epsilon\right)\nonumber\\&\hspace{7cm}+ \log(n+k) 
    \sup_{V\in\cV(\rho)} \ \sup_{\underline{\alpha}\in\Delta_k(V)} \ P\left(-\underbar{$\alpha$}\nu_j\log\frac{\underbar{$\alpha$}\nu_j}{\kappa_n(T_n)(j)}>\epsilon\right).\label{eq:mod_split}
\end{align}
\noindent Considering the first term in the right-side above and noting by~\eqref{eq:kappa} that $\kappa_n(T_n)(j)\geq1/(n+k),$ clearly the probability equals $0$ if $\epsilon\geq \underbar{$\alpha$}\nu_j\log(n+k)$ so that it suffices to consider
\begin{equation}
\label{eq:eps_lb}
\underbar{$\alpha$}\nu_j\log(n+k)>\epsilon
\ \ \ \text{or} \ \ \ \underbar{$\alpha$}\nu_j>\frac{\epsilon}{\log(n+k)}.
\end{equation}
\noindent Then, in~\eqref{eq:mod_split},
\begin{align}
    P\left(\underbar{$\alpha$}\nu_j\log\frac{\underbar{$\alpha$}\nu_j}{\kappa_n(T_n)(j)}>\epsilon\right)&=P\left(\kappa_n(T_n)(j)<\underbar{$\alpha$}\nu_j2^{-\frac{\epsilon}{\underline{\alpha}\nu_j}}\right)\nonumber\\
    &\leq P\left(\kappa_n(T_n)(j)<\underbar{$\alpha$}\nu_j2^{-\epsilon}\right)\nonumber\\
    &\leq P\left(\widetilde{Q}(T_n)\nu_j<\underbar{$\alpha$}\nu_j2^{-\epsilon}\left(1+\frac{k}{n}\right)\right), \  \ \ \text{by~\eqref{eq:kappa}}\nonumber\\
    &\leq P\left(\left|\widetilde{Q}(T_n)\nu_j-\underline{\alpha}\nu_j\right|\geq\underline{\alpha}\nu_j\left(1-2^{-\epsilon}\left(1+\frac{k}{n}\right)\right)\right)\nonumber\\
    &= P\left(\left|\widetilde{Q}(T_n)\nu_j-\underline{\alpha}\nu_j\right|\geq\underline{\alpha}\nu_j\left(1-2^{-\epsilon}\left(1+\frac{k}{n}\right)\right)^+\right)\label{eq:prob_conv_init_pf}
\end{align}
\noindent where $x^+\triangleq\max\{0,x\}, \ x\in\mathbb{R}.$ Denote for $\underline{\alpha}\in\mathbb{R}^k, \ \lVert\underline{\alpha}\rVert_1\triangleq \sum\limits_{j=1}^k \ |\alpha_j|;$ and for $V\in\cV(\rho)$,  $\left\lVert V^{-1}\right\rVert_1\triangleq \max\limits_{1\leq j\leq k} \ \lVert \nu_j \rVert_1.$ Then, in~\eqref{eq:prob_conv_init_pf},
\begin{equation}
    \left|\widetilde{Q}(T_n)\nu_j-\underbar{$\alpha$}\nu_j\right|\leq \left\lVert \widetilde{Q}(T_n)-\underbar{$\alpha$} \right \rVert_1 \left\lVert \nu_j\right\rVert_1\leq \left\lVert \widetilde{Q}(T_n)-\underbar{$\alpha$} \right \rVert_1 \left\lVert V^{-1}\right\rVert_1 \leq \left\lVert \widetilde{Q}(T_n)-\underbar{$\alpha$} \right \rVert_1 \frac{k}{2\rho-1}\label{eq:onenorm_bound}
\end{equation}
\noindent since for a strictly diagonally-dominated $V\in\cV(\rho),$ we have by~{\cite[Theorem $1$]{Varah75}} and~{\cite[Section 3, Theorem $1$]{Mor08}} that $\left\lVert V^{-1}\right\rVert_1\leq k/(2\rho-1),$ noting that $\rho>0.5.$ Furthermore,
\begin{align}
    \left\lVert \widetilde{Q}(T_n)-\underbar{$\alpha$} \right \rVert_1
    &\leq \left\lVert \widetilde{Q}(T_n)-T_n \right \rVert_1+\left\lVert T_n-\underbar{$\alpha$} \right \rVert_1\nonumber\\
    &\leq \sqrt{2\ln 2}\left(\sqrt{  D\left(T_n\big|\big|\widetilde{Q}(T_n)\right)}+\sqrt{  D\left(T_n\big|\big|\alpha\right)}\right), \ \ \ \text{by~{\cite[Lemma $11.6.1$]{Cover06}}}\nonumber\\
    &\leq 2\sqrt{2\ln 2 \  D\left(T_n\big|\big|\alpha\right)}, \ \ \text{by~\eqref{eq:Reverse_Iproj}.}\label{eq:main_idea}
\end{align}
\noindent Hence, in~\eqref{eq:prob_conv_init_pf}, by~\eqref{eq:onenorm_bound},~\eqref{eq:main_idea},
\begin{align}
    P\left(\underbar{$\alpha$}\nu_j\log\frac{\underbar{$\alpha$}\nu_j}{\kappa_n(T_n)(j)}>\epsilon\right)&\leq
    P\left(\sqrt{D\left(T_n\big|\big| \underbar{$\alpha$}\right)}\geq
    \frac{\left(2\rho-1\right)}{2k\sqrt{2\ln 2}}\underbar{$\alpha$}\nu_j
    \left(1-2^{-\epsilon}\left(1+\frac{k}{n}\right)\right)^+
    \right)\nonumber\\
    &\leq
    P\left(D\left(T_n\big|\big| \underbar{$\alpha$}\right)\geq
    \frac{(2\rho-1)^2}{8k^2\ln 2}\frac{\epsilon^2}{\log^2(n+k)}
    \left(\left(1-2^{-\epsilon}\left(1+\frac{k}{n}\right)\right)^+\right)^2
    \right)\label{eq:div_eps}
\end{align}
using~\eqref{eq:eps_lb}. Denoting the threshold above by $\tau_1(\rho,\epsilon,n),$ straightforward manipulation yields that
\[
\tau_1(\rho,\epsilon,n)\geq\frac{ c_1(\rho,\epsilon)}{\log^2(n+k)}>0
\]
\noindent with
\[
c_1(\rho,\epsilon)=\frac{(2\rho-1)^2\epsilon^2 c^2}{8k^2 \ln 2}, \ \ \ \left(1-2^{-\epsilon}(1+k)\right)^+<c<1-2^{-\epsilon} 
\]
\noindent for all $n\geq N_1(\epsilon,k,c)=k/\left((1-c)2^{\epsilon}-1\right).$  Then, in~\eqref{eq:div_eps} for all $n\geq N_1(\epsilon,k,c)$, by~{\cite[Theorem $11.2.1$]{Cover06}}, the first term in the right-side of~\eqref{eq:mod_split} is
\begin{align}
\leq \log(n+k)    P\left(D\left(T_n\big|\big| \underbar{$\alpha$}\right)\geq\frac{c_1(\rho,\epsilon)}{\log^2(n+k)}\right)&\leq  \log(n+k)   \exp\left[-n\left(\frac{c_1(\rho,\epsilon)}{\log^2(n+k)}-k\frac{\log(n+1)}{n}\right)\right]\nonumber\\&\leq\exp\left[-\left(c_1(\rho,\epsilon)\frac{n}{\log^2(n+k)}-(k+1)\log(n+k)\right)\right]\label{eq:div_thre}
\end{align}
\noindent since $\log(n+k)\leq (n+k)$ whereby because \[\lim\limits_n \ \  c_1(\rho,\epsilon)\frac{n}{\log^2(n+k)}-(k+1)\log(n+k)=\infty,\]
\noindent we get
\begin{equation}
    \label{eq:prob_split_res1}
\lim_n \ \log(n+k) \sup_{V\in\cV(\rho)} \ \sup_{\underline{\alpha}\in\Delta_k(\rho)} \ P\left(\underbar{$\alpha$}\nu_j\log\frac{\underbar{$\alpha$}\nu_j}{\kappa_n(T_n)(j)}>\epsilon\right) = 0.
\end{equation}

Turning to the second term in the right-side of~\eqref{eq:mod_split}, and considering
\[
\sup_{V\in\cV(\rho)} \ \sup_{\underbar{$\alpha$}\in\Delta_k(V)} \ P\left(-\underbar{$\alpha$}\nu_j\log\frac{\underbar{$\alpha$}\nu_j}{\kappa_n(T_n)(j)}>\epsilon\right),
\]
\noindent note that the probability is $0$ for $\epsilon\geq -\underline{\alpha}\nu_j\log\underline{\alpha}\nu_j.$ Then, since $-\underline{\alpha}\nu_j\log\underline{\alpha}\nu_j\leq 0.5,$ considering 
\begin{equation}
\label{eq:eps_star}
-\underline{\alpha}\nu_j\log\underline{\alpha}\nu_j> \epsilon \ \ 
\text{for} \ \ 0<\epsilon<0.5,
\end{equation}
\noindent we get
\begin{flalign}
  & \hspace{0.2cm} P\left(-\underbar{$\alpha$}\nu_j\log\frac{\underbar{$\alpha$}\nu_j}{\kappa_n(T_n)(j)}>\epsilon\right)\leq 
     P\left(\kappa_n(T_n)(j)>\underbar{$\alpha$}\nu_j2^{\epsilon}\right)&\nonumber\\
    &=P\Bigg(\widetilde{Q}(T_n)\nu_j>
    \underbar{$\alpha$}\nu_j2^{\epsilon}\left(1+\frac{k}{n}\right)-\frac{1}{n}\Bigg)&\nonumber\\
    &\leq P\Bigg(\big|\widetilde{Q}(T_n)\nu_j-\underline{\alpha}\nu_j\big|\geq 
    \underbar{$\alpha$}\nu_j\left(2^{\epsilon}\left(1+\frac{k}{n}\right)-1\right)-\frac{1}{n}\Bigg)&\nonumber\\
    &\leq P\Bigg(\left\lVert\widetilde{Q}(T_n)-\underline{\alpha}\right\rVert_1\left\lVert V^{-1}\right
    \rVert_1\geq\underbar{$\alpha$}\nu_j\left(2^{\epsilon}\left(1+\frac{k}{n}\right)-1\right)-\frac{1}{n}\Bigg),  \ \ \ \text{by the first inequality in~\eqref{eq:onenorm_bound}}&\nonumber\\
    &    \leq P\Bigg(\left\lVert\widetilde{Q}(T_n)-\underline{\alpha}\right\rVert_1\geq\left(\underbar{$\alpha$}\nu_j\left(2^{\epsilon}\left(1+\frac{k}{n}\right)-1\right)-\frac{1}{n}\right)\frac{2\rho-1}{k}\Bigg), \ \ \  \text{by the last inequality in~\eqref{eq:onenorm_bound}}&\nonumber\\
        &\leq P\left(\sqrt{D\left(T_n\big|\big|\underline{\alpha}\right)}\geq \frac{\left(2\rho-1\right)}{2k\sqrt{2\ln 2}}\left(t^*(\epsilon)\left(2^{\epsilon}\left(1+\frac{k}{n}\right)-1\right)-\frac{1}{n}\right)^+\right),\ \ \ \text{by~\eqref{eq:main_idea} and~\eqref{eq:eps_star}}&\label{eq:prob_thr2}\\
           &= P\left(D\left(T_n\big|\big|\underline{\alpha}\right)\geq \frac{\left(2\rho-1\right)^2}{8k^2\ln 2}\left(\left(t^*(\epsilon)\left(2^{\epsilon}\left(1+\frac{k}{n}\right)-1\right)-\frac{1}{n}\right)^+\right)^2\right)&\label{eq:prob_thr3}
\end{flalign}
\noindent where $t^*(\epsilon)$ in~\eqref{eq:prob_thr2} is the solution of $-t\log t=\epsilon$ for $t\in[0,0.5).$ Denoting by $\tau_2\left(\rho,\epsilon,n\right)$ the threshold in~\eqref{eq:prob_thr3}, we observe that $
\tau_2(\rho,\epsilon,n)\geq c_2(\rho,\epsilon)>0$
\noindent with
\[
c_2(\rho,\epsilon)=\frac{(2\rho-1)^2 d^2}{8k^2 \ln 2}, \ \ \  0<d<t^*(\epsilon)\left(2^{\epsilon}-1\right) 
\]
\noindent for all $n\geq N_2(\epsilon,k,d)=\left(1-kt^*(\epsilon)2^{\epsilon}\right)^+/\left(t^*(\epsilon)\left(2^{\epsilon}-1\right)-d\right).$ Then, bounding above the second term in~\eqref{eq:mod_split} upon treating the probability in~\eqref{eq:prob_thr3} in the manner of~\eqref{eq:div_thre}, and observing that
 \[\lim\limits_n \ \  c_2(\rho,\epsilon)n-(k+1)\log(n+k)=\infty,\]
 \noindent we have
\begin{equation}
\label{eq:prob_split_res2}
\lim_n \ \log(n+k) \sup_{V\in\cV(\rho)} \ \sup_{\underline{\alpha}\in\Delta_k(V)} \ P\left(-\underbar{$\alpha$}\nu_j\log\frac{\underbar{$\alpha$}\nu_j}{\kappa_n(T_n)(j)}>\epsilon\right)=0.\end{equation}
\noindent Upon combining~\eqref{eq:prob_split_res1} and~\eqref{eq:prob_split_res2}, we get~\eqref{eq:conv_final}.

\section{Lower Bound for~\eqref{eq:unif_form},~\eqref{eq:V2lb} and Proof of~\eqref{eq:Q'n-pmf},~\eqref{eq:N0}}
\label{app:achiev_pf}

\noindent {\bf Lower bound for~\eqref{eq:unif_form}}:\vspace{0.1cm}\\

We first bound~\eqref{eq:unif_form} below by restricting the supremum further according to
\begin{align}
&\inf_{\widehat{P}_n^{\beta^{(n)}}\in\mathcal{S}_n(k')} \ \sup_{\substack{\underline{\alpha}\in\Delta_{k'}:\\ 
\underline{\alpha}(j')=0, \ j'=k'-l',\ldots,k'-1}} \nonumber\\ &\hspace{3.8cm}\sum_{j'=0}^{k'-l'-1} \ \frac{\underline{\alpha}(j')-l\left(\frac{1-l\rho}{k-l'-l}\right)}{l\rho-l\left(\frac{1-l\rho}{k-l'-l}\right)}\log \left(\frac{\underline{\alpha}(j')-l\left(\frac{1-l\rho}{k-l'-l}\right)}{l\rho-l\left(\frac{1-l\rho}{k-l'-l}\right)}\frac{\big|f'^{-1}(j')\big|}{\widehat{P}_n^{\beta^{(n)}}\left(Q'^{(n)}(\underline{\alpha})\right)\left(f'^{-1}(j')\right)} \right)
\nonumber\\ &\geq
\inf_{\widehat{P}_n^{\beta^{(n)}}\in\mathcal{S}_n(k')} \ \sup_{\substack{\underline{\alpha}\in\Delta_{k'}:\\
l\left(\frac{1-l\rho}{k-l'-l}\right)\leq \underline{\alpha}(j')\leq\min\left\{\frac{\left\lceil nl\left(\frac{1-l\rho}{k-l'-l}\right)\right\rceil}{n}, \ l\rho\right\}, \ j'=1,\ldots,k'-l'-1 \\
\underline{\alpha}(j')=0, \ j'=k'-l',\ldots,k'-1}}\nonumber \\ &\hspace{3.6cm}\sum_{j'=0}^{k'-l'-1} \ \frac{\underline{\alpha}(j')-l\left(\frac{1-l\rho}{k-l'-l}\right)}{l\rho-l\left(\frac{1-l\rho}{k-l'-l}\right)}\log \left(\frac{\underline{\alpha}(j')-l\left(\frac{1-l\rho}{k-l'-l}\right)}{l\rho-l\left(\frac{1-l\rho}{k-l'-l}\right)}\frac{\big|f'^{-1}(j')\big|}{\widehat{P}_n^{\beta^{(n)}}\left(Q'^{(n)}(\underline{\alpha})\right)\left(f'^{-1}(j')\right)} \right).\label{eq:fl_bound}
\end{align}
\noindent Observe from~\eqref{eq:qna} that $Q'^{(n)}(\underline{\alpha})\in\cQ'^{(n)}$ is the same for all $\underline{\alpha}\in\Delta_{k'}$ satisfying the constraints in the right-side of~\eqref{eq:fl_bound}. Hence, with
\[
\widehat{P}_n^{\beta^{(n)}}\left(Q'^{(n)}(\underline{\alpha})\right)\left(f'^{-1}(j')\right)=\beta_{j'},\ \ \ \ j'\in\cZ',
\]
\noindent for some $\underline{\beta}=\{\beta_0,\beta_1,\ldots,\beta_{k'-1}\}\in\Delta_{k'}$, the right-side of~\eqref{eq:fl_bound}
\begin{equation}
\label{eq:C2}
 \geq   \inf_{\underline{\beta}\in\Delta_{k'}} \ \sup_{\underline{\alpha}\in\Delta_{k'}\sim\text{right-side of }\eqref{eq:fl_bound}} \left\{ \sum_{j'=0}^{k'-l'-1} \ \frac{\underline{\alpha}(j')-l\left(\frac{1-l\rho}{k-l'-l}\right)}{l\rho-l\left(\frac{1-l\rho}{k-l'-l}\right)}\log \left(\frac{\underline{\alpha}(j')-l\left(\frac{1-l\rho}{k-l'-l}\right)}{l\rho-l\left(\frac{1-l\rho}{k-l'-l}\right)}\frac{\big|f'^{-1}(j')\big|}{\beta_{j'}} \right)\right\}.
\end{equation}
\noindent For each fixed $\underline{\beta}\in\Delta_{k'},$ we further bound below the expression in~\eqref{eq:C2} by limiting the supremum to a maximum over a finite set $\{\underline{\alpha}_t:t=1,\ldots,k'-l'\}$ made up of $k'-l'=\left\lfloor\frac{k}{l}\right\rfloor\geq 2$  elements (see~\eqref{eq:k/l-lb},~\eqref{eq:red_set}) specified by
\begin{equation}
\label{eq:alpha-1}
\underline{\alpha}_1=\left(l\rho,l\left(\frac{1-l\rho}{k-l'-l}\right),\ldots,l\left(\frac{1-l\rho}{k-l'-l}\right),
0,\ldots,0
\right)
\end{equation}
\noindent with $k'-l'$ nonzero elements; and for $t=2,\ldots,k'-l',$ $\underline{\alpha}_t=\left(\underline{\alpha}_t(0),\underline{\alpha}_t(1),\ldots,\underline{\alpha}_t(k'-1)\right)$ specified by
\begin{equation}
\label{eq:alpha-others}
\underline{\alpha}_t(j')=
\begin{cases}
l\rho-\left(\min\left\{\frac{\left\lceil nl\left(\frac{1-l\rho}{k-l'-l}\right)\right\rceil}{n},\ l\rho\right\}-l\left(\frac{1-l\rho}{k-l'-l}\right)\right), \ \ &j'=0\\
l\left(\frac{1-l\rho}{k-l'-l}\right), \ \ &j'=1,\ldots,t-2,t,\ldots,k'-l'-1\\
\min\left\{\frac{\left\lceil nl\left(\frac{1-l\rho}{k-l'-l}\right)\right\rceil}{n},\ l\rho\right\}, \ \ &j'=t-1\\
0,\ \ &j'\geq k'-l'.
\end{cases}
\end{equation}
\noindent A description of the $k'-l'-1$ values above is as follows: For $t=2,\ldots,k'-l', \ \underline{\alpha}_{t}(t-1)=\min\left\{\frac{\left\lceil nl\left(\frac{1-l\rho}{k-l'-l}\right)\right\rceil}{n},\ l\rho\right\},$ $ \underline{\alpha}_t(1)=\cdots=\underline{\alpha}_t(t-2)=\underline{\alpha}_t(t)=\cdots=\underline{\alpha}_t(k'-l'-1)=l\left(\frac{1-l\rho}{k-l'-l}\right)$ and $\underline{\alpha}_t(j')=0, \ j'\geq k'-l'.$ We now check that when $\underline{\alpha}$ is from the set $\{\underline{\alpha}_t:t=1,\ldots,k'-l'\},$ the coefficients of the log terms in~\eqref{eq:C2}, i.e., the right-side of~\eqref{eq:PXV'_condn1} under~\eqref{eq:alpha-1},~\eqref{eq:alpha-others}, take values in $[0,1].$ For this, it suffices to verify that $l\left(\frac{1-l\rho}{k-l'-l}\right)\leq\underline{\alpha}_t(0)\leq l\rho, \ t=2,\ldots,k'-l'.$ Observe that
\begin{align}
    l\left(\frac{1-l\rho}{k-l'-l}\right)&\leq l\frac{(1-l\rho)}{l}, \ \ \ \text{since from~\eqref{eq:k/l-lb}, $k-l'-l\geq l$}\nonumber\\
   & \leq l\rho, \ \ \ \text{using $l\rho\geq l/(l+1)\geq 1/2$ from~\eqref{eq:rho_bd},}\label{eq:lrho-inequality}
\end{align}
\noindent  so that $\min\left\{\frac{\left\lceil nl\left(\frac{1-l\rho}{k-l'-l}\right)\right\rceil}{n},l\rho\right\}-l\left(\frac{1-l\rho}{k-l'-l}\right)\geq 0.$ Therefore,
\[\underline{\alpha}_t(0)= l\rho-\left(\min\left\{\frac{\left\lceil nl\left(\frac{1-l\rho}{k-l'-l}\right)\right\rceil}{n},l\rho\right\}-l\left(\frac{1-l\rho}{k-l'-l}\right)\right) \leq l\rho, \ \ t=2,\ldots,k'-l',\]
and 
\begin{multline*}\underline{\alpha}_t(0)= l\rho-\left(\min\left\{\frac{\left\lceil nl\left(\frac{1-l\rho}{k-l'-l}\right)\right\rceil}{n},l\rho\right\}-l\left(\frac{1-l\rho}{k-l'-l}\right)\right)\\ \geq l\rho-\left(l\rho-l\left(\frac{1-l\rho}{k-l'-l}\right)\right)= l\left( \frac{1-l\rho}{k-l'-l}\right), \ \  t=2,\ldots,k'-l'.
\end{multline*}
\noindent This completes the verification.

Now set
\begin{equation}
\label{eq:mu-n}
\mu_n\triangleq \frac{\min\left\{\frac{\left\lceil nl\left(\frac{1-l\rho}{k-l'-l}\right)\right\rceil}{n},\ l\rho\right\}-l\left(\frac{1-l\rho}{k-l'-l}\right)}{l\rho-l\left(\frac{1-l\rho}{k-l'-l}\right)}.
\end{equation}
\noindent Then the supremum in~\eqref{eq:C2} reduces to a maximum over $k'-l'$ choices of $\underline{\alpha}_t, \ t=1,\ldots,k'-l',$ given by~\eqref{eq:alpha-1},~\eqref{eq:alpha-others}. Then, with $\mu_n$ as in~\eqref{eq:mu-n}, the right-side of~\eqref{eq:C2} is bounded below as
\begin{multline}
\label{eq:sup-max}
    \geq \inf_{\underline{\beta}\in\Delta_{k'}} \ \max\Bigg\{\log\frac{\big|f'^{-1}(0)\big|}{\beta_0},\mu_n\log\left(\mu_n\frac{\big|f'^{-1}(1)\big|}{\beta_1}\right)+(1-\mu_n)\log\left((1-\mu_n)\frac{\big|f'^{-1}(0)\big|}{\beta_0}\right),\\ \ldots,\mu_n\log\left(\mu_n\frac{\big|f'^{-1}(k'-l'-1)\big|}{\beta_{k'-l'-1}}\right)+(1-\mu_n)\log\left((1-\mu_n)\frac{\big|f'^{-1}(0)\big|}{\beta_0}\right)\Bigg\}
   \end{multline}
   \noindent where the $t$th term, $1\leq t\leq k'-l'$ is obtained by evaluating the expression within $\{\cdot\}$ in~\eqref{eq:C2} at $\underline{\alpha}_t.$ Moreover, the right-side of~\eqref{eq:sup-max} is
   \begin{equation}
   \label{eq:sup-maxred}
   =\inf_{\underline{\beta}\in\Delta_{k'}} \ \max\left\{\log\frac{\big|f'^{-1}(0)\big|}{\beta_0}, \max_{1\leq j'\leq k'-l'-1} \ \left\{\mu_n\log\left(\mu_n\frac{\big|f'^{-1}(j')\big|}{\beta_{j'}}\right)+(1-\mu_n)\log\left((1-\mu_n)\frac{\big|f'^{-1}(0)\big|}{\beta_0}\right)\right\}\right\}.
\end{equation}
\noindent Hence, for the case $l\leq \left\lfloor \frac{k}{2}\right\rfloor,$ the user-selected pmf $P_X$ is one among a set of $k'-l'$  $k'$-sparse pmfs with associated mapping $f'$ \eqref{eq:f'_map} and $P_X\left(f'^{-1}\right)=\underline{\alpha}_t, \ t=1,\ldots,k'-l'.$

Since for each $\underline{\beta}\in\Delta_{k'}, \ \beta_{k'-l'},\ldots,\beta_{k'-1}$ do not appear in~\eqref{eq:sup-maxred}, it is sufficient to consider $\underline{\beta}\in\Delta_{k'}$ such that
\begin{equation}
    \label{eq:beta0}
    \beta_{k'-l'}=\cdots=\beta_{k'-1}=0.
\end{equation}
\noindent For every $0\leq\beta_0\leq 1,$ we solve the inner $\inf \ \max$ in~\eqref{eq:sup-maxred}, which, using~\eqref{eq:beta0}, is equivalent to
\begin{equation}
\label{eq:innermax}
    \inf_{\substack{\beta_1,\ldots,\beta_{k'-l'-1}:\\\beta_{j'}\geq 0, \ j'=1,\ldots,k'-l'-1\\
    \sum\limits_{j'=1}^{k'-l'-1}\beta_{j'}=1-\beta_0}\\} \ \max_{1\leq j'\leq k'-l'-1} \ \frac{\big|f'^{-1}(j')\big|}{\beta_{j'}}.
\end{equation}
\noindent We claim that
\begin{equation}
\label{eq:innermax_lb}
\max_{1\leq j'\leq k'-l'-1} \ \frac{\big|f'^{-1}(j')\big|}{\beta_{j'}} \geq \frac{\sum\limits_{j''=1}^{k'-l'-1}\big|f'^{-1}(j'')\big|}{1-\beta_0}
\end{equation}
\noindent since, if
\begin{align*}
  \max_{1\leq j'\leq k'-l'-1} \  \frac{\big|f'^{-1}(j')\big|}{\beta_{j'}} &< \frac{\sum\limits_{j''=1}^{k'-l'-1}\big|f'^{-1}(j'')\big|}{1-\beta_0}\\
    \text{i.e., }\frac{\big|f'^{-1}(j')\big|}{\beta_{j'}} &< \frac{\sum\limits_{j''=1}^{k'-l'-1}\big|f'^{-1}(j'')\big|}{1-\beta_0}, \ \ \  j'=1,\ldots,k'-l'-1\\
    \big|f'^{-1}(j')\big| &< \frac{\beta_{j'}}{1-\beta_0}\sum\limits_{j''=1}^{k'-l'-1}\big|f'^{-1}(j'')\big|,\ \ \ j'=1,\ldots,k'-l'-1,
\end{align*}
\noindent and by summing over $j'=1,\ldots,k'-l'-1$ on both sides
\[    \sum_{j'=1}^{k'-l'-1}\big|f'^{-1}(j')\big| < \frac{\sum\limits_{j'=1}^{k'-l'-1}\beta_{j'}}{1-\beta_0}\sum\limits_{j''=1}^{k'-l'-1}\big|f'^{-1}(j'')\big|=\sum\limits_{j''=1}^{k'-l'-1}\big|f'^{-1}(j'')\big|
\]
\noindent which is a contradiction. Also, by choosing
\begin{equation*}
\beta_{j'}=\frac{\big|f'^{-1}(j')\big|}{\sum\limits_{j''=1}^{k'-l'-1}\big|f'^{-1}(j'')\big|
}(1-\beta_0), \ \ \ j'=1,\ldots,k'-l'-1,
\end{equation*}
\noindent we get
\begin{equation}
\label{eq:innermax_ub}
     \inf_{\substack{\beta_1,\ldots,\beta_{k'-l'-1}:\\\beta_{j'}\geq 0, \ j'=1,\ldots,k'-l'-1\\
    \sum\limits_{j'=1}^{k'-l'-1}\beta_{j'}=1-\beta_0}\\} \ \max_{1\leq j'\leq k'-l'-1} \ \frac{\big|f'^{-1}(j')\big|}{\beta_{j'}}\leq 
    \frac{\sum\limits_{j''=1}^{k'-l'-1}\big|f'^{-1}(j'')\big|}{1-\beta_0}.
\end{equation}
\noindent Thus, the expression in~\eqref{eq:innermax} equals the (common) right-sides of~\eqref{eq:innermax_lb},~\eqref{eq:innermax_ub},  using which the lower bound in~\eqref{eq:sup-maxred} becomes
\begin{equation}\label{eq:max-2red}
    \inf_{0\leq\beta_0\leq 1} \ \max\left\{
    \log\frac{\big|f'^{-1}(0)\big|}{\beta_0}, \mu_n\log\left(\mu_n\frac{\sum\limits_{j'=1}^{k'-l'-1}\big|f'^{-1}(j')\big|}{1-\beta_{0}}\right)+(1-\mu_n)\log\left((1-\mu_n)\frac{\big|f'^{-1}(0)\big|}{\beta_0}\right)\right\}.
\end{equation}
\noindent The first term in~\eqref{eq:max-2red} is decreasing in $0\leq\beta_0\leq 1$ and the second term is convex in $0\leq\beta_0\leq 1$ since it can be written as
\begin{equation}
\label{eq:div_convex}
    D\left(\text{Ber}(\mu_n)||\text{Ber}(1-\beta_0)\right)+\mu_n\log\sum\limits_{j'=1}^{k'-l'-1}\big|f'^{-1}(j')\big|+(1-\mu_n)\log\big|f'^{-1}(0)\big|
\end{equation}
\noindent with the minimum being attained at $\beta_0=\beta_0'\triangleq 1-\mu_n.$ Straightforward but tedious calculations show that the terms intersect exactly once
at
\[
\beta_0=\beta_0''\triangleq\left(1+\frac{\sum\limits_{j'=1}^{k'-l'-1}\big|f'^{-1}(j')\big|}{\big|f'^{-1}(0)\big|}(1-\mu_n)^{\frac{1-\mu_n}{\mu_n}}\mu_n\right)^{-1},
\]
\noindent and\footnotemark\footnotetext{When $\mu_n=0, \ \beta_0''=1.$} the first term is larger than the second when $\beta_0<\beta_0''$ and smaller when $\beta_0>\beta_0''.$ We now distinguish between the cases $\big|f'^{-1}(0)\big|\geq\sum\limits_{j'=1}^{k'-l'-1}\big|f'^{-1}(j')\big|$ and $\big|f'^{-1}(0)\big|<\sum\limits_{j'=1}^{k'-l'-1}\big|f'^{-1}(j')\big|.$\\
(i) $\big|f'^{-1}(0)\big|\geq\sum\limits_{j'=1}^{k'-l'-1}\big|f'^{-1}(j')\big|$: It holds that
\[
\beta_0''\geq\beta_0'
\]
\noindent i.e.,
\[
\left(1+\frac{\sum\limits_{j'=1}^{k'-l'-1}\big|f'^{-1}(j')\big|}{\big|f'^{-1}(0)\big|}(1-\mu_n)^{\frac{1-\mu_n}{\mu_n}}\mu_n\right)^{-1}\geq 1-\mu_n
\]
\noindent which is
\[
\big|f'^{-1}(0)\big|\geq\sum\limits_{j'=1}^{k'-l'-1}\big|f'^{-1}(j')\big|(1-\mu_n)^{\frac{1}{\mu_n}}
\]
\noindent because $(1-\mu_n)^{\frac{1}{\mu_n}}\leq 1$ and $\big|f'^{-1}(0)\big|\geq\sum\limits_{j'=1}^{k'-l'-1}\big|f'^{-1}(j')\big|.$ Then $\inf\limits_{0\leq\beta_0\leq 1}$ in~\eqref{eq:max-2red} is attained as a minimum at $\beta_0=\beta_0'',$ and becomes
\begin{equation}
\label{eq:i}
    \log \big|f'^{-1}(0)\big| + \log\left(1+\frac{\sum\limits_{j'=1}^{k'-l'-1}\big|f'^{-1}(j')\big|}{\big|f'^{-1}(0)\big|}(1-\mu_n)^{\frac{1-\mu_n}{\mu_n}}\mu_n\right),
\end{equation}
\noindent with $\mu_n$ as in~\eqref{eq:mu-n}.\\
(ii) $\big|f'^{-1}(0)\big|<\sum\limits_{j'=1}^{k'-l'-1}\big|f'^{-1}(j')\big|:$ We have that~\eqref{eq:max-2red} is bounded below as
\begin{align}
    &\geq \inf_{0\leq\beta_0\leq 1} \  D\left(\text{Ber}(1-\mu_n)||\text{Ber}(\beta_0)\right)+\mu_n\log\sum\limits_{j'=1}^{k'-l'-1}\big|f'^{-1}(j')\big|+(1-\mu_n)\log\big|f'^{-1}(0)\big|\nonumber\\
   &= \log\big|f'^{-1}(0)\big|+\mu_n\log\frac{\sum\limits_{j'=1}^{k'-l'-1}\big|f'^{-1}(j')\big|}{\big|f'^{-1}(0)\big|}\label{eq:ii}.
\end{align}
\noindent Note that the second $\log$ term in~\eqref{eq:ii} is positive.

From~\eqref{eq:i} along with the observation that $(1-\mu_n)^{\frac{1-\mu_n}{\mu_n}}\geq 1/e$ and~\eqref{eq:ii}, we obtain the desired lower bound for~\eqref{eq:unif_form} which along with~\eqref{eq:pi_n_hoeffding} and~\eqref{eq:mu-n} gives~\eqref{eq:1},~\eqref{eq:2},~\eqref{eq:3}.
\vspace{0.1cm}\\ \\
{\bf Lower bound for~\eqref{eq:V2lb}}:\vspace{0.1cm}\\ \\
Turning next to the task of bounding~\eqref{eq:V2lb} below, we have
\begin{align}
&\inf_{\widehat{P}_n^{\beta^{(n)}}\in\mathcal{S}_n(k')} \              \sup_{\substack{\underline{\alpha}\in\Delta_{k'}:\\ \underline{\alpha}(j')=0, \ j'=2,\ldots,k'-1}} \ \frac{\underline{\alpha}(0)}{l\rho}\log \left(\frac{\underline{\alpha}(0)}{l\rho}\frac{\big|f'^{-1}(0)\big|}{ \widehat{P}^{\beta^{(n)}}_n\left(Q'^{(n)}(\underline{\alpha})\right)\left(f'^{-1}(0)\right)} \right)\nonumber
 \\
&\hspace{5cm}+ \frac{\underline{\alpha}(1)-(1-l\rho)}{l\rho}\log \left(\frac{\underline{\alpha}(1)-(1-l\rho)}{l\rho}\frac{\big|f'^{-1}(1)\big|}{ \widehat{P}^{\beta^{(n)}}_n\left(Q'^{(n)}(\underline{\alpha})\right)\left(f'^{-1}(1)\right)} \right)\nonumber\\&\geq
\inf_{\widehat{P}_n^{\beta^{(n)}}\in\mathcal{S}_n(k')} \              \sup_{\substack{\underline{\alpha}\in\Delta_{k'}:\\
        1-l\rho\leq\underline{\alpha}(1)\leq\frac{\left\lceil n(1-l\rho)\right\rceil}{n}\nonumber\\
        \underline{\alpha}(j')=0, \ j'=2,\ldots,k'-1}} \ \frac{\underline{\alpha}(0)}{l\rho}\log \left(\frac{\underline{\alpha}(0)}{l\rho}\frac{\big|f'^{-1}(0)\big|}{ \widehat{P}^{\beta^{(n)}}_n\left(Q'^{(n)}(\underline{\alpha})\right)\left(f'^{-1}(0)\right)} \right)
\nonumber \\
&\hspace{5.5cm}+ \frac{\underline{\alpha}(1)-(1-l\rho)}{l\rho}\log \left(\frac{\underline{\alpha}(1)-(1-l\rho)}{l\rho}\frac{\big|f'^{-1}(1)\big|}{ \widehat{P}^{\beta^{(n)}}_n\left(Q'^{(n)}(\underline{\alpha})\right)\left(f'^{-1}(1)\right)} \right).\label{eq:fl_bound1}
\end{align}
\noindent In this case, too, observe from~\eqref{eq:qna} that $Q'^{(n)}(\underline{\alpha})\in\cQ'^{(n)}$ remains unchanged for all $\underline{\alpha}\in\Delta_{k'}$ satisfying the constraints in the right-side of~\eqref{eq:fl_bound1}. Hence, with
\[
\widehat{P}_n^{\beta^{(n)}}\left(Q'^{(n)}(\underline{\alpha})\right)\left(f'^{-1}(j')\right)=\beta_{j'},\ \ \ \ j'\in\cZ',
\]
\noindent for some $\underline{\beta}=\{\beta_0,\beta_1,\ldots,\beta_{k'-1}\}\in\Delta_{k'}$, the right-side of~\eqref{eq:fl_bound1}
\begin{multline}
\label{eq:C3}
 \geq  \inf_{\underline{\beta}\in\Delta_{k'}} \              \sup_{\substack{\underline{\alpha}\in\Delta_{k'}:\\
        1-l\rho\leq\underline{\alpha}(1)\leq\frac{\left\lceil n(1-l\rho)\right\rceil}{n}\\
        \underline{\alpha}(j')=0, \ j'=2,\ldots,k'-1}} \ \Bigg\{\frac{\underline{\alpha}(0)}{l\rho}\log \left(\frac{\underline{\alpha}(0)}{l\rho}\frac{\big|f'^{-1}(0)\big|}{\beta_0} \right)
\\
+ \frac{\underline{\alpha}(1)-(1-l\rho)}{l\rho}\log \left(\frac{\underline{\alpha}(1)-(1-l\rho)}{l\rho}\frac{\big|f'^{-1}(1)\big|}{\beta_1} \right)\Bigg\}.
\end{multline}
\noindent We further bound~\eqref{eq:C3} below by replacing the supremum by a maximum of the expression within $\{\cdot\}$ in~\eqref{eq:C3} evaluated at two points $\underline{\alpha}_1,\underline{\alpha}_2$ given by
\begin{align}
    \underline{\alpha}_1&=\left(l\rho,1-l\rho,0,\ldots,0\right)\nonumber\\
    \underline{\alpha}_2&=\left(1-\frac{\left\lceil n(1-l\rho)\right\rceil}{n},\frac{\left\lceil n(1-l\rho)\right\rceil}{n},0,\ldots,0\right). \label{eq:alpha1-2}
\end{align}
\noindent When $\underline{\alpha}=\underline{\alpha}_1,$ the coefficients of the $\log$ terms in~\eqref{eq:C3} are in $[0,1];$ and when $\underline{\alpha}=\underline{\alpha}_2,$ the same holds upon observing that  $0\leq\underline{\alpha}_2(0)\leq l\rho.$  Then with
\begin{equation}
\label{eq:theta-n}
\theta_n=\frac{\frac{\left\lceil n(1-l\rho)\right\rceil}{n}-(1-l\rho)}{l\rho},
\end{equation}
\noindent the expression in~\eqref{eq:C3} is bounded below by
\begin{equation}
\label{eq:sup-maxred1}
\inf_{\underline{\beta}\in\Delta_{k'}} \ \max \ \left\{
\log\frac{\big|f'^{-1}(0)\big|}{\beta_0},\theta_n\log\left(\theta_n\frac{\big|f'^{-1}(1)\big|}{\beta_{1}}\right)+(1-\theta_n)\log\left((1-\theta_n)\frac{\big|f'^{-1}(0)\big|}{\beta_0}\right)
\right\},
\end{equation}
\noindent where the two terms within $\{\cdot\}$ in~\eqref{eq:sup-maxred1} are obtained by evaluating the term within $\{\cdot\}$ in~\eqref{eq:C3} at $\underline{\alpha}_1$ and $\underline{\alpha}_2,$ respectively. Hence, for the case $l> \left\lfloor \frac{k}{2}\right\rfloor,$ the user-selected pmf $P_X$ is one among two  $k'$-sparse pmfs with associated mapping $f'$ \eqref{eq:f'_map} and $P_X\left(f'^{-1}\right)=\underline{\alpha}_t, \ t=1,2.$

For every $\underline{\beta}\in\Delta_{k'}, \ \beta_{2},\ldots,\beta_{k'-1}$ do not appear in~\eqref{eq:sup-maxred1}, so that it  suffices to consider $\underline{\beta}\in\Delta_{k'}$ that satisfies
\[
    \beta_{2}=\cdots=\beta_{k'-1}=0
\]
\noindent whereby~\eqref{eq:sup-maxred1} becomes
\begin{equation}
\label{eq:sup-maxred2}
\inf_{0\leq\beta_0\leq 1} \ \max \ \left\{\log
\frac{\big|f'^{-1}(0)\big|}{\beta_0},\theta_n\log\left(\theta_n\frac{\big|f'^{-1}(1)\big|}{1-\beta_{0}}\right)+(1-\theta_n)\log\left((1-\theta_n)\frac{\big|f'^{-1}(0)\big|}{\beta_0}\right)
\right\}.
\end{equation}
\noindent Observe that~\eqref{eq:sup-maxred2} is the same as~\eqref{eq:max-2red} with $\mu_n$ replaced by $\theta_n$ and $\sum\limits_{j'=1}^{k'-l'-1}\big|f'^{-1}(j')\big|$ by $\big|f'^{-1}(1)\big|.$ Since, by the assumption in~\eqref{eq:priv_nonasymp_lbb}, $\big|f'^{-1}(0)\big|\geq\big|f'^{-1}(1)\big|$, applying the same steps from~\eqref{eq:max-2red} - \eqref{eq:i} for the case $l\leq\left\lfloor\frac{k}{2}\right\rfloor$ and $\big|f'^{-1}(0)\big|\geq \sum\limits_{j'=1}^{k'-l'-1}\big|f'^{-1}(j')\big|,$ we get that~\eqref{eq:sup-maxred2} equals
\begin{equation}
\label{eq:iL}
    \log \big|f'^{-1}(0)\big| + \log\left(1+\frac{\big|f'^{-1}(1)\big|}{\big|f'^{-1}(0)\big|}(1-\theta_n)^{\frac{1-\theta_n}{\theta_n}}\theta_n\right).
\end{equation}
From\footnotemark\footnotetext{When $\theta_n=0,$ the expression in~\eqref{eq:iL} reduces to $\log\big|f^{-1}(0)\big|.$}~\eqref{eq:iL} and the fact that $(1-\theta_n)^{\frac{1-\theta_n}{\theta_n}}\geq 1/e,$ we get the desired lower bound for~\eqref{eq:V2lb} which along with~\eqref{eq:pi_n_hoeffding1} and~\eqref{eq:theta-n} gives~\eqref{eq:1},~\eqref{eq:22},~\eqref{eq:33}.\vspace{0.2cm}\\ \\

\noindent {\bf Proof of~\eqref{eq:Q'n-pmf},~\eqref{eq:N0}}:\\

To show\eqref{eq:Q'n-pmf},~\eqref{eq:N0} for the case $l\leq\left\lfloor\frac{k}{2}\right\rfloor,$ we have  from~\eqref{eq:alpha-1},~\eqref{eq:alpha-others} that
\begin{align*}
\min_{t=1,\ldots,k'-l'} \ \underline{\alpha}_t(0)&= l\rho-\left(\min\left\{\frac{\left\lceil nl\left(\frac{1-l\rho}{k-l'-l}\right)\right\rceil}{n},\ l\rho\right\}-l\left(\frac{1-l\rho}{k-l'-l}\right)\right)\\&\geq l\rho-\left(\frac{\left\lceil nl\left(\frac{1-l\rho}{k-l'-l}\right)\right\rceil}{n}-l\left(\frac{1-l\rho}{k-l'-l}\right)\right)\\&\geq l\rho-\frac{1}{n}\geq\frac{1}{2}-\frac{1}{n} \ \ \ \text{since from~\eqref{eq:rho_bd}, $l\rho\geq\frac{l}{l+1}\geq\frac{1}{2},$}
\end{align*}
\noindent so that the last inequality in~\eqref{eq:Q'n-pmf} holds if 
\begin{equation}
\label{eq:last}
\frac{1}{2}-\frac{1}{n}-\frac{k'-1}{n}=\frac{1}{2}-\frac{k'}{n}\geq 0 \ \ \ \text{i.e.,} \ \ \ n\geq N_0(k')= 2k'.
\end{equation}
\noindent Next, for the case $l>\left\lfloor\frac{k}{2}\right\rfloor,$ from~\eqref{eq:alpha1-2},
\[
    \min_{t=1,2} \ \underline{\alpha}_t(0)=1-\frac{\left\lceil n(1-l\rho)\right\rceil}{n}\geq 1-\frac{ n(1-l\rho)+1}{n}=l\rho-\frac{1}{n}\geq\frac{1}{2}-\frac{1}{n}
\]
\noindent and, in this case too,~\eqref{eq:last} holds.
\qeed

\section{Proof of~\eqref{eq:asymp_rho0.5}}
\label{app:asymp-rho0.5}

First, choosing $g:\Delta_k\rightarrow\Delta_r$ given by
\[
g\left(P_X\left(f^{-1}\right)\right)(x)=\frac{P_X\left(f^{-1}(j)\right)}{\big|f^{-1}(j)\big|}, \ \ x\in f^{-1}(j), \ \ j\in\cZ,
\]
\noindent the left-side of~\eqref{eq:asymp_rho0.5} is maximized by $P_X$ being a point-mass on any $x\in f^{-1}(0).$ Then,~\eqref{eq:asymp_rho0.5} holds with ``$\leq.$''

Next, the reverse inequality ``$\geq$'' in~\eqref{eq:asymp_rho0.5} obtains from mimicking the steps in~\eqref{eq:lemm-final-eqn} with $l=1.$\qeed

\section*{Acknowledgments}
The authors are grateful to: Peter Kairouz, Himanshu Tyagi and Shun Watanabe for their helpful critique of our problem formulation; Lorenzo Finesso for his informative pointers that led to Remark (ii) after Definition 6; and the anonymous referees and associate editor for their thoughtful comments which led to material improvements in presentation.

\end{document}

%% file: preamble.tex
\newcommand{\cQ}{{\mathcal Q}}

\newcommand{\cT}{{\mathcal T}}

\newcommand{\cV}{{\mathcal V}}

\newcommand{\cX}{{\mathcal X}}

\newcommand{\cZ}{{\mathcal Z}}

\usepackage{color}

\theoremstyle{remark}
\newtheorem*{remark*}{Remark}
\newtheorem*{remarks*}{Remarks}

